\documentclass[letterpaper]{IEEEtran}

\newcommand{\RCO}{R_{\op{CO}}}

\newcommand{\RS}{R_{\op{S}}}
\newcommand{\CS}{C_{\op{S}}}
\newcommand{\NR}{\op{NR}}
\newcommand{\CW}{C_{\op{W}}}

\title{On the Optimality of Secret Key Agreement\\ via Omniscience}

\author{Chung Chan, Manuj Mukherjee, Navin Kashyap and Qiaoqiao Zhou
	\thanks{Parts of this work were presented at the 2016 IEEE International Symposium on Information Theory (ISIT 2016), Barcelona, Spain, and at the 2016 IEEE Information Theory Workshop (ITW 2016), Cambridge, UK.}
	\thanks{C.\ Chan (email: chung.chan@cityu.edu.hk) is with the Department of Computer Science, City University of Hong Kong. His work was supported by a grant from the University Grants Committee of the Hong Kong Special Administrative Region, China (Project No. 14200714).}
	\thanks{Q.\ Zhou is with the Department of Information Engineering and the Institute of Network Coding, the
          Chinese University of Hong Kong. His work was supported by a grant from the University Grants Committee of Hong Kong Special Administrative Region, China (Project No. AoE/E-02/08).}
        \thanks{N.\ Kashyap (nkashyap@iisc.ac.in) and M.\ Mukherjee (manuj@iisc.ac.in) are with the Department of Electrical Communication Engineering, Indian Institute of Science, Bangalore 560012. Their work was supported in part by a Swarnajayanti Fellowship awarded to N.\ Kashyap by the Department of Science \& Technology, Government of India.}}

\usepackage[numbers,sort&compress]{natbib}

\makeatletter

\usepackage{etex}

\usepackage{zref-savepos}

\newcounter{mnote}%[page]

\def\xmarginnote{%
  \xymarginnote{\hskip -\marginparsep \hskip -\marginparwidth}}

\def\ymarginnote{%
  \xymarginnote{\hskip\columnwidth \hskip\marginparsep}}

\long\def\xymarginnote#1#2{%
\vadjust{#1%
\smash{\hbox{{%
        \hsize\marginparwidth
        \@parboxrestore
        \@marginparreset
\footnotesize #2}}}}}

\def\mnoteson{%
\gdef\mnote##1{\refstepcounter{mnote}\label{##1}%
  \zsavepos{##1}%
  \ifnum20432158>\number\zposx{##1}%
  \xmarginnote{{\color{blue}\bf $\langle$\arabic{mnote}$\rangle$}}% 
  \else
  \ymarginnote{{\color{blue}\bf $\langle$\arabic{mnote}$\rangle$}}%
  \fi%
}
  }
\gdef\mnotesoff{\gdef\mnote##1{}}
\mnoteson
\mnotesoff

\usepackage{framed}
\usepackage{subcaption}
\usepackage{comment}
\usepackage[svgnames,dvipsnames]{xcolor}
\usepackage[all]{xy}
\usepackage{tikz}
\usetikzlibrary{positioning,matrix,through,calc,arrows,fit,shapes,decorations.pathreplacing,decorations.markings,}

\tikzstyle{block} = [draw,fill=blue!20,minimum size=2em]

\usepackage{qsymbols,amssymb,mathrsfs}
\usepackage{amsmath}
\usepackage[standard,thmmarks]{ntheorem}
\theoremstyle{plain}
\theoremsymbol{\ensuremath{_\vartriangleleft}}
\theorembodyfont{\itshape}
\theoremheaderfont{\normalfont\bfseries}
\theoremseparator{}

\theoremstyle{nonumberplain}
\theoremheaderfont{\scshape}
\theorembodyfont{\normalfont}
\theoremsymbol{\ensuremath{_\blacktriangleleft}}

\theoremnumbering{arabic}
\theoremstyle{plain}
\usepackage{latexsym}
\theoremsymbol{\ensuremath{_\Box}}
\theorembodyfont{\itshape}
\theoremheaderfont{\normalfont\bfseries}
\theoremseparator{}
\newtheorem{Conjecture}{Conjecture}

\theorembodyfont{\upshape}
\theoremprework{\bigskip\hrule}
\theorempostwork{\hrule\bigskip}
\usepackage[overload]{empheq} % no \intertext and \displaybreak

\let\iftwocolumn\if@twocolumn
\g@addto@macro\@twocolumntrue{\let\iftwocolumn\if@twocolumn}
\g@addto@macro\@twocolumnfalse{\let\iftwocolumn\if@twocolumn}

\mathtoolsset{showonlyrefs=false,showmanualtags}
\let\underbrace\LaTeXunderbrace % adapt spacing to font sizes
\let\overbrace\LaTeXoverbrace
\renewcommand{\eqref}[1]{\textup{(\refeq{#1})}} % eqref was not allowed in
\newtagform{brackets}[]{(}{)}   % new tags for equations
\usetagform{brackets}
\usepackage[Smaller]{cancel}

\PassOptionsToPackage{breaklinks,letterpaper,hyperindex=true,backref=false,bookmarksnumbered,bookmarksopen,linktocpage,colorlinks,linkcolor=BrickRed,citecolor=OliveGreen,urlcolor=Blue,pdfstartview=FitH}{hyperref}
\usepackage{hyperref}

\usepackage{graphicx,psfrag}
\graphicspath{{figure/}{image/}} % Search path of figures

\usepackage{diagbox} % \backslashbox{}{} for slashed entries
\usepackage{algorithm2e}
\usepackage{listings} 
\lstdefinelanguage{Maple}{
  morekeywords={proc,module,end, for,from,to,by,while,in,do,od
    ,if,elif,else,then,fi ,use,try,catch,finally}, sensitive,
  morecomment=[l]\#,
  morestring=[b]",morestring=[b]`}[keywords,comments,strings]
\DeclareMathAlphabet{\mathpzc}{OT1}{pzc}{m}{it}
\usepackage{upgreek} % \upalpha,\upbeta, ...
\usepackage{dsfont}  % \mathds

\def\multi@nostar#1#2{%
  \expandafter\def\csname multi#1\endcsname##1{%
    \if ##1.\let\next=\relax \else
    \def\next{\csname multi#1\endcsname}     
    \expandafter\newcommand\csname #1##1\endcsname{#2}
    \fi\next}}

\def\multi@star#1#2{%
  \expandafter\def\csname #1\endcsname##1{#2}
  \multi@nostar{#1}{#2}
}

\newcommand{\multi}{%
  \@ifstar \multi@star \multi@nostar}

\multi*{rm}{\mathrm{#1}}
\multi*{mc}{\mathcal{#1}}
\multi*{op}{\mathop {\operator@font #1}}
\multi*{ds}{\mathds{#1}}
\multi*{set}{\mathcal{#1}}
\multi*{rsfs}{\mathscr{#1}}
\multi*{pz}{\mathpzc{#1}}
\multi*{M}{\boldsymbol{#1}}
\multi*{R}{\mathsf{#1}}
\multi*{RM}{\M{\R{#1}}}
\multi*{bb}{\mathbb{#1}}
\multi*{td}{\tilde{#1}}
\multi*{tR}{\tilde{\mathsf{#1}}}
\multi*{trM}{\tilde{\M{\R{#1}}}}
\multi*{tset}{\tilde{\mathcal{#1}}}
\multi*{tM}{\tilde{\M{#1}}}
\multi*{baM}{\bar{\M{#1}}}
\multi*{ol}{\overline{#1}}

\multirm  ABCDEFGHIJKLMNOPQRSTUVWXYZabcdefghijklmnopqrstuvwxyz.
\multiol  ABCDEFGHIJKLMNOPQRSTUVWXYZabcdefghijklmnopqrstuvwxyz.
\multitR   ABCDEFGHIJKLMNOPQRSTUVWXYZabcdefghijklmnopqrstuvwxyz.
\multitd   ABCDEFGHIJKLMNOPQRSTUVWXYZabcdefghijklmnopqrstuvwxyz.
\multitset ABCDEFGHIJKLMNOPQRSTUVWXYZabcdefghijklmnopqrstuvwxyz.
\multitM   ABCDEFGHIJKLMNOPQRSTUVWXYZabcdefghijklmnopqrstuvwxyz.
\multibaM   ABCDEFGHIJKLMNOPQRSTUVWXYZabcdefghijklmnopqrstuvwxyz.
\multitrM   ABCDEFGHIJKLMNOPQRSTUVWXYZabcdefghijklmnopqrstuvwxyz.
\multimc   ABCDEFGHIJKLMNOPQRSTUVWXYZabcdefghijklmnopqrstuvwxyz.
\multiop   ABCDEFGHIJKLMNOPQRSTUVWXYZabcdefghijklmnopqrstuvwxyz.
\multids   ABCDEFGHIJKLMNOPQRSTUVWXYZabcdefghijklmnopqrstuvwxyz.
\multiset  ABCDEFGHIJKLMNOPQRSTUVWXYZabcdefghijklmnopqrstuvwxyz.
\multirsfs ABCDEFGHIJKLMNOPQRSTUVWXYZabcdefghijklmnopqrstuvwxyz.
\multipz   ABCDEFGHIJKLMNOPQRSTUVWXYZabcdefghijklmnopqrstuvwxyz.
\multiM    ABCDEFGHIJKLMNOPQRSTUVWXYZabcdefghijklmnopqrstuvwxyz.
\multiR    ABCDEFGHIJKL NO QR TUVWXYZabcd fghijklmnopqrstuvwxyz.
\multibb   ABCDEFGHIJKLMNOPQRSTUVWXYZabcdefghijklmnopqrstuvwxyz.
\multiRM   ABCDEFGHIJKLMNOPQRSTUVWXYZabcdefghijklmnopqrstuvwxyz.

\newcommand{\dotleq}{\buildrel \textstyle  .\over {\smash{\lower
      .2ex\hbox{\ensuremath\leqslant}}\vphantom{=}}}
\newcommand{\dotgeq}{\buildrel \textstyle  .\over {\smash{\lower
      .2ex\hbox{\ensuremath\geqslant}}\vphantom{=}}}

\newcommand{\bM}{\begin{bmatrix}}
\newcommand{\eM}{\end{bmatrix}}
\newcommand{\bSM}{\left[\begin{smallmatrix}}
\newcommand{\eSM}{\end{smallmatrix}\right]}
\renewcommand*\env@matrix[1][*\c@MaxMatrixCols c]{%
  \hskip -\arraycolsep
  \let\@ifnextchar\new@ifnextchar
  \array{#1}}

\newqsymbol{`N}{\mathbb{N}}
\newqsymbol{`R}{\mathbb{R}}
\newqsymbol{`P}{\mathbb{P}}
\newqsymbol{`Z}{\mathbb{Z}}

\newqsymbol{`|}{\mid}
\newqsymbol{`8}{\infty}
\newqsymbol{`1}{\left}
\newqsymbol{`2}{\right}
\newqsymbol{`6}{\partial}
\newqsymbol{`0}{\emptyset}
\newqsymbol{`-}{\leftrightarrow}
\newqsymbol{`<}{\langle}
\newqsymbol{`>}{\rangle}

\DeclarePairedDelimiter\abs{\lvert}{\rvert}

\DeclarePairedDelimiter\Set{\{}{\}}
\newcommand{\imod}[1]{\allowbreak\mkern10mu({\operator@font mod}\,\,#1)}

\newcommand{\threecols}[3]{
\hbox to \textwidth{%
      \normalfont\rlap{\parbox[b]{\textwidth}{\raggedright#1\strut}}%
        \hss\parbox[b]{\textwidth}{\centering#2\strut}\hss
        \llap{\parbox[b]{\textwidth}{\raggedleft#3\strut}}%
    }% hbox 
}

\newcommand{\reason}[2][\relax]{
  \ifthenelse{\equal{#1}{\relax}}{
    \left(\text{#2}\right)
  }{
    \left(\parbox{#1}{\raggedright #2}\right)
  }
}

\newcommand{\utag}[2]{\mathop{#2}\limits^{\text{(#1)}}}
\newcommand{\uref}[1]{(#1)}

\let\SavedDoubleVert\relax
\let\protect\relax
{\catcode`\|=\active
  \xdef\extendvert{\protect\expandafter\noexpand\csname extendvert \endcsname}
  \expandafter\gdef\csname extendvert \endcsname#1{\mskip-5mu \left.%
      \ifx\SavedDoubleVert\relax \let\SavedDoubleVert\|\fi
     \:{\let\|\SetDoubleVert
       \mathcode`\|32768\let|\SetVert
     #1}\:\right.\mskip-5mu}
}
\def\SetVert{\@ifnextchar|{\|\@gobble}% turn || into \|
    {\egroup\;\mid@vertical\;\bgroup}}
\def\SetDoubleVert{\egroup\;\mid@dblvertical\;\bgroup}

\begingroup
 \edef\@tempa{\meaning\middle}
 \edef\@tempb{\string\middle}
\expandafter \endgroup \ifx\@tempa\@tempb
 \def\mid@vertical{\middle|}
 \def\mid@dblvertical{\middle\SavedDoubleVert}
\else
 \def\mid@vertical{\mskip1mu\vrule\mskip1mu}
 \def\mid@dblvertical{\mskip1mu\vrule\mskip2.5mu\vrule\mskip1mu}
\fi

\makeatother

\usepackage{ctable}
\usepackage{fouridx}
\usepackage{framed}
\usetikzlibrary{positioning,matrix}

\usepackage{paralist}
\usepackage{enumerate}

\usepackage[normalem]{ulem}

\numberwithin{equation}{section}
\makeatletter
\@addtoreset{equation}{section}
\renewcommand{\theequation}{\arabic{section}.\arabic{equation}}
\@addtoreset{Theorem}{section}
\renewcommand{\theTheorem}{\arabic{section}.\arabic{Theorem}}
\@addtoreset{Lemma}{section}
\renewcommand{\theLemma}{\arabic{section}.\arabic{Lemma}}
\@addtoreset{Corollary}{section}
\renewcommand{\theCorollary}{\arabic{section}.\arabic{Corollary}}
\@addtoreset{Example}{section}
\renewcommand{\theExample}{\arabic{section}.\arabic{Example}}
\@addtoreset{Remark}{section}
\renewcommand{\theRemark}{\arabic{section}.\arabic{Remark}}
\@addtoreset{Proposition}{section}
\renewcommand{\theProposition}{\arabic{section}.\arabic{Proposition}}
\@addtoreset{Definition}{section}
\renewcommand{\theDefinition}{\arabic{section}.\arabic{Definition}}
\@addtoreset{Claim}{section}
\renewcommand{\theClaim}{\arabic{section}.\arabic{Claim}}
\@addtoreset{Subclaim}{Theorem}
\renewcommand{\theSubclaim}{\theTheorem\Alph{Subclaim}}
\makeatother

{\endMakeFramed}

\newenvironment{ybox}{
	\setlength{\FrameSep}{1.5mm}
	\setlength{\FrameRule}{0mm}
  \MakeFramed {\FrameRestore}}%
{\endMakeFramed}

\newenvironment{gbox}{
	\setlength{\FrameSep}{1.5mm}
\setlength{\FrameRule}{0mm}
  \MakeFramed {\FrameRestore}}%
{\endMakeFramed}

\newenvironment{bbox}{
	\setlength{\FrameSep}{1.5mm}
\setlength{\FrameRule}{0mm}
  \MakeFramed {\FrameRestore}}%
{\endMakeFramed}

 {\endMakeFramed}

\usepackage{enumitem}

\usepackage{etoolbox}

\let\theparentequation\theequation
\patchcmd{\theparentequation}{equation}{parentequation}{}{}

\renewenvironment{subequations}[1][]{%              optional argument: label-name for (first) parent equation
	\refstepcounter{equation}%
	\setcounter{parentequation}{\value{equation}}%    parentequation = equation
	\setcounter{equation}{0}%                         (sub)equation  = 0
	\def\theequation{\theparentequation\alph{equation}}% 
	\let\parentlabel\label%                           Evade sanitation performed by amsmath
	\ifx\\#1\\\relax\else\label{#1}\fi%               #1 given: \label{#1}, otherwise: nothing
	\ignorespaces
}{%
	\setcounter{equation}{\value{parentequation}}%    equation = subequation
	\ignorespacesafterend
}

\newcommand*{\nextParentEquation}[1][]{%            optional argument: label-name for (first) parent equation
	\refstepcounter{parentequation}%                  parentequation++
	\setcounter{equation}{0}%                         equation = 0
	\ifx\\#1\\\relax\else\parentlabel{#1}\fi%         #1 given: \label{#1}, otherwise: nothing
}

\begin{document}

%\IEEEoverridecommandlockouts
%\nocite{add}
\maketitle
\begin{abstract}
  For the multiterminal secret key agreement problem under a private source model, it is known that the maximum key rate, i.e., the secrecy capacity, can be achieved through communication for omniscience, but the omniscience strategy can be strictly suboptimal in terms of minimizing the public discussion rate. While a single-letter characterization is not known for the minimum discussion rate needed for achieving the secrecy capacity, we derive single-letter lower and upper bounds that yield some simple conditions for omniscience to be discussion-rate optimal. These conditions turn out to be enough to deduce the optimality of omniscience for a large class of sources including the hypergraphical sources. Through conjectures and examples, we explore other source models to which our methods do not easily extend.
\end{abstract} 


\section{Introduction}
\label{sec:introduction}

We consider the secret key agreement problem of \cite{csiszar04}, possibly with trusted and untrusted helpers, as well as silent users as in \cite{amin10a}. Two or more users want to agree on a secret key after observing some discrete memoryless correlated private sources that take values from finite alphabet sets. The users are allowed to discuss (possibly interactively) with other users publicly over a noiseless authenticated broadcast channel. After the discussion, each active user (who is not a helper) attempts to compute a common secret key that is asymptotically uniformly random and independent of the public discussion as well as the private sources of the untrusted helpers. The maximum achievable key rate is called the \emph{secrecy capacity} $\CS$, and the minimum public discussion rate required to achieve the capacity is called the \emph{communication complexity} $\RS$. While $\CS$ was characterized in \cite{csiszar04}, a single-letter characterization for $\RS$ remains open, and is the main focus of this work.\mnote{a:test}

For the general source model with possibly trusted helpers, it was shown in~\cite{csiszar04} that $\RS$ can be upper bounded by the smallest rate $\RCO$ of communication for omniscience (CO), the state where every active user can asymptotically recover the entire private source. More precisely, the proposed capacity-achieving scheme is through omniscience, i.e., by having users communicate in public until every user recovers the entire private source and then extract a common secret key as a function of the recovered source that is asymptotically independent of the public discussion. While this omniscience strategy was shown to be capacity-achieving, it was also pointed out in \cite{csiszar04} to be suboptimal in the sense that strict inequality $\RS<\RCO$ is possible. 

For the general source model with two users but no helpers, there is a multi-letter characterization of $\RS$ in \cite{tyagi13}, and an example was also given where non-interactive discussion, i.e., the usual independent source coding scheme over a source network~\cite{csiszar2011information}, was shown to be suboptimal. When the number of discussion rounds is bounded, their characterization becomes a single-letter expression. \cite{MKS16} extended the framework of \cite{tyagi13} to the multiterminal case and obtained a  lower bound of $\RS$. The lower bound is a multi-letter even when the number of rounds is bounded. A special hypergraphical private source model~\cite{chan10md} was also considered in \cite{courtade16} in the multi-user case but without helpers, and $\RS$ was characterized when the discussion is non-asymptotic and restricted to be linear functions over a finite field. However, the expression was NP-hard to compute, and it was shown to be a loose upper bound for $\RS$ in the asymptotic model~\cite{courtade16}.

While a single-letter characterization remains unknown even for the two-user case, simpler questions about the communication complexity may be asked. 
In the no-helper case, \cite{mukherjee14} considered the refined condition of \emph{omnivocality}, which is the scenario when every user must discuss at strictly positive rate to achieve the secrecy capacity. The result was further refined by \cite{zhang15} to a set of vocality conditions that describes whether a particular user needs to discuss at strictly positive rate to achieve the capacity. These conditions were  conjectured to be necessary and sufficient, but the conjectures turn out to be easy to resolve~(see \cite{chan15mi,chan15so,chan16so}) using 
\begin{compactenum}
\item the characterization of the secrecy capacity in \cite{amin10a} in the no-helper case under the additional vocality constraints that a given proper subset of the users, called the \emph{silent users}, are not allowed to discuss, and
\item the properties of the multivariate mutual information (MMI)~\cite{chan15mi} that was shown in~\cite{chan10md,chan10phd} to be equal to the secrecy capacity in the no-helper case.
\end{compactenum}

%\subsection{Contributions}

In this work, we consider a different question that turns out to be easier to address than the problem of characterizing $\RS$: 
% appears more challenging but also easier to solve than characterizing $\RS$: 
When is omniscience optimal for achieving secrecy capacity, i.e., when is $\RS=\RCO$?
This question was raised in \cite{mukherjee15} in the no-helper case, and a sufficient condition for the optimality of omniscience was given in the special case of the pairwise independent network (PIN) model defined in \cite{nitinawarat-ye10,nitinawarat10}. The sufficient condition was later shown to be necessary in \cite{MKS16}. % using the idea of the decremental secret key agreement in \cite{chan16isit}. 
However, the result does not apply to more general source models beyond PIN, such as the hypergraphical model. Moreover, the problem formulation in \cite{MKS16} precludes additional randomization in the public discussion; it was conjectured (but not proved) there that randomization does not affect $\RS$. In this work, we overcome the above weaknesses and the following are the contributions:
\begin{ybox}
  \begin{enumerate}
  \item Derive single-letter lower and upper bounds for a general source model possibly with helpers and silent users, and with private randomization allowed.
  \item Obtain easily computable sufficient as well as necessary conditions for the optimality of omniscience.
  \item Discover more scenarios beyond PIN for which $\RS$ can be characterized by $\RCO$.
  \item Give concrete examples where the sufficient/necessary conditions can fail to be necessary/sufficient respectively, which may inspire further improvement on the bounds.
  \end{enumerate} 
\end{ybox}
% words on organization and notations
The results in the no-helper case will be stated more meaningfully using the MMI in~\cite{chan15mi} that extends Shannon's mutual information to the multivariate case. $\RS$ can be viewed as a measure of discord of the mutual information, and the public discussion viewed as an irreversible process of making the mutual information among the users less and less discordant until a consensus is achieved wherein the mutual information among the users is consolidated as a common secret key without further discussion.

\begin{bbox}
The paper is organized as follows: 
\begin{compactitem}
	\item The main ideas of the paper are motivated in Section~\ref{sec:motivation} with some simple examples. Some background knowledge in secret key agreement is assumed.
	\item Section~\ref{sec:problem} formulates the problem by introducing \begin{inparaenum} \item the secret key agreement problem with different types of users in Section~\ref{sec:RS}, and \item the capacity-achieving omniscience strategy in Section~\ref{sec:OO}. \end{inparaenum} 
	\item For ease of understanding, the main results are introduced in two stages. The basic scenario with no helpers or silent users is first tackled in Section~\ref{sec:nohelper}, where the fundamental proof techniques can be conveyed without much notational complexity.
	\item In the second stage, the proof techniques are extended to the general scenarios with helpers and silent users. We first derive single-letter upper bounds on the communication complexity in Section~\ref{sec:upperbound}, which follows directly from the achievability result of the omniscience strategy in Section~\ref{sec:UB_RCO} or indirectly by a change of scenario in Section~\ref{sec:UB_scenario_change}.
	\item Single-letter lower bounds for the general scenario are derived in Section~\ref{sec:lowerbound}. We extend the proof techniques in an information-theoretically meaningful manner, by introducing in Section~\ref{sec:I`l} some properties of a fractional partition information measure useful for proving converse results. The general lower bound is then derived in Section~\ref{sec:LB} using the converse proof techniques. The tightness of the bound is investigated in Section~\ref{sec:helpers}, \ref{sec:silent}, \ref{sec:hypsilent} and \ref{sec:untrusted_users}, where the general lower bound is specialized and strengthened to different forms under different scenarios and for the hypergraphical source model.
	\item Section~\ref{sec:challenge} explain the challenges that remain. The current techniques was shown to be limited for a non-hypergraphical source in Section~\ref{sec:limitation}, resolving the conjecture in \cite{chan16itw}. Potential improvements of the results are conjectured and illustrated in Section~\ref{sec:LB:scenario_change}.
	%\item Section~\ref{sec:conclusion} summarizes the current approach and mentions alternative approaches that can lead to stronger results in some cases.
\end{compactitem}
Proofs of the results are included in the appendices.
\end{bbox}
%and identify easily computable conditions and meaningful examples for the optimality of omniscience.  

\section{Motivation}
\label{sec:motivation}

The purpose of this section is to present some simple motivating examples. It is assumed that the reader is familiar with the basic problem of multiterminal secret key agreement, as introduced in \cite{csiszar04}.

We first introduce the idea of secret key agreement informally by the following example where omniscience is strictly suboptimal $\RS<\RCO$.

\begin{Example}
  \label{eg:XJ}
  Let $\RX_0,\RX_1$ and $\RJ$ be uniformly random and independent bits. Suppose users~$1$ and $2$ observe the private sources
  \begin{subequations}
    \label{eq:XJ}
    \begin{align*}
      \RZ_1 &:=(\RX_0,\RX_1)\kern1em \text{and} \\
      \RZ_2 &:=(\RX_\RJ,\RJ)
    \end{align*}
  \end{subequations}
  respectively, where $\RX_\RJ$ is equal to $\RX_0$ if $\RJ=0$, and equal to $\RX_1$ otherwise.
  A secret key agreement scheme with block length $n=1$ is to have
  \begin{align*}
    \RF&:=\RF_2=\RJ\kern1em \text{and}\\
    \RK&:=\RX_{\RJ},
  \end{align*}	
  i.e., have user~$2$ reveal $\RJ$ in public so that both users can compute and use $\RX_\RJ$ as the secret key, which can be shown to be independent of $\RF$ as desired. This is capacity-achieving because the secrecy capacity in the two-user case is the mutual information~\cite{csiszar04}
  \begin{align*}
    \CS&=I(\RZ_1\wedge\RZ_2)=1\kern1em
  \end{align*}
  and so the communication complexity $\RS$ is at most $H(\RJ)=1$. Note that omniscience has not been attained because $H(\RZ_1|\RZ_2)>0$ (and so user~$2$ cannot recover $\RZ_{1-\RJ}$ unless user~$1$ also communicates). More precisely, from \cite{csiszar04}, the minimum rate of communication for omniscience is 
  \begin{align*}
    \RCO=H(\RZ_1|\RZ_2)+H(\RZ_2|\RZ_1)=2>1\geq \RS.
  \end{align*}
  %It can also be shown that $\RS>0$ because $J_{\op{GK}}=0<\CS$.
  In particular, to achieve omniscience, user $1$ needs to discuss at rate at least $H(\RZ_1|\RZ_2)$ while user $2$ needs to discuss at rate at least $H(\RZ_2|\RZ_1)$, hence the $\RCO$ formula above.
\end{Example}

$\RS$ is difficult to compute even for the above example. Nevertheless, there is a simple condition for  omniscience to be optimal in the general two-user case, which is obvious from~\cite{ahlswede93,maurer93,tyagi13}:
\begin{Proposition}
  \label{pro:2user}
  For the two-user case, $\RS=\RCO$ iff $\RCO=0$, i.e., $H(\RZ_1|\RZ_2)=H(\RZ_2|\RZ_1)=0$ where $\RZ_i$ is the private source observed by user~$i\in \Set{1,2}$.
\end{Proposition}

\begin{Proof}
  The ``if" case is trivial and follows from the bound  $\RS\leq \RCO$. To prove the ``only if" case, note that the capacity-achieving scheme of \cite{ahlswede93,maurer93} has a discussion rate of $\min\Set{H(\RZ_1|\RZ_2),H(\RZ_2|\RZ_1)}\in [\RS,\RCO]$.  $\RS=\RCO$ implies that the minimum is $\RCO=H(\RZ_1|\RZ_2)+H(\RZ_2|\RZ_1)$~\cite{csiszar04}, which happens iff $H(\RZ_1|\RZ_2)=H(\RZ_2|\RZ_1)=0$,
  or equivalently, $\RCO=0$.
\end{Proof}

One of our goals is to extend the above condition to the multiterminal case to discover new scenarios where omniscience is optimal: % such as the following example. 

\begin{Example}
  \label{eg:interf}
  Suppose user~$3$ observes the private source 
  \begin{align}
    \RZ_3:=\RZ_1\oplus \RZ_2,\label{eq:interf}
  \end{align} 
  which is the XOR of two uniformly random and independent bits $\RZ_1$ and $\RZ_2$ observed by users~$1$ and $2$ respectively.
  In the no-helper case, a secret key agreement scheme is to have each user $i\in \Set{1,2,3}$ observe $n=2$ i.i.d.\ samples, $\RZ_{i1}$ and $\RZ_{i2}$, of its private source, and then choose
  \begin{align*}
    \RF&:=(\RF_1,\RF_2,\RF_3)=(\RZ_{11}\oplus \RZ_{12},\RZ_{22},\RZ_{31})\kern1em \text {and}\\
    \RK&:=\RZ_{11}.
  \end{align*}	
  It can be shown that $\RK$ is independent of $(\RF_1,\RF_2,\RF_3)$ and therefore secure. User~$1$ can recover the key trivially, while users~$2$ and $3$ can recover it from their observations and the public discussion by computing respectively
  \begin{align*}
    \RF_3\oplus \RZ_{21}&=\RK\kern1em \text {and}\\
    \RF_1\oplus \RF_2\oplus \RZ_{32}&=\RK
  \end{align*}	
  by \eqref{eq:interf}. This is capacity-achieving because the secrecy capacity is upper bounded by~\cite[(26)]{csiszar04} as 
  \begin{align*}
    \CS\leq \frac12 `1[\sum_{i=1}^3H(\RZ_i)-H(\RZ_1,\RZ_2,\RZ_3)`2]=\frac12,
  \end{align*}
  which is achieved by the current scheme. Omniscience is also attained because $H(\RK,\RF)=4$, which is the randomness of the entire source sequence $(\RZ_1^n,\RZ_2^n,\RZ_3^n)$. Since every user can observe $\RF$ and recover $\RK$, they can also recover the entire source sequence.
\end{Example}

The above example belongs to a more general finite linear source model~\cite{chan10phd} instead of the PIN or hypergraphical source model considered in the existing works of \cite{MKS16,mukherjee16}. Our result will imply $\RS=\RCO$ for this example. %, even with a trusted or untrusted helper.

\section{Problem Formulation}
\label{sec:problem}

While the no-helper case provides much intuition into the problem of communication complexity, we will consider the more general scenario with helpers and silent users, which unveils new challenges and inspires new techniques. More precisely, we will extend the secret key agreement protocol of \cite{csiszar04} without silent users and that of \cite{amin10a} without helpers to study the problem of communication complexity in the general case with both helpers and silent users. It will be seen that the secret key agreement scheme via omniscience from \cite{csiszar04} needs to be modified, in particular, to minimize the discussion of the untrusted users, and to incorporate silent users as in \cite{amin10a}. 

\subsection{Communication Complexity}
\label{sec:RS}

The following specifies all the user sets involved in the secret key agreement problem:
\begin{ybox}
  \noindent\underline{User sets}\\[-.8em]
  \begin{compactdesc}
  \item[$V$:] The ordered finite set of all users, where $\abs{V}\geq 2$. Unless stated otherwise, we assume $V=[\abs{V}]$ where 
    \begin{align}
      [m] &:=\Set{1,\dots,m} \label{eq:[m]}
    \end{align}
    for any positive integer $m\geq 2$.
  \item [$A\subseteq V$:] The subset of $\abs{A}\geq 2$ users, called the active users (who want to share a common secret key among themselves). $V`/A$ is called the set of helpers (who help the active users share the secret key).
  \item [$D\subseteq V`/A$:] The subset of untrusted helpers (whose observations are wiretapped). The subset\footnotemark $V`/A`/D$ consists of the trusted helpers.
  \item [$S\subseteq A\cup D$:] The subset of silent users (who cannot speak in public). $V`/S$ consists of the vocal users. Without loss of generality, we assume $V`/S:=[\abs {V`/S}]$ unless stated otherwise.
  \end{compactdesc}
\end{ybox}
 \footnotetext{For sets $E,F,G$, we will use the notation $E `/ F`/ G$ to denote the set difference $(E `/ F) `/ G$.}

The users have access to a private (discrete memoryless multiple) source denoted by the random vector
\begin{subequations}
  \label{eq:ZV}
  \begin{align}
    \RZ_V&:=(\RZ_i\mid i\in V)\sim P_{\RZ_V} \kern1em \text{taking values from}\\
    Z_V &:=\prod_{i\in V} Z_i,
  \end{align}
\end{subequations}
which is assumed to be finite. Note that, for notational convenience, we use capital letter in sans serif font for random variables and the same capital letter in the usual math italic font for the alphabet sets. $P_{\RZ_V}$ denotes the joint distribution of $\RZ_i$'s. 

The vector $(A,S,D,V,\RZ_V)$ of user sets and private source is called a scenario. Given a scenario, the vocal users discuss in public until the active users can recover a secret key of their choice that is secured against a wiretapper who can listen to the public discussion and wiretap the private source of the untrusted users. The protocol can be divided into the following phases for ease of exposition:
\begin{ybox}
  \noindent\underline{Secret key agreement protocol}\\[-.8em]
  \begin{compactitem}[]
  \item\textbf{Private observation:} Each user~$i \in V$ observes an i.i.d.\ sequence 
    \begin{align*}
      \RZ_i^n:=(\RZ_{it}\mid t\in [n])=(\RZ_{i1},\dots,\RZ_{in})\label{eq:Z^n}
    \end{align*}
    of its private source $\RZ_i$ for some block length $n$.
  \item\textbf{Private randomization:} Each user~$i \in V`/D`/S$ generates a random variable $\RU_i$ independent of the private source, i.e.,
    \begin{align}
      H(\RU_{V`/D`/S}|\RZ_V^n)=\sum_{i\in V`/D`/S} H(\RU_i).\label{eq:U}
    \end{align}
	(We will show in Proposition~\ref{pro:USD} that the silent and untrusted users need not randomize for the problem of interest.)
    For convenience, we let
    \begin{align}
      \tRZ_i&:=\begin{cases}
      (\RU_i,\RZ_i^n) & i\in V`/D`/S\\
      \RZ_i^n & i\in S\cup D \kern1em \text {(otherwise)}
      \end{cases} \label{eq:tRZi}
    \end{align}
    be the entire private observation of user $i\in V$.
  \item\textbf{Public discussion:} Using a public authenticated noiseless channel, the vocal users broadcast some messages in a round-robin fashion interactively for a finite number of rounds. More precisely, at times $t=1,\dots,r$ for some positive integer $r$, the vocal user $i\in V`/S$ broadcasts to everyone a function of its accumulated observations, denoted as
    \begin{align}
      \RF_{it}&:= f_{it}(\tRZ_i,\tRF_{it}) \label{eq:F} \kern1em \text {where}
    \end{align}
    \vspace{-1em}
    \begin{subequations}
      \label{eq:FiF}
      \begin{align}
	\tRF_{it}&:=(\RF_{[i-1]\,t},\RF_{V`/S}^{t-1}),
      \end{align}
      which includes the previous messages $\RF_{[i-1]\,t}:=(\RF_{jt}\mid j<i)$ broadcast in the same round and the messages $\RF_{V`/S}^{t-1}:=(\RF_{V`/S\,`t}\mid `t<t)=(\RF_{i`t}\mid i\in V`/S,`t<t)$ broadcast in previous rounds. Note that, unless otherwise stated, we assumed without loss of generality that the discussion in each round is in the ascending order of $i\in V$ and that $[i-1]\subseteq V`/S$. We also use
      \begin{align}
	\RF_i&:=(\RF_{it}\mid t\in [r]) \text{ and} \label{eq:FiF:Fi}\\
	\RF &:= (\RF_i\mid i\in V`/S)\label{eq:FiF:F}
      \end{align}
    \end{subequations}
    to denote, respectively, the vector of all messages from user~$i\in V`/S$ and all vocal users. 
  \item\textbf{Key generation:}
    Each user $i \in A$ is required to recover a common secret key from his accumulated observations in the sense that
    \begin{align}
      \lim_{n\to `8} \Pr`1(\exists i\in A, \RK\neq `q_i(\tRZ_i,\RF)`2) &= 0   \label{eq:recover}
    \end{align}
    for a random variable $\RK$, called the secret key, and some function $`q_i$ that recovers the key from the entire observation of user $i\in A$.
    The secret key $\RK$ must also be nearly uniformly random and independent of the wiretapper's observations $(\RF,\tRZ_D)$, i.e.,
    \begin{align}
      \lim_{n\to `8} \frac1n`1[\log{\abs{K}} - H(\RK|\RF,\tRZ_D)`2] &=0, \label{eq:secrecy}
    \end{align}
    where $K$ denotes the finite alphabet set of possible key values.
  \end{compactitem}
\end{ybox}

The secrecy capacity is defined as
\begin{align}  \label{eq:CS}
  \CS:=\sup \liminf_{n\to `8} \frac1n\log{\abs{K}}
\end{align}
where the supremum is taken over all key rates achievable for the given scenario $(A,S,D,V,\RZ_V)$ but with any sequence (in $n$) of choices of other parameters respecting the constraints on private randomization~\eqref{eq:U}, interactive public discussion~\eqref{eq:F} as well as recoverability~\eqref{eq:recover} and secrecy~\eqref{eq:secrecy} of the secret key. A $\CS$-achieving scheme corresponds to a sequence of choices with achievable key rate equal to the capacity. If the supremum in \eqref{eq:CS} and the constraints \eqref{eq:recover} and \eqref{eq:secrecy} can be achieved for a finite $n$, the capacity is said to be achievable non-asymptotically.

The communication complexity is the minimum public discussion rate required to achieve the secrecy capacity, i.e.,
\begin{align}  \label{eq:rsk}
  \RS:=\inf \limsup_{n\to`8} \frac1n \log \abs{F},
\end{align}
where $F$ denotes the finite alphabet set of possible values of $\RF$ and the infimum is taken over all the discussion rates of $\CS$-achieving schemes. 

\begin{Remark}
  \label{rem:generality}
Our problem formulation covers \cite{csiszar04,amin10a} as special cases:
  \begin{compactitem}
  \item Without silent active users, i.e., $S\subseteq D$, our formulation reduces to that in \cite{csiszar04}; 
  \item Without trusted helpers, i.e., $A=V`/D$, but at least one vocal active user $A`/S\neq `0$, we obtain the formulation in \cite{amin10a}.
  \end{compactitem}
  The wiretapper's side information in \cite{csiszar04,amin10a} can be covered equivalently as the private source $\RZ_i$ of a silent untrusted user $i\in S\cap D$.
\end{Remark}

We will focus on the case without silent untrusted users, i.e., $S\cap D=`0$, because with silent untrusted users, even the secrecy capacity is largely unknown, let alone the communication complexity. Indeed, our case of interest will be further restricted to the following for a similar reason:
\begin{ybox} $S\subsetneq A$ with at least one vocal active user.
\end{ybox} 
\noindent The secrecy capacity when all active users are silent remains unknown except in the special case with only two trusted users~\cite{csiszar00} or without helpers.\footnote{In the case when all users are active and silent, i.e., $V=A=S$, it is straightforward to show that $\CS=J_{\op {GK}}(\RZ_V):=\max\Set{H(\RU)\mid H(\RU|\RZ_i)=0,\forall i\in A}$, which is the multivariate extension of G\'acs-K\"orner common information~\cite{gac72}. We would like to point out here that there is a subtle issue with our preliminary work in \cite{chan16itw}, in which it was claimed but not proved that the G\'acs-K\"orner common information is equal to the secrecy capacity at zero rate of public discussion. We are not able to extend the converse result~\cite{gac72} from no discussion to sub-linear amount (in $n$) of discussion. Hence, in \cite{chan16itw}, $\CS>J_{\op {GK}}(\RZ_V)$ can only be conjectured as a sufficient condition for $\RS>0$.} We also remark that certain user types need not be considered in the problem formulation.

\begin{Remark}
  \label{rem:redundant_users}
  Without loss of optimality, one \emph{need not} consider the presence of the following users:
  \begin{compactitem}
  \item Untrusted active users, i.e., $A\nsubseteq V`/D$: The secrecy capacity is zero trivially because the recoverability condition~\eqref{eq:recover} for such users means that the wiretapper can also recover the key, hence violating the secrecy condition~\eqref{eq:secrecy}. 
  \item Silent trusted helpers, i.e., $S\nsubseteq A\cup D$: Their presence affect neither the recoverability condition~\eqref{eq:recover} (by being silent) nor the secrecy condition~\ref{eq:secrecy} (by being trusted).
  \end{compactitem}
\end{Remark}

It was conjectured in \cite{MKS16} that private randomization does not reduce $\RS$ in the case when all users are vocal and active. In the general case with helpers and silent users, the conjecture also appears very plausible, with no apparent counter-example that suggests otherwise. Indeed, as the following result shows, private randomization by any silent or untrusted user is not necessary, and so our formulation precluded them without loss of optimality.

\begin{Proposition}
  \label{pro:USD}
  Allowing private randomization by any silent or untrusted user~$j\in S\cup D$, i.e., modifying \eqref{eq:tRZi} with
  \begin{align}
    \tRZ_j=(\RU_j,\RZ_j^n)\kern1em \text{where $I(\RU_j\wedge \tRZ_{V`/\Set{j}},\RZ_j^n)=0$}
    \label{eq:USD}
  \end{align}
  neither increases $\CS$ nor decreases $\RS$.
\end{Proposition}

\begin{Proof}
	See Appendix~\ref{sec:proof:USD}.
\end{Proof}

\subsection{Optimality of Omniscience}
\label{sec:OO}

Next, we take a step back to formulate the easier problem of the optimality of a general class of $\CS$-achieving strategies (in terms of minimizing the public discussion rate, i.e., achieving $\RS$).
In both the case \cite{csiszar04} (with helpers but no active users) and the case \cite{amin10a} (with active users but no helpers), it can be seen that the proposed $\CS$-achieving schemes require the active users to recover the private sources of the vocal users after public discussion. We will extend this idea to the following $\CS$-achieving scheme for the general case of interest described with helpers and silent users:
\begin{Definition}
  \label{def:O}
  For $S\subsetneq A$, the \emph{omniscience strategy for secret key agreement} requires each vocal user $i\in V`/S$ to broadcast in public a function 
  
  \begin{align}
    \RF_i:=f_i(\tRZ_i)=f_i(\RZ_i^n) \label{eq:F:O}
  \end{align}
  of its source such that each active user can first recover the private sources of the (vocal) untrusted users in the sense that
  \begin{subequations}
    \label{eq:recover:O}
    \begin{align}
      \lim_{n\to `8} \Pr`1(\exists i\in A,\RZ_{D}^n\neq `f_i(\tRZ_i,\RF_{D})`2)&= 0\label{eq:recover:O1}
    \end{align}
    for some function $`f_i$'s,	and then recover the private sources of all other vocal users, i.e.,
    \begin{align}
    	\begin{split}
    		\lim_{n\to `8} \Pr\big(&\exists i\in A,\\
    		&\RZ_{V`/D`/S}^n\neq \psi_i(\tRZ_i,\RF_{V`/D`/S},\RZ_D^n)\big)= 0
    	\end{split}
       \label{eq:recover:O2}
    \end{align}
    for some function $\psi_i$'s. Note that the omniscience strategy does not require private randomness. Furthermore, a natural question to ask is whether it is important that $\RZ_D^n$ be recovered before the other private sources are. This will be addressed in Example~\ref{eg:chain} and the remark preceding it.
  \end{subequations}	
  We also require the omniscience strategy to minimize the total discussion rate, denoted by
  \begin{align}
  	\begin{split}
    \RCO &:= \inf\limsup_{n\to`8} \frac1n \abs {F}\\
    &= \inf \limsup_{n\to`8} \frac1n \sum_{i\in V`/S} \abs {F_i},
    \end{split}\label{eq:RCO}
  \end{align}
  the infimum being taken over all functions $f_i$, $i \in V `/ S$, that satisfy \eqref{eq:F:O}--\eqref{eq:recover:O}.
  The two recoverability constraints in \eqref{eq:recover:O} will be called the \emph{omniscience constraints}, to distinguish them from the 
 \emph{recoverability constraint}~\eqref{eq:recover} for the secret key. For the omniscience strategy to be $\CS$-achieving, we will also limit the discussion rates of the untrusted users to satisfy\footnote{Although the proof of Theorem~\ref{thm:CSRCO} relies on \eqref{eq:rD}, we conjecture that \eqref{eq:rD} is not required for the omniscience strategy to be $\CS$-achieving.}
  \begin{align}
    &`1(\lim_{n\to`8} \frac1n \log\abs {F_i} \Big| i\in D`2) \in \rsfsR(\RZ_D) \kern1em \text {where} \label{eq:rD}\\
    &\rsfsR(\RZ_D):=\Set {r_D\in `R^D \mid r(B)\leq H(\RZ_B), \forall B\subseteq D}. \label{eq:pzRZD}
  \end{align}
  The secret key is then chosen as a function
  \begin{align}
    \RK=`q(\RZ_{V`/S}^n) \label{eq:K:O}
  \end{align}
  of the entire private source of the vocal users at the maximum rate subject to the secrecy constraint~\eqref{eq:secrecy}. (Note that \eqref{eq:recover} immediately follows from \eqref{eq:recover:O}.)
\end{Definition}

We will show in Section~\ref{sec:upperbound} that the omniscience strategy in Definition~\ref{def:O} is $\CS$-achieving in the general case of interest, and that $\RCO$ has a single-letter linear-programming characterization. Therefore, $\RCO$ serves as a computable upper bound on $\RS$. We say that omniscience is optimal for secret key agreement if the bound is tight, i.e., $\RS=\RCO$, in which case $\RS$ has a single-letter characterization given by $\RCO$. Our goal is to discover general classes of scenarios under which omniscience is or is not optimal, i.e., the sufficient or necessary conditions for the optimality of omniscience. In particular, we will specialize/strengthen the results to the hypergraphical source model:
\begin{Definition}[\mbox{\cite[Definition~2.4]{chan10md}}]
  \label{def:BN}
  $\RZ_V$ is a hypergraphical source with respect to a hypergraph $(V,E,`x)$ with edge function $`x:E\to 2^V`/\Set{`0}$ (which maps from an edge label in $E$ to a non-empty subset of $V$) iff
   \begin{align}
  	\RZ_i=(\RX_e\mid e\in E, i\in `x(e))\kern1em\forall i\in V.
  	\label{eq:X2Z}
  \end{align}
   for some independent (hyper-)edge variables $\RX_e$ for $e\in E$ with $H(\RX_e)>0$.
  %(The edge variables can be further restricted to take values from a finite field for the study of the problems of delay~\cite{chan11delay} and coding~\cite{chan12ud}.)
\end{Definition}
The above source model also covers the PIN model in~\cite{nitinawarat10,nitinawarat-ye10} as a special case:
\begin{Definition}[\cite{nitinawarat10}]
  \label{def:PIN}
  $\RZ_V$ is a PIN iff it is hypergraphical with respect to a graph $(V,E,`x)$ with edge function $`x:E\to {{V}\choose{2}}$ (no self-loops).
\end{Definition}
An example of a hypergraphical source and a PIN is given at the end of this section (Example~\ref{eg:chain}).

We remark that the omniscience strategy above differs from that in \cite{csiszar04} even in the case without silent users:
\begin{Remark}
  Instead of \eqref{eq:recover:O1}, \cite{csiszar04} require the entire source of the untrusted user to be revealed in public in the sense that
  \begin{align}
    \lim_{n\to`8} \Pr`1(\RZ_{D}^n\neq `f(\RF_D)`2)=0, \label{eq:recover:O:CN04}
  \end{align}
  i.e., the source of the untrusted users can be recovered not only by the active users but also by anyone who gets to listen to the discussion $\RF_D$ by the untrusted users. As will be shown by the following example, $\RCO$ can be strictly larger with this requirement, resulting in a looser upper bound on $\RS$. The example also shows that \eqref{eq:recover:O1} and \eqref{eq:recover:O2} should not be combined into the constraint
  \begin{align}
    \lim_{n\to`8} \Pr`1(\exists i\in A,\RZ_{V`/S}^n\neq `f_i(\RZ_i^n,\RF)`2)=0 \label{eq:recover:OC}
  \end{align}
  because even an optimal discussion $\RF$ under this constraint can leak too much information to the wiretapper. Hence omniscience through~\eqref{eq:recover:OC} no longer guarantees achieving $\CS$.
\end{Remark}

%cc% A figure to illustrate the following example.

\begin{Example}
  \label{eg:chain}
  Let $\RX_{a}$ and $\RX_{b}$ be two uniformly random and independent bits, and
  \begin{align*}
    \label{eq:chain}
    \begin{split}
      \RZ_1&:=\hphantom{(}\RX_{a}\\
      \RZ_2&:=(\RX_{a},\RX_{b})\\
      \RZ_3&:=\hphantom{(\RX_{a},}\kern.2em\RX_{b}\\
      \RZ_4&:=(\RX_{a},\RX_{b})
    \end{split}
  \end{align*}
  With $V=[3]$, the source $\RZ_V=(\RZ_1,\RZ_2,\RZ_3)$ is a PIN with vertex set $[3]$, edge set $E=\Set{a,b}$ and the edge function
  \begin{align*}
    `x(e)=\begin{cases}
    \Set{1,2} & e=a\\
    \Set{2,3} & e=b.
    \end{cases}
  \end{align*}
  With $V=[4]$ instead, the source $\RZ_V$ is not a PIN but a hypergraphical source with the edge function modified to
  \begin{align*}
    `x(e)=\begin{cases}
    \Set{1,2,4} & e=a\\
    \Set{2,3,4} & e=b.
    \end{cases}
  \end{align*}

  Consider the scenario $(A,S,D,V)=(\Set{2,4},`0,\Set{3},[4])$. It can be shown that 
  \begin{align*}
    \CS=1\kern1em \text {and} \kern1em \RS=\RCO=0,
  \end{align*}
  achieved non-asymptotically with 
  \begin{align*}
    n=1, \kern1em \RK:=\RX_{a}\kern1em \text {and $\RF$ deterministic.}
  \end{align*}
  Hence, omniscience is optimal in this case. Now, if the recoverability condition~\eqref{eq:recover:O:CN04} in \cite{csiszar04} were imposed instead of \eqref{eq:recover:O1}, then $\RCO\geq H(\RZ_3)=H(\RX_{b})=1>0=\RS$, and so the omniscience scheme would not be optimal.
  
  Consider the scenario  $(A,S,D,V)=(\Set{1,2,4},`0,\Set{3},[4])$ instead. It can be shown that 
  \begin{align*}
    \CS=1\kern1em \text {and} \kern1em \RS=0,
  \end{align*}
  achieved non-asymptotically with 
  \begin{align*}
    n=1, \kern1em \RK=\RX_{a}\kern1em \text {and $\RF$ deterministic.}
  \end{align*}
  However, since the active user~$1$ does not observe $\RX_{b}$ directly from its private source,
  \begin{align*}
    \RCO\geq H(\RZ_V|\RZ_1)\geq H(\RX_{b})=1,
  \end{align*}
  which is achieved by choosing $\RF:=\RF_3:=\RX_{b}$. It follows that $\RS=0<1=\RCO$, and so omniscience is not optimal. Now, if \eqref{eq:recover:OC} were imposed instead of \eqref{eq:recover:O}, then $\RCO=1$ as before but it could be achieved with $\RF:=\RF_2:=\RX_{a}\oplus \RX_{b}$, from which user~$1$ can recover $\RX_{b}$ as $\RF\oplus \RZ_1$. However, the wiretapper can also recover $\RZ_1$ as $\RF\oplus \RZ_3$ by wiretapping the source of the untrusted user~$3$. Since the entire source, i.e., $\RX_{a}$ and $\RX_{b}$, can be recovered by the wiretapper, any secret key $\RK$ satisfying  \eqref{eq:K:O} and \eqref{eq:secrecy} must have zero rate. In other words, the current discussion for omniscience, despite being optimal in achieving $\RCO$, leaks too much information to the wiretapper.
\end{Example}

\section{With No helpers or Silent Users}
\label{sec:nohelper}

In this section, we will introduce the main ideas through the basic scenario $A=V$ and $S=`0$. Unless stated otherwise, the basic scenario will be assumed for all the results in this section.

\subsection{Preliminaries on MMI and Fundamental Partition}

$\CS$ in the current case is characterized by $\RCO$ as: 
\begin{Proposition}[\cite{csiszar04}]
  \label{pro:CSRCO:CN04}
  The omniscience strategy achieves
  \begin{align}
    \CS=H(\RZ_{V})-\RCO, \label{eq:CSRCO:CN04}
  \end{align}
  and so $\RS\leq \RCO$.
\end{Proposition}
$\RCO$ was also characterized in \cite{csiszar04} as a linear program using standard techniques of independent source coding~\cite{csiszar2011information}. In fact, $|RCO$ is easily computable since the expression for $\RCO$ in \eqref{eq:RCO} was argued to be solvable in polynomial time\footnote{This is assuming that the entropy function $B\mapsto H(\RZ_B)$ for each $B\subseteq V$ can be evaluated in polynomial time.} with respect to the size of the network~\cite{chan11isit,milosavljevic11}.

To study the tightness of the $\RCO$ upper bound, we will make use of the following (conditional) multivariate mutual information (MMI) measure and its properties studied in \cite{chan15mi}: For a finite set $U$ and a random vector $(\RZ'_U,\RW')$,
\begin{subequations}
  \label{eq:mi}
  \begin{align}
    I(\RZ'_U|\RW') 
    &:= \min_{\mcP\in \Pi'(U)} I_{\mcP}(\RZ'_U|\RW'), \kern2em \text{with}\label{eq:I}\\
    \kern-1em I_{\mcP}(\RZ'_U|\RW')
    &:= \tfrac{1}{\abs{\mcP}-1}D`1(P_{\RZ'_U|\RW'}`1\|\prod\nolimits_{C\in \mcP} P_{\RZ'_C|\RW'}`2|P_{\RW'}`2)\notag \kern-1em \\	
    &:=\tfrac{1}{\abs{\mcP}-1}`1[\sum_{C\in \mcP} H(\RZ'_C|\RW') - H(\RZ'_U|\RW')`2]\kern-.25em,\label{eq:IP}\kern-1em
  \end{align}
  where $\Pi'(U)$ is the collection of partitions of $U$ into at least two non-empty disjoint parts, and
  $D(\cdot\|\cdot\mid\cdot)$ is the conditional Kullback--Leibler divergence.
\end{subequations}
We also define the unconditional MMI measures $I(\RZ'_U)$ and $I_{\mcP}(\RZ'_U)$ by dropping the conditioning on $\RW'$ throughout \eqref{eq:mi}.

The MMI appeared as an upper bound on the secrecy capacity in \cite[(26)]{csiszar04} in the special case without helpers. In \cite{chan2008tightness}, the bound \cite[(26)]{csiszar04} was shown to be loose in the more general case with helpers but identified to be tight in the no-helper case and therefore proposed as a measure of mutual information among multiple random variables:
\begin{Proposition}[\mbox{\cite[Theorem~1.1]{chan10md}}]
  \label{pro:I}
  %For $A=V$ and $S=`0$, we have 
  $\CS=I(\RZ_V)$ in the case without helpers or silent users.
\end{Proposition}
The proof uses the submodularity~\cite{fujishige78} of the entropy function $B\mapsto H(\RZ'_B|\RW')$ for $B\subseteq U$ (a class of Shannon-type inequalities~\cite{yeung91,yeung08}) to show that the linear-programming characterization of $\CS$ in \cite{csiszar04} is equal to the MMI.
%The MMI was also called ``shared information" in \cite{CIT-072}.
A simple proof using the Dilworth truncation was given in~\cite{chan15mi}. %where it was also pointed out why the divergence bound is slack using the connection of secret key agreement to network coding discovered in~\cite{chan11isit,chan12ud}. 
Like Shannon's mutual information, the MMI has various fundamental information-theoretic properties including the data processing inequality~\cite{chan15mi} (which will be refined in Lemma~\ref{lem:DPI}).

Denote the set of all optimal partitions to \eqref{eq:I} as
\begin{align}
  \Uppi^*(\RZ'_U|\RW'):=\Set{\mcP\in \Pi'(U)\mid I_{\mcP}(\RZ'_U|\RW') = I(\RZ'_U|\RW')}.\label{eq:Pi*}
\end{align}
The set $\Pi'(U)$ is endowed with a partial order, denoted by $\preceq$, with $\mcP \preceq \mcP'$ having the meaning 
\begin{align}
  \forall C\in \mcP, \exists C'\in \mcP'\text{ such that } C\subseteq C'.\label{eq:<P}
\end{align}
% (`$\prec$' denotes the strict inequality when $\mcP\neq \mcP'$.)
In other words, $\mcP$ can be obtained from $\mcP'$ by further partitioning some parts of $\mcP'$; we then say that $\mcP$ is \emph{finer} than $\mcP'$. We will consider the finest partition in $\Uppi^*(\RZ'_U | \RW')$, the existence of which is guaranteed by the following proposition.
\begin{Proposition}[\mbox{\cite[Lemma 5.1 and Theorem~5.2]{chan15mi}}]
  \label{pro:fundamental}	 
  $\Uppi^*(\RZ'_U|\RW')$ forms a lower semi-lattice with respect to\ the partial order~\eqref{eq:<P}. In particular, there is a unique finest partition in $\Uppi^*(\RZ'_U|\RW')$.
\end{Proposition}
The unique finest partition in $\Uppi^*(\RZ'_U|\RW')$ is called the \emph{fundamental partition}, and is denoted as $\mcP^*(\RZ'_U|\RW')$. Again, the unconditional versions of these definitions, namely, $\Uppi^*(\RZ'_U)$ and $\mcP^*(\RZ'_U)$, are obtained by dropping the conditioning on $\RW'$ throughout. The fundamental partition has various meaningful interpretations in the problems of vocality~\cite{mukherjee14,zhang15}, successive omniscience~\cite{chan16so}, data clustering~\cite{chan15allerton,chan16cluster} and feature selection~\cite{chan16allerton}.

%\subsection{Lower bound by Wyner common information \& dual total correlation}

The condition for the optimality of omniscience in~\cite{mukherjee15,MKS16} for the PIN model in Definition~\ref{def:PIN} is expressed in terms of the fundamental partition.
\begin{Proposition}[\mbox{\cite[Theorem~8, Corollary~23]{MKS16}}]
  \label{pro:mukherjee15}
  For the PIN model, we have $\RS=\RCO$ iff $\mcP^*(\RZ_V)=\{\{i\}\mid i\in V\}$, namely, the partition into singletons. 
\end{Proposition}
The result was based on a lower bound on $\RS$ in~\cite{MKS16} that extends the result of~\cite{tyagi13} to the multiterminal setting using the multi-letter multivariate Wyner common information: %defined as
\begin{subequations}
  \label{eq:A=V:CW}
  \begin{align} 
    \CW &:=\inf \limsup_{n\to`8} \frac{1}{n}H(\RL) \kern1em \text{such that}\label{eq:A=V:CW1}\\
    &\lim_{n\to`8}\frac{1}{n}I_{\mcP^*(\RZ_V)}(\RZ_V^n|\RL)=0\label{eq:A=V:CW2}
  \end{align}
\end{subequations}
where the infimum is for a given $\RZ_V$. Note that $\mcP^*(\RZ_V)$ is used instead of $\mcP^*(\RZ_V^n|\RL)$. Furthermore, \cite{MKS16} required $\RL$ to be a function of $\RZ_V^n$, i.e., $H(\RL|\RZ_V^n)=0$.
\begin{Proposition}[\mbox{\cite[Theorem~2]{MKS16}}]%\footnote{The lower bound to $\RS^{\textup{NR}}(\RZ_V)$ presented in Proposition~\ref{pro:rsk_pre_lb} differs slightly from the original version in \cite[Theorem~2]{MKS16}. To be precise, Proposition~\ref{pro:rsk_pre_lb} combines the results of Theorem~2 and Proposition~1 of \cite{MKS16}.}
  \label{pro:rsk_pre_lb}	 
  The communication complexity $\RS^{\NR}$ with private randomization~\eqref{eq:U} precluded in the problem formulation is lowered bounded as
  \begin{align}\RS^{\NR} \geq \CW-I(\RZ_V),\label{eq:RSNR:LB}
  \end{align}
  which holds also with the additional constraint that $H(\RL|\RZ_V^n)=0$.
\end{Proposition}

The use of the above lower bound is somewhat limited by the difficulty in evaluating the multi-letter expression $\CW$ and the problem formulation that precludes randomization. The derivation of Proposition~\ref{pro:rsk_pre_lb} requires quite a bit of machinery to evaluate $\CW$, and to extend the result to allow randomization. We will improve the bound (in Theorem~\ref{thm:LB:A=V} in Section~\ref{sec:A=V:results}) with a single-letter expression, for which we need the following definition:
\begin{Definition}
  \label{def:JWP}
  For a finite set $U$ with size $\abs {U}>1$ and random vector $(\RZ'_U,\RW')$, the (conditional) \emph{partition Wyner common information} of $\RZ'_U$ given $\RW'$ with respect to\ the partition $\mcP\in \Pi'(U)$ is
  \begin{subequations}
    \label{eq:JWP}
    \begin{align}
      J_{\opW,\mcP}(\RZ'_U|\RW')&:=\inf \Set{I(\RW \wedge \RZ'_U|\RW')\mid \label{eq:JWP1}\\
	& \kern3em I_{\mcP}(\RZ'_U|\RW,\RW')=0}, \label{eq:JWP2}
    \end{align}
  \end{subequations}
  where the minimum is taken over all possible choices of the random variable $\RW$ (or $P_{\RW|\RZ'_U,\RW'}$). $J_{\opW}(\RZ'_i\wedge\RZ'_j|\RW')$ denotes the bivariate case $U=\Set{i,j}$ where $i\neq j$. (The version without conditioning reduces to the usual Wyner common information introduced by~\cite{wyner75}.)
\end{Definition}
If $\mcP$ is the partition into singletons, and $\RW'$ is determinisitic, then $J_{\opW,\mcP}$ is the extension in~\cite{liu10} of the Wyner common information~\cite{wyner75} from the bivariate case $J_{\opW}(\RZ_i\wedge \RZ_j)$, to the multivariate case. Following the same argument as in~\cite{wyner75}, the expression~\eqref{eq:JWP} is computable with the following bound on support size:
\begin{Proposition}
  \label{pro:JWP}
  For the partition Wyner common information~\eqref{eq:JWP}, it is admissible to impose 
  \begin{align}
    \abs {W} \leq \abs {Z'_U}\abs {W'},\label{eq:A=V:W}
  \end{align}
  and $\inf$ can be replaced by $\min$, i.e., the infimum can be achieved by a choice of $\RW$ satisfying \eqref{eq:A=V:W} in addition. 
\end{Proposition}
\begin{Proof}
  This follows from the same argument as in \cite{wyner75} and will be proved for the more general setting in Proposition~\ref{pro:|w|}.
\end{Proof}

Despite the above result, $J_{W,\mcP}$ is not easy to compute even for the bivariate case~\cite{wyner75}. Fortunately, it has non-trivial entropic~\cite{chan15mi} bounds that are easy to compute from the entropy function of the given random vector:
\begin{Proposition}
	\label{pro:DTC}
\begin{align}
  &\kern-.5em H(\RZ'_U|\RW')\geq J_{\opW,\mcP}(\RZ'_U|\RW')\geq J_{D,\mcP}(\RZ'_U|\RW')\kern1em \text{where}
  \label{eq:DJH}\\
  %\end{align}
  %\begin{align}
  &\kern-.5em J_{D,\mcP}(\RZ'_U|\RW'):=H(\RZ'_U|\RW') - \kern-.5em\sum_{C\in \mcP}\kern-.3em H(\RZ'_C|\RZ'_{U`/C},\RW'),\label{eq:JD}\kern-.5em
\end{align}
which will be called the \emph{partition dual total correlation.}
\end{Proposition}
\begin{Proof}
Since $\RW = \RZ'_U$ is always a feasible solution to~\eqref{eq:JWP}, $J_{\opW,\mcP}(\RZ'_U|\RW')\leq H(\RZ'_U|\RW')$, which gives the first inequality in \eqref{eq:DJH}. To prove the second inequality, it suffices to show
\begin{align*}
	I(\RW\wedge \RZ'_U|\RW') \geq J_{D,\mcP}(\RZ'_U|\RW')
\end{align*}
for all feasible solution $\RW$. To do so, notice that
the constraint~\eqref{eq:JWP2} means that $\RZ'_{C}$ for $C\in \mcP$ are mutually independent given $(\RW,\RW')$, and so
\begin{align*}
	I(\RW\wedge \RZ'_U|\RW')
	&=H(\RZ'_U|\RW')-H(\RZ'_U|\RW',\RW)\\
	&\utag{a}=H(\RZ'_U|\RW')-\sum_{C\in \mcP} H(\RZ'_{C}|\RW',\RW)\\
	&\utag{b}=H(\RZ'_U|\RW')-\sum_{C\in \mcP} H(\RZ'_{C}|\RW',\RW, \RZ'_{U\setminus C})\\
	&\geq H(\RZ'_U|\RW')-\sum_{C\in \mcP}H(\RZ'_{C}|\RW',\RZ'_{U`/C})\\
	&=J_{D,\mcP}(\RZ'_U|\RW'),
\end{align*}
where we have applied the independence of $\RZ'_C$'s in \uref{a} to rewrite $H(\RZ'_{U}|\RW',\RW)$ as the sums $\sum_{C\in \mcP}H(\RZ'_{C}|\RW',\RW)$ and in \uref{b} to rewrite $\sum_{C\in \mcP} H(\RZ'_{C}|\RW',\RW)$ as $\sum_{C\in \mcP}H(\RZ'_{C}|\RW',\RW,\RZ'_{U\setminus C})$ respectively. 
\end{Proof}
When $\mcP$ is the partition into singletons, $J_{\opD,\mcP}$ is Han's dual total
correlation~\cite{han75}, which has been shown to be the best entropic lower bound for $J_{\opW,\mcP}$ even after incorporating non-Shannon-type inequalities~\cite{chen16}.

\subsection{Main results}
\label{sec:A=V:results}

We give a single-letter lower bound on $\RS$ that improves upon the result of Proposition~\ref{pro:rsk_pre_lb} by allowing private randomization.

\begin{Theorem}\label{thm:LB:A=V}%\label{lemma:lb}
  For any source $\RZ_V$,
  \begin{subequations}
    \label{eq:LB:A=V}
    \begin{align}
      \RS &\geq J_{\opW,\mcP^*}(\RZ_V)-I(\RZ_V)\label{eq:LB:A=V:JW}\\
      &\geq J_{\opD,\mcP^*}(\RZ_V) - I(\RZ_V)\label{eq:LB:A=V:JD}	
    \end{align}
  \end{subequations}
  where $\mcP^*$ denotes $\mcP^*(\RZ_V)$ for convenience, and $J_{\opW,\mcP^*}$ and $J_{\opD,\mcP^*}$ are the partition Wyner common information~\eqref{eq:JWP} and the partition dual total correlation~\eqref{eq:JD}.
\end{Theorem}
\begin{Proof}
  See Appendix~\ref{sec:LB:A=V:proof}.
\end{Proof}
It was shown in~\cite[Theorem 6.3]{chan15mi}\mnote{a:r7} that $J_{\opD,\mcP}(\RZ_V)$ is no smaller than $I(\RZ_V)$ for all $\mcP\in \Pi'(V)$, therefore, the lower bounds above are non-negative.
\begin{Corollary}
  \label{cor:LB}
  $\RS=\RCO$ if $J_{\opW,\mcP^*}(\RZ_V)=H(\RZ_V)$.\kern-1em
\end{Corollary}
\begin{Proof}
  This follows from Theorem~\ref{thm:LB:A=V} by virtue of Proposition~\ref{pro:CSRCO:CN04} and \ref{pro:I}, i.e., substituting  $J_{\opW,\mcP^*}(\RZ_V)=H(\RZ_V)$ and $I(\RZ_V)=\CS$ to the right hand side (r.h.s.)\ of \eqref{eq:LB:A=V:JW} gives $\RCO$.
\end{Proof}

Compared to Proposition~\ref{pro:rsk_pre_lb}, \eqref{eq:LB:A=V:JW} is single-letter rather than multi-letter. Furthermore, \eqref{eq:LB:A=V:JD} is a simple linear function of the entropy vector of $\RZ_V$ given $\mcP^*(\RZ_V)$, which is easier to evaluate than \eqref{eq:LB:A=V:JW}.

From Corollary~\ref{cor:LB}, we obtain the following sufficient condition for the optimality of omniscience under a general source model:
\begin{Theorem}
  \label{thm:OO:A=V}
  $\RS=\RCO$ if 
  \begin{align}
    H(\RZ_C|\RZ_{V\setminus C})=0 \kern1em \forall C\in \mcP^*(\RZ_V),
    \label{eq:OO:A=V}
  \end{align}
  where $\mcP^*$ is the fundamental partition in Proposition~\ref{pro:fundamental}, namely, the finest optimal partition for the MMI~\eqref{eq:I}.
\end{Theorem}

\begin{Proof}
The condition in \eqref{eq:OO:A=V} implies that $J_{\opD,\mcP^*}(\RZ_V)= H(\RZ_V)$, and therefore, by \eqref{eq:DJH}, we also have $J_{\opW,\mcP^*}(\RZ_V)= H(\RZ_V)$. The theorem now follows from Corollary~\ref{cor:LB}.
\end{Proof}
Condition \eqref{eq:OO:A=V} means that, \emph{for all $C \in \mcP^*(\RZ_V)$, no randomness of $\RZ_C$ is independent of $\RZ_{V\setminus C}$. This condition covers all the existing results}: 
\begin{compactitem}
\item \eqref{eq:OO:A=V} covers the condition for the $2$-user case in Proposition~\ref{pro:2user} because $\mcP^*(\RZ_{\Set{1,2}})=\Set{\Set{1},\Set{2}}$. 
\item \eqref{eq:OO:A=V} also extends the sufficiency part of the condition in Proposition~\ref{pro:mukherjee15} because \eqref{eq:OO:A=V} holds for $\mcP^*(\RZ_V)=\{\{i\}: i\in V\}$ trivially, as every edge variable $\RX_e$ ($e\in E$) is a component of $\RZ_{j}$ and $\RZ_{k}$ for the distinct pair $\Set{j,k}=`x(e)$ of incident nodes. % of the edge $e$. 
\end{compactitem}

Despite its generality, \eqref{eq:OO:A=V} can be checked easily because $\mcP^*(\RZ_V)$ can be computed in strongly polynomial-time. The following is an example for which \emph{the optimality of omniscience can be easily derived by \eqref{eq:OO:A=V} but not by the existing results.}
\begin{Example}
  \eqref{eq:OO:A=V} holds for the source in Example~\ref{eg:interf} as
  \begin{align*}
    \mcP^*(\RZ_{\Set{1,2,3}}) &= \Set{\Set{1},\Set{2},\Set{3}},\kern1em \text{and}\\
    H(\RZ_1|\RZ_2,\RZ_3)&=H(\RZ_2|\RZ_1,\RZ_3)=H(\RZ_3|\RZ_1,\RZ_2)=0.
  \end{align*}
  Hence, $\RS=\RCO$ by Theorem~\ref{thm:OO:A=V}. This example is not covered by Proposition~\ref{pro:mukherjee15} because the private source belongs to the more general finite linear source model~\cite{chan10phd} rather than the PIN model (Definition~\ref{def:PIN}) (or the hypergraphical source model in Definition~\ref{def:BN}).
\end{Example}

\subsection{Stronger Results for Hypergraphical Sources}

The necessity of the condition in Proposition~\ref{pro:mukherjee15} can be extended to the more general hypergraphical source model in Definition~\ref{def:BN}:
\begin{Theorem}
  \label{thm:hypergraph}
  For hypergraphical sources with respect to\ the hypergraph $(V,E,`x)$, we have $\RS=\RCO$ iff  
  \begin{align}
    \not\exists e\in E\kern0.5em \text{such that}\kern0.5em`x(e)\subseteq C \kern0.5em\text{for some}\kern0.5em C\in \mcP^*(\RZ_V),\kern-0.5em\label{eq:hyp:cross}
  \end{align}
 which means that there does not exists a hyperedge entirely contained by a part of the fundamental partition, i.e., every hyperedge crosses the fundamental partition.
  %	Thus, \eqref{eq:OO:A=V} is necessary and sufficient for $\RS$-maximality.
\end{Theorem}
\begin{Proof}
  See Section~\ref{sec:hypergraph:proof}.
\end{Proof}
\begin{Example}
  Let $\RX_{a},\RX_{b}$ and $\RX_{\opc}$ be uniformly random and independent bits. With $V:=[5]$, define the private source as 
  \begin{align*}
    \RZ_1&:=\RX_{a}\\
    \RZ_2&:=\RX_{b}\\
    \RZ_3&:=\RX_{\opc}\\
    \RZ_4&:=(\RX_{a},\RX_{b})\\
    \RZ_5&:=(\RX_{a},\RX_{b},\RX_{\opc}).
  \end{align*}
  It is hypergraphical with edge function
  \begin{align*}
    `x(e)=
    \begin{cases}
      \Set{1,4,5} & e=a\\
      \Set{2,4,5} & e=b\\
      \Set{3,5} & e=c.
    \end{cases}
  \end{align*}

  To check condition~\eqref{eq:hyp:cross}, we can first obtain
  \begin{align*}
    I(\RZ_V)=1\kern1em\text {and}\kern1em  \mcP^*(\RZ_V)=\Set{\Set{1},\Set{2},\Set{3},\Set{4,5}}.
  \end{align*}
  Then, \eqref{eq:hyp:cross} holds because every hyperedge crosses $\mcP^*(\RZ_V)$. \eqref{eq:OO:A=V} also holds because, for every $C\in \mcP^*(\RZ_V)$, every edge variable in $\RZ_C$ also appears in $\RZ_{V`/C}$. By Theorem~\ref{thm:OO:A=V},
  \begin{align*}
    \RS=\RCO=H(\RZ_V)-I(\RZ_V)=2 \kern1em\text{by \eqref{eq:CSRCO:CN04}}.
  \end{align*}
  This can be achieved non-asymptotically with $n=1$, $\RK:=\RZ_1=\RX_{a}$ and $\RF_5:=(\RX_{a}\oplus\RX_{b},\RX_{a}\oplus \RX_{\opc})$.
\end{Example}

$J_{\opW,\mcP^*}(\RZ_V)$ can be evaluated for hypergraphical sources because its lower bound by \eqref{eq:DJH} is tight:
\begin{Proposition}
  \label{pro:hyp:JW}
  For hypergraphical sources with respect to\ the hypergraph $(V,E,`x)$, we have
  \begin{subequations}
    \begin{align}
      J_{\opW,\mcP^*}(\RZ_V)&=H(\RX_{E^*})\kern1em \text {where}\\
      E^*&:=\Set{e\in E\mid \not\exists C\in \mcP^*(\RZ_V),`x(e)\subseteq C}
    \end{align}
  \end{subequations}
  is the set of hyperedges that cross $\mcP^*(\RZ_V)$. Furthermore, an optimal solution to \eqref{eq:JWP} is $\RW:=(\RX_e\mid e\in E^*)$. 
\end{Proposition}
\begin{Proof}
  See Appendix~\ref{sec:hyp:JW:proof}.
\end{Proof}
This means that the lower bound~\eqref{eq:LB:A=V:JW} can be easily computed for hypergraphical sources. Interestingly, while the lower bound leads to a complete characterization of the optimality of omniscience for the hypergraphical model, it may be loose in general when condition \eqref{eq:hyp:cross} is not satisfied. A counter example can be found even for the PIN model as follows. 
\begin{Example}
  \label{eg:slack}
  Let $\RX_{a}$, $\RX_{b}$ and $\RX_{\opc}$ be uniformly random and independent bits. With $V=[3]$,  define
  \begin{align}
    \label{eq:slack}		
    \begin{split}
      \RZ_1&:=\RX_{a}\\
      \RZ_2&:=(\RX_{a},\RX_{b},\RX_{\opc})\\
      \RZ_3&:=\hphantom{\RX_{a},}\kern.2em(\RX_{b},\RX_{\opc}),
    \end{split}
  \end{align}
  which is a PIN. It can be shown that 
  \begin{align*}
    I(\RZ_V)=1 \kern1em \text {and} \kern1em \mcP^*(\RZ_V)=\Set{\Set{1},\Set{2,3}}.
  \end{align*}
  The edge $a$ is the only edge that crosses $\mcP^*(\RZ_V)$. Therefore, $J_{\opW,\mcP^*}(\RZ_V)=H(\RX_{a})=1$, and so \eqref{eq:LB:A=V:JW} gives the trivial lower bound $\RS\geq 1-1=0$. However, it was proved in \cite{chan17csr} that $\RS=1$ for this example, and so the bound is loose. %cc% simple proof?
  %However, the bound is loose because it can be shown that the multivariate G\'acss and K\"orner's common information is $0$, and so $\RS>0$.
\end{Example}

\section{Single-Letter Upper Bounds and\\ Necessary Conditions}
\label{sec:upperbound}

In this section, we consider the general case $S\subsetneq A$, with possibly helpers and silent users. The single-letter upper bound on $\RS$ by $\RCO$ %in Proposition~\ref{pro:CSRCO:CN04} remains 
continues to hold in the more general case because the omniscience strategy in Definition~\ref{def:O} can be shown to be $\CS$-achieving.  %Consequently, $\CS$ can also be characterized in terms of (a component of) the $\RCO$. We will also introduce new techniques that potentially improve the $\RCO$ upper bound. Note that, omniscience is not optimal if there is an improvement in the bound, which is therefore a necessary condition for the optimality of omniscience.

\subsection{Smallest Rate of CO}
\label{sec:UB_RCO}

The following result establishes the $\RCO$ upper bound on $\RS$ and characterizes $\CS$ and $\RCO$.
\begin{Theorem}
  \label{thm:CSRCO}
  With $S\subsetneq A$, the omniscience strategy in Definition~\ref{def:O} is $\CS$-achieving, with
  \begin{align}
    \CS &= H(\RZ_{V`/D`/S}\mid \RZ_D)-`r\label{eq:CSRCO}\\
    \RS &\leq \RCO = \bar{`r}+`r \label{eq:RS<=RCO}
  \end{align}
  where $`r$ and $\bar{`r}$ are defined as the following linear programs:
  \begin{subequations}
    \label{eq:`r}
    \begin{align}
      \kern-1em `r &:=\min \big\{r(V`/D`/S)\mid r_{V`/D`/S} \in `R^{V`/D`/S}, \label{eq:`rmin}\\
      & \kern-.2em r(B)\geq H(\RZ_B|\RZ_{V`/S`/B},\RZ_j)\;\forall j\in A, B\subseteq V\kern-.2em`/\kern-.2em D\kern-.2em`/\kern-.2em S\} \kern-.5em\label{eq:`rSW}
    \end{align}
  \end{subequations}
  \vspace{-1em}
  \begin{subequations}
    \label{eq:`rbar}
    \begin{align}
      \kern-1em \bar{`r} &:=\min \big\{r(D)\mid r_{D} \in \rsfsR(\RZ_D),\label{eq:`rbarmin}\\
      & \kern1em r(B)\geq H(\RZ_B|\RZ_{D`/B},\RZ_j)\; \forall j\in A, B\subseteq D \}.\kern2.2em\label{eq:`rbarSW}
    \end{align}
  \end{subequations}
  $\rsfsR(\RZ_D)$ is defined in \eqref{eq:pzRZD}, and we have used the notation $r_B:=(r_i\mid i\in B)$ and $r(B):=\sum_{i\in B} r_i$ for any set $B$.
\end{Theorem}

\begin{Proof}
  See Appendix~\ref{sec:CSRCO:proof}.
\end{Proof}
The single-letter characterizations for $\rho$ and $\bar{\rho}$ in \eqref{eq:`r} and \eqref{eq:`rbar} can be computed in polynomial time,\footnote{This can be argued as in \cite{chan11isit} by noting that the separation oracle corresponds to performing a polynomial number of submodular function minimizations, which can be done in polynomial time.} and hence, so can $\CS$ and $\RCO$. \eqref{eq:CSRCO} covers the results of \cite{csiszar04,amin10a} as the following special cases:
\begin{Corollary}[\mbox{\cite[Theorem~2]{csiszar04}}]
  \label{cor:CSRCO:CN04}
  For $S=`0$,
  \begin{align*}
    &\CS = H(\RZ_{V`/D}|\RZ_D)-`r \kern1em \text{where}\\
    &`r =\min \big\{r(V`/D)\mid r(B)\geq H(\RZ_B|\RZ_{V`/B}), \forall B\in \mcH \}
  \end{align*}
  and $\mcH:=\Set{B\subseteq V`/D\mid `0\neq B\nsupseteq A}$.
\end{Corollary}
\begin{Proof}
  When $S=`0$, \eqref{eq:`rSW} becomes
  \begin{align*}
    r(B)\geq H(\RZ_B|\RZ_{V`/B},\RZ_j), \forall j\in A, B\subseteq V`/D.
  \end{align*}
  This yields the expression in the corollary after removing the redundant constraints where $B=`0$ or $B\ni j$. 
\end{Proof}
\begin{Corollary}[\mbox{\cite[Theorem~6]{amin10a}}]
  \label{cor:CSRCO:Amin10}
  For $S\subsetneq A=V$,
  \begin{align*}
    &\CS = H(\RZ_{V`/S})-`r \kern1em \text{where}\\
    &`r =\min \big\{r(V`/S)\mid r(B`/S)\geq H(\RZ_{B`/S}|\RZ_{V`/B}), \forall B\in \mcH \}
  \end{align*}
  and $\mcH:=\Set{B\subseteq V\mid `0\neq B\nsupseteq A}$.
\end{Corollary}
\begin{Proof}
  With $S\subsetneq A=V$, \eqref{eq:`rSW} becomes
  \begin{align*}
    r(B)\geq H(\RZ_B|\RZ_{V`/S`/B},\RZ_j) \kern1em \forall j\in A, B\subseteq V`/S.
  \end{align*}
  The constraints with $B\ni j$ are again redundant and so we can impose $j\not\in B$. With $B'=B\cup S`/\Set{j}$, the constraints can be rewritten as
  \begin{align*}
    r(B'`/S)\geq H(\RZ_{B'`/S}|\RZ_{V`/B'}).
  \end{align*}
  The constraints can only be weaker if some element in $S$ is removed from $B'$, as the r.h.s.\ cannot increase but the left hand side (l.h.s.)\ remains unchanged. This yields the expression in the corollary.
\end{Proof}
As illustrated by Example~\ref{eg:chain}, $\bar{`r}$ can be strictly smaller than $H(\RZ_D)$, i.e., the omniscience strategy is an improved version of that \cite{csiszar04} when $S=`0\neq D$.   
Consequently, the $\RCO$ upper bound~\eqref{eq:RS<=RCO} is also improved.

\subsection{Change of Scenario}
\label{sec:UB_scenario_change}
%c%
%\subsection{Via change of scenario}\label{sec:ub:ch}

In this section we will introduce some general techniques to strengthen the upper bound on $\RS$. In particular, we will make use of the monotonicity of $(\CS,\RS)$ with respect to\ certain changes of scenario, namely the vector $(A,S,D,V,\RZ_V)$ of user sets and the private source. We first consider changes in the user sets.

\begin{Theorem}\label{thm:user}
  Suppose $(\CS,\RS)$ becomes $(\CS',\RS')$ by one of the following changes in the user sets:
  \begin{enumerate}[label=(\roman*)]
  \item A vocal active user is turned into a silent active user, and a new trusted helper with the same private source as the original vocal active user is added. That is to say, $(S,V)$ becomes $(S\cup\{i\},V\cup\{i'\})$ for some $i\in A`/S$, with $i'\notin V$ being a new user with private source $\RZ_{i'}=\RZ_i$.
  \item A trusted helper is removed, i.e., $V$ becomes $V`/ \{i\}$ for some $i\in V`/(A\cup D)$.
  \end{enumerate}
  Then, we have $\CS'\leq\CS$. If equality holds, then $\RS'\geq\RS$.
\end{Theorem} 
\begin{Proof}
	See Appendix~\ref{sec:user:proof}.
\end{Proof}

Therefore, using Theorem~\ref{thm:user}, if $\CS'=\CS$, then the $\RCO$ of the new scenario can serve as an upper bound on the $\RS$ of the original scenario. This leads to the following application.

\begin{Corollary}\label{cor:ub:s}
  With $S\subsetneq A$, if $\CS$ remains unchanged after
  \begin{enumerate}[label=(\roman*)]
  \item turning a proper subset of vocal active users into silent active users, and 
  \item removing all the trusted helpers,
  \end{enumerate}
  i.e., $(S,V)$ becomes $(S',V')$ causing $(\CS,\RS,\RCO)$ to change to $(\CS',\RS',\RCO')$, such that $\CS=\CS'$, $V'=A\cup D$, $S\subseteq S'\subsetneq A$. Then, 
  \begin{gather}
    \RS\leq\RS'\leq\RCO'\leq\RCO. \label{ub:s}
  \end{gather}
  It follows that $\RS=\RCO$ only if $\CS\neq\CS'$ or 
  \begin{equation}
    H(\RZ_{V'`/S'})=H(\RZ_{V`/S}), \label{OO:S:NC}
  \end{equation}
  i.e., $H(\RZ_{(S' `/S)\cup(V`/V')}|\RZ_{V'`/S'})=0$.
\end{Corollary}
\begin{Proof}
	See Appendix~\ref{sec:user:proof}.
\end{Proof}

The following is another application of Theorem~\ref{thm:user} when the entire set of vocal active users is turned into silent active users.
\begin{Corollary}\label{cor:ub:o}
  With $S\subsetneq A$, if 
  \begin{equation}
    \CS\leq H(\RU|\RZ_D) \label{ub:o}
  \end{equation}
  for any common function $\RU$ such that
  \begin{equation}
    H(\RU|\RZ_i)=0\kern1em \forall i\in A, \label{ub:o:cf}
  \end{equation}
  then $\RS=0$. In this case, $\RS=\RCO$ iff $\RCO=0$, i.e., 
  \begin{equation}
    H(\RZ_{V`/S}|\RZ_i)=0, \kern1em \forall i\in A. \label{OO:O:NC}
  \end{equation}
\end{Corollary}
\begin{Proof}
	See Appendix~\ref{sec:user:proof}.
\end{Proof}

\begin{Example}\label{eg:PIN:NC:1}
To illustrate Corollary~\ref{cor:ub:s}, consider Example~\ref{eg:XJ} with $A=V=\{1,2\}, D=S=\emptyset, \RZ_1=(\RX_0,\RX_1)$ and $\RZ_2=(\RX_{\RJ},\RJ)$. If we choose $S'=\{1\}$ and everything else the same, then condition \eqref{OO:S:NC} fails because $H(\RZ_2)=2<3=H(\RZ_{\{1,2\}})$, or equivalently, $H(\RZ_1|\RZ_2)=1>0$, but $\CS'=I(\RZ_1\wedge\RZ_2)=\CS$, which follows from Proposition~\ref{prop:CS:s} and \eqref{eq:mi}. Hence, by Corollary~\ref{cor:ub:s}, $\RS<\RCO$ as expected.
\end{Example}

\begin{Example}\label{eg:PIN:NC}
The necessary condition~\eqref{OO:S:NC} may not be sufficient in general. For instance, consider Example~\ref{eg:slack} with $A=V=[3]$ but with $S=\{1,3\}$. Note that the only possible choice of $S'$ in \eqref{OO:S:NC} is $S$, and so \eqref{OO:S:NC} holds trivially. However, by result of \cite{chan16isit}, it can be shown that the randomness of $\RX_{\opc}$ can be reduced without diminishing the capacity. In this example, $\CS=\min\{I(\RZ_1\wedge\RZ_2),I(\RZ_2\wedge\RZ_3)\}=1$ by Proposition~\ref{prop:CS:s}, which remains unchanged even if $\RX_{\opc}$ is eliminated (doing so will only reduce $I(\RZ_2\wedge\RZ_1)$ from 2 to 1).
Consequently, $\RCO' < \RCO$, and hence, $\RS \leq\RS' < \RCO$.
\end{Example}

The following is a single-letter bound that generalizes the idea beyond the hypergraphical source.

\begin{Theorem}\label{thm:ub:sl}
  For any finite set $Q$, let 
  \begin{equation}
    \RZ_i^{(q)}:=\zeta_i^{(q)}(\RZ_i)\kern1em \forall i\in V, q\in Q, \label{eq:zvq}
  \end{equation}
  and for some functions $\zeta_i^{(q)}$ such that 
  \begin{gather}
    I(\RZ_{V`/D}^{(q)}\wedge\RZ_D|\RZ_D^{(q)})=0\kern1em \forall q\in Q. \label{ub:sl:secr}
  \end{gather}
  If, for some random variable $\RQ$ independent of $\RZ_V$, we have
  \begin{equation}
    \CS\leq H(\RZ_{V`/S}^{(\RQ)}|\RQ)-\RCO', \label{ub:sl:cs}
  \end{equation}
  where $\RCO'$ is the smallest rate of CO for $\RZ_V^{(\RQ)}$ given $\RQ$ (i.e., with $\RQ$ observerd a priori), then
  \begin{equation}
    \RS\leq\RCO'\leq H(\RZ_{V`/S}^{(\RQ)}|\RQ)-\CS.\TheoremSymbol \label{ub:sl}
  \end{equation}
\end{Theorem}
\begin{Proof}
 See Appendix~\ref{sec:ul:sl:proof}.
\end{Proof}

This result covers the PIN model in  Example~\ref{eg:PIN:NC}, with $\RQ$ chosen to be deterministic and $\RZ_V$ processed to $\RZ_V'$, where $\RZ_1':=\RZ_1=\RX_{a}$, $\RZ_2':=(\RX_{a},\RX_{b})$,  $\RZ_3':=\RX_{b}$. The following example shows that \eqref{ub:sl:secr} is useful in handling the case with untrusted helpers as well.

\begin{Example}\label{eg:PIN:NC:2}
  Consider the same source as in Example~\ref{eg:PIN:NC} (Example~\ref{eg:slack}) but with $(A,S,D)=(\{2,3\},\emptyset,\{1\})$ instead. Then, $\CS=I(\RZ_2\wedge\RZ_3|\RZ_1)=2$. We process $\RZ_V$ to $\RZ_V'$ where $\RZ_2'=\RZ_2=(\RX_{b},\RX_{\opc})$, $\RZ_3'=(\RX_{b},\RX_{\opc})$, and $\RZ_1'$ is determinisitic. Then, the secrecy capacity remains unchanged, i.e., equal to $I(\RZ'_2\wedge\RZ'_3|\RZ'_1)=2$, and $I(\RZ_{V`/D}'\wedge\RZ_D|\RZ_D')=I(\RZ_{\{2,3\}}'\wedge\RZ_1)=I(\RX_{b},\RX_{\opc}\wedge\RX_{a})=0$ satisfy \eqref{ub:sl:secr}. $\RCO'=0$ since $\RZ_{\{1,2,3\}}'=\RZ_2'=\RZ_3'$, and so $\RS=0<\RCO=H(\RZ_1)=1$ by Theorem~\ref{thm:ub:sl}, and so, omniscience is not optimal.
\end{Example}
Note that, in the above example, the edge variable $\RX_{\opc}$ observed by the untrusted user $3$ can be removed without affecting $\RS$. This can be proved more generally:
%We conjecture more generally that:
%\begin{Conjecture}
%For hypergraphical sources, $\RS$ remains unchanged by removing any hyperedge observed only by the terminals in D. %However, a proof remains elusive.
%\end{Conjecture}
\begin{Proposition} 
	\label{pro:DX}
	For any random variable $\RX$ independent of $\RZ_V$, consider the new scenario with $\RZ_V$ changed to $\RZ'_V$ where
	\begin{align}
		\RZ'_i = \begin{cases}
			(\RZ_{i}, \RX) & i\in T\\
			\RZ_i & \text{otherwise},
		\end{cases}
	\end{align}
	for some $T\subseteq V$ such that $T\cap D\neq `0$, i.e., $\RX$ is observed by the wiretapper. Then, both $\CS$ and $\RS$ remain unchanged.
\end{Proposition}
\begin{Proof}
	To prove Proposition~\ref{pro:DX}, note that the proof of Proposition~\ref{pro:USD} in Appendix~\ref{sec:proof:USD} remains valid even if $\tRU_i$ for an untrusted user $i\in D$ is observed by other user $j\in V$, i.e., with~\eqref{eq:F} modified to have $\RF_i$ depend on $\tRU_i$ directly. Hence, with $\tRU_i=\RX^n$, the proof of Proposition~\ref{pro:USD} shows that $\RX^n$ neither increases $\CS$ nor decreases $\RS$, as desired.
\end{Proof}
\begin{Corollary}
	\label{cor:DX}
	For any hypergraphical source, the hyperedges $e\in E$ with $`x(e)\cap D\neq `0$ can be removed without changing $\CS$ and $\RS$.
\end{Corollary}
\begin{Proof}
	The corollary follows from Proposition~\ref{pro:DX} with $\RZ'_V$ being the original hypergraphical source and $\RZ_V$ being the source after removing the edge variable $\RX:=\RX_e$.
\end{Proof}

While $\RQ$ was chosen to be deterministic for the previous example, it is sometimes useful to make $\RQ$ random as shown by the following example.

\begin{Example}\label{eg:ub:sl:1}
  Let $\RX_{a},\RX_{b},\RX_{\opc},\RX_{\opd}$ and $\RX_{\ope}$ be uniformly random and independent bits, and define
  \begin{align*}
    \RZ_1:= & \;(\RX_{a},\RX_{b},\phantom{\RX_{\opc},\RX_{\opd},}\RX_{\ope})\\
    \RZ_2:= & \;(\RX_{a},\RX_{b},\RX_{\opc})\\
    \RZ_3:= & \;\phantom{\RX_{a},\RX_{b},}\;(\RX_{\opc},\RX_{\opd})\\
    \RZ_4:= & \;\phantom{\RX_{a},\RX_{b},\RX_{\opc},}\;(\RX_{\opd},\RX_{\ope})
  \end{align*}
  With $A=V=[4], S=D=\emptyset$, we have
  \begin{align*}
  	\CS&=I(\RZ_V)=1.5 \kern1em \text {with}\\
  	\mcP^*(V)&=\{\{1,2\},\{3\},\{4\}\}\\
  	\RCO &=H(\RZ_V)-I(\RZ_V)\\
  	&=5-1.5=3.5.
  \end{align*}

  Let $\RQ$ be a uniformly random bit independent of $\RZ_V$ and process $\RZ_V$ to $\RZ_V^{(\RQ)}$ with $\RZ_i^{(\RQ)}:=\RZ_i$ for $i\in\{2,3\}$ but
  \begin{align*}
  	\RZ_1^{(\RQ)}&:=
  	\begin{cases}
  	(\RX_{a},\RX_{b}, \RX_{\ope}) & \text {if $\RQ=1$}\\
  	(\RX_{a},\RX_{b}) & \text {otherwise, and}
  	\end{cases}\\
  	\RZ_4^{(\RQ)}&:=
  	\begin{cases}
  		(\RX_{\opd}, \RX_{\ope}) & \text {if $\RQ=1$}\\
  		\RX_{\opd} & \text {otherwise}.
  	\end{cases}
  \end{align*}
  It follows that 
  \begin{align*}
  	H(\RZ_{\Set {1,4}}^{(\RQ)}|\RQ)
  	&= 0.5 H(\RX_{\Set {a,b,\opd,\ope}}) + 0.5H(\RX_{\Set {a,b,\opd}})\\
  	&= 3.5 < 4= H(\RZ_{\Set {1,4}}).
  \end{align*}
  By Proposition~\ref{pro:CSRCO:CN04} and \ref{pro:I}, we have $\RCO'=H(\RZ_V^{(\RQ)}|\RQ)-I(\RZ_V^{(\RQ)}|\RQ)=4.5-1.5=3$, because 
  \begin{align*}
  H(\RZ_V^{(\RQ)}|\RQ)&=\frac{4+5}{2}=4.5 \kern1em \text{ and}\\
  I(\RZ_{V}^{(\RQ)}|\RQ)&=\frac{2.5+2.5+2+2-4.5}{3}=1.5.
  \end{align*}
  Hence, $\RS\leq\RCO'<\RCO$, and so omniscience is not optimal. 
  
  It can be seen the benefit of making $\RQ$ random is that it allows the edge $\ope$ to be removed a fraction (half) of the time. Note that a complete removal of the edge, i.e., with $\RQ=0$ deterministically, is suboptimal, because it diminishes the secrecy capacity, i.e.,
  \begin{align*}
  	I(\RZ_{V}^{(\RQ)}|\RQ=0) &= \frac{2+2+2+2-4}{3} = \frac43<1.5.\TheoremSymbol
  \end{align*}
\end{Example}

The following example shows that Theorem~\ref{thm:ub:sl} is useful for more general sources that are not necessarily hypergraphical.

\begin{Example}\label{eg:ub:sl:2}
  Let $\RX_0,\RX_1$ and $\RJ$ be uniformly random and independent bits, and define
  \begin{align*}
  \RZ_1&:=(\RJ,\RX_0\oplus\RX_1)\\
  \RZ_2&:=(\RX_0,\RX_1)\\
  \RZ_3&:=\RX_{\RJ}.
  \end{align*}
  With $A=V=[3]$ and $S=D=\emptyset$, we have $\CS=I(\RZ_V)=1$ and $\RCO=H(\RZ_V)-I(\RZ_V)=2$. Now, with $\RZ_i':=\RZ_i$ for $i\in \Set {2,3}$ and
  \begin{align*}
  	\RZ_1' &:= 
  	\begin{cases}
  		(\RJ,\RX_0\oplus \RX_1) & \text {if $\RX_0\neq \RX_1$, i.e., $\RX_0\oplus \RX_1=1$,}\\
  		\RX_0\oplus \RX_1 & \text {otherwise}, 
  	\end{cases}
  \end{align*}
  (or, alternatively, $\RZ'_1:=(2\RJ-1)\cdot (\RX_0\oplus \RX_1)$ which takes value from $\Set {-1,0,1}$.) It follows that
  \begin{align*}
  	H(\RZ'_1)&\utag{a}=H(\RX_0\oplus \RX_1,\RZ'_1)\\
  	&= H(\RX_0\oplus \RX_1) + H(\RZ'_1|\RX_0\oplus \RX_1)\\
  	&\utag{b}= 1+0.5  = 1.5<2=H(\RZ_1)
  \end{align*}
  where \uref{a} is because $\RZ'_1$ determines $\RX_0\oplus \RX_1$; \uref{b} is because $H(\RZ'_1|\RX_0\oplus \RX_1=0)=0$ while $H(\RZ'_1|\RX_0\oplus \RX_1=1)=H(\RJ)=1$.
  Using this, it can be shown that $(\CS',\RCO')$ is given by $\CS'=I(\RZ_V')=1$ and $\RCO'=H(\RZ_V')-I(\RZ_V')=2.5-1=1.5$. By Theorem~\ref{thm:ub:sl}, we have $\RS\leq\RCO'<\RCO$, and so the omniscience strategy is not optimal. Indeed, it can be shown that $\RS=1.5$ by the result of \cite{chan17csr}. 
  
  As an interesting side note, although the omniscience strategy is not optimal, it can be non-asymptotic, for instance, by setting $n=1$, $\RK=\RX_{\RJ}$, $\RF_1=\RJ$, $\RF_2=\RX_{1-\RJ}$ and $\RF_3$ deterministic. However, it seems impossible to achieve $\RS\leq 1.5$ non-asymptotically. To construct an asymptotic scheme, note that the fraction of time $\RX_0\oplus \RX_1=0$ is $1/2$ almost surely as $n\to `8$ by the law of large number. Whenever $\RX_0\oplus \RX_1=0$, both user $1$ and $2$ knows. In particular, user~$2$ can recover $\RX_{1-\RJ}$ even without knowing $\RJ$ since $\RX_0=\RX_1$. Hence,  $\RX_{\RJ}$ can potentially be used as a secret key bit without omniscience of the source, i.e., without user $2$ knowing $\RJ$ all the time. To do so, however, the public discussion must be chosen carefully in order not to let the wiretapper know the time instances when $\RX_0=\RX_1$. This can be done by an asymptotic scheme, where the realizations of $\RJ$ for the time instances when $\RX_0\neq \RX_1$ are concatenated and then truncated/zero-padded by user $1$ to form a sequence of length $n/2+\sqrt {n}$. Then, the sequence can be revealed in public as $\RF_1$, which does not leak any information about the time instances where $\RX_0=\RX_1$. Since user $2$ can recover $\RX_0\oplus \RX_1$ from his private observation, he can recover the sequence of realizations of $\RZ_{\RJ-1}$ almost completely (close to a fraction of $1$ by the law of large number) and reveal it in public as $\RF_2$. Hence, almost the entire sequence of $\RX_{\RJ}$ can be recovered by everyone and used as the secret key.
\end{Example}

%We conclude this section with a conjecture, the proof of which remains elusive.
%\begin{Conjecture}
%	For Example~\ref{eg:ub:sl:2}, $\RS=1.5$ and cannot be achieved non-asymptotically. 
%\end{Conjecture}

%For the hypergraphical source, the result may be further strengthened due to the following observation.
%
%For more general source model, the lower bound on $\RS$ and the sufficient condition for the optimality of omniscience may be slack and not necessary respectively, as  shown by the following example. 
%\begin{Example}
%	Consider the source in Example~\ref{eg:fls}. Let $(A,S,D)=(\Set{1,2},`0,\Set{3})$. It can be shown that $\CS=I(\RZ_1\wedge\RZ_2|\RZ_3)=1$ and $\RCO=1$, achieved with $n=1$, $(\RK,\RF)=(\RZ_1,\RZ_3)$. However, the lower bound provided by~\eqref{eq:rs:lb} is trivial because $1=I(\RZ_1\wedge\RZ_2|\RZ_3)\leq J_{\opW}(\RZ_1\RZ_2|\RZ_3)\leq H(\RZ_1)=1$. On the other hand, it can be shown that $\RS=\RCO=1$, and so the trivial lower bound is slack.
%\end{Example} 
%
%We conclude this section by providing the proofs of Theorem~\ref{thm:user}-\ref{thm:ub:sl}, and Corollary~\ref{cor:ub:s}-\ref{cor:ub:o}.

\section{Single-Letter Lower Bounds and\\ Sufficient Conditions}
\label{sec:lowerbound}

In this section, we derive general single-letter bounds on $\RS$. We will first extend the definitions in \eqref{eq:mi} to characterize $\CS$. 

\subsection{Fractional Partition Information}
\label{sec:I`l}

We will use the following generalization of the notion of partitions. 
For a finite set $U$, a \emph{fractional partition} is a non-negative set function $\lambda: 2^U\to `R_+$ that satisfies 
\begin{align}
  \sum_{B\subseteq U: i\in B} `l(B)= 1 \kern1em \forall i\in U. \label{eq:`l}
\end{align}
For a set family $\mcH\subseteq 2^U`/\Set{`0}$, we use $`L(U,\mcH)$ to denote the set of fractional partitions $\lambda$ whose support lies within $\mcH$, i.e.,
\begin{align}
  \op{supp}(\lambda):=\{B \subseteq U \mid \lambda(B) >0\}  \subseteq \mcH. \label{eq:`L}
\end{align}
For instance, the indicator function $\chi_{\mcP}$ of a partition $\mcP\in \Pi(U)$ is a fractional partition, i.e.,
\begin{align}\label{eq:partition}
  `l(B)=\chi_{\mcP}(B)=\begin{cases}
  1 & B\in \mcP\\
  0 & \text{otherwise.}
  \end{cases}
\end{align}
However, the notion of fractional partition is more general. An important case of interest is 
\begin{align}\label{eq:co-partition}
  `l(C)=\frac{\chi_{\mcP}(U`/ C)}{\abs{\mcP}-1}=\begin{cases}
  \frac1{\abs{\mcP}-1} & U`/ C\in \mcP\\
  0 & \text{otherwise,}
  \end{cases}
\end{align}
for some $\mcP\in \Pi'(U)$. This is called a \emph{co-partition}.
\begin{Definition}[\mbox{\cite[(4.4b)]{chan15mi}}] 
  \label{def:I_`l} 
  For a finite set $U$ with size $\abs{U}>1$, $`l\in `L(U,2^U`/\Set {`0,U})$ and a random vector $(\RZ'_U,\RW')$, define the (conditional) \emph{fractional partition information} as
  \begin{align}\label{eq:I_`l}
    \kern-.2em I_{`l}(\RZ'_U|\RW')&:=H(\RZ'_U|\RW') 
    %\notag\\ &\kern3em
    \kern-.2em - \kern-.2em \sum_{\mathclap{B\in 2^U`/\Set{`0,U}}} \kern-.2em`l(B)H(\RZ'_B|\RZ'_{U`/ B}\kern-.1em, \kern-.1em \RW').\kern-.2em
  \end{align}
  For $\mcP\in \Pi'(U)$, $I_{\mcP}(\RZ'_U|\RW')$~\eqref{eq:IP} and $J_{D,\mcP}(\RZ'_V|
  \RW')$~\eqref{eq:JD} are the special cases of $I_{`l}(\RZ'_U|\RW')$ when $`l$ satisfies~\eqref{eq:co-partition} and \eqref{eq:partition} respectively. 
\end{Definition}
The secrecy capacity was first characterized using fractional partitions in \cite{csiszar08}. $I_{`l}$ for different values of $`l$ was introduced in \cite{chan15mi} as the space of information measures relating various multivariate information measures.

The secrecy capacity in the case without silent users can be characterized by $I_{`l}$ as follows:
\begin{Proposition}[\mbox{\cite[Theorem~3.1]{csiszar08}}] 
  \label{pro:Cs:CN08}
  For $S=\emptyset$, 
  \begin{align}
    \CS=\min_{`l\in `L(V`/ D,\mcH)} I_{`l}(\RZ_{V`/ D}|\RZ_D) \label{eq:Cs:CN08}
  \end{align}	
  where $\mcH:=\Set{B\subseteq V`/D:`0\neq B\nsupseteq A}$.
\end{Proposition}
%The divergence upper bound~\cite[(26)]{csiszar04} on $\CS$ is the special case by restricting $`l$ to be the co-partition~\eqref{eq:partition}. Chan and Zheng~\cite{chan10md} shown that such restriction on $`l$ is not admissible to~\eqref{eq:Cs:CN08} in general unless $A=V`/ D$ (no trusted helpers).

Like $I_{\mcP}(\RZ_V)$~\eqref{eq:IP}, $I_{`l}(\RZ'_V)$~\eqref{eq:I_`l} is also non-negative~\cite{csiszar08}, which is a consequence of the Shearer-type lemma in~\cite{madiman10}. We will need the stronger statement below (with an equality condition): % (and also extend to include conditioning).
\begin{Proposition}[\mbox{\cite[Lemma~6.1]{chan15mi}}] 
  \label{pro:shearer}
  For any random vector $(\RZ'_U,\RW')$ and $`l\in `L(U,2^U`/\Set{`0,U})$, we have $I_{`l}(\RZ'_{U}|\RW')\geq 0$ with equality iff 
  \begin{align}
    I(\RZ'_B\wedge\RZ'_{U`/ B}|\RW')=0  \kern1em \forall B\in \op{supp}(`l),  \label{eq:I_`l=0}
  \end{align}
  which is the condition in terms of Shannon's mutual information for the fractional partition information to be zero.
\end{Proposition}
For completeness, we will prove a stronger version of the result in Appendix~\ref{sec:shearer:proof}.

As pointed out in~\cite[Footnote~17]{chan15mi}, $I_{`l}$ \eqref{eq:I_`l} also satisfies the data processing inequality~\cite[(5.20b)]{chan15mi}. We will use the following more elaborate version:
\begin{Lemma}
  \label{lem:DPI}
  For any random vector $(\RZ'_U,\RW',\RY')$, $`l\in `L(U,2^U`/\Set{`0,U})$ and $i\in U$, we have
\begin{equation}
  I_{`l}(\RZ'_U|\RW') \geq I_{`l}(\RZ''_U|\RW')-`d, \label{eq:DPI1}
 \end{equation}
 where
 \begin{align*}
    \RZ''_j &:=\begin{cases}
    \RY', & j=i\\
    \RZ'_j, & j\in U`/\Set{i}
    \end{cases}\kern1em \text {and}\\
     `d &:={`1(\sum_{B\in 2^U `/ \{\emptyset,U\}} \kern-1.5em`l(B)-1`2)}\, I(\RY'\wedge\RZ'_{U`/\{i\}}|\RW',\RZ'_i).  %\label{eq:`d}.
\end{align*}
  Furthermore, 
  \begin{equation}
    I_{`l}(\RZ'_U|\RW')\geq I_{`l}(\RZ'_U|\RW',\RY')-`d+`g, \label{eq:DPI2}
   \end{equation}
    %`d&:={(\textstyle\sum_{B\in \mcH}`l(B)-1)} \,I(\RY'\wedge\RZ'_{U`/\{i\}}|\RW',\RZ'_i)  \kern-2em\notag\\
    where 
    $$
    `g:=\min_{\substack{B\in \op{supp}(`l) :\\ i\in B}} \, \max_{j\in U`/ B}I(\RY'\wedge\RZ'_j|\RW')
    $$
  and $`d$ is as defined for \eqref{eq:DPI1}.
  %n.b. $\sum_{B}`l(B)\in`1[1, \abs{U}`2]$ and $\bigcup_{B\in \op{supp}(`l)} U`/B\neq \emptyset$~\cite{chan10phd}.
\end{Lemma}
\begin{Proof}
	See Appendix~\ref{sec:DPI:proof}.
\end{Proof}
\eqref{eq:DPI1} and~\eqref{eq:DPI2} can be viewed as the extensions of the following well-known data processing inequality in the bivariate case $U=\{1,2\}$ for the Markov chain $\RZ'_1-\RZ'_2-\RY'$ (i.e., $I(\RZ'_1\wedge\RY'|\RZ'_2)=0$):
\begin{subequations}\label{eq:DPI:2}
  \begin{align}
    &I(\RZ'_1\wedge\RZ'_2)\geq I(\RZ'_1\wedge\RY')\kern1em \text{and}   \label{eq:DPI:2:1}\\
    &I(\RZ'_1\wedge\RZ'_2)\geq I(\RZ'_1\wedge\RZ'_2|\RY')+I(\RZ'_1\wedge\RY').   \label{eq:DPI:2:2}
  \end{align}
\end{subequations}
More precisely, $`L(U, 2^U`/\Set{`0,U})$ contains only the partition (co-partition) $`l$ with $`l(\Set{1})=`l(\Set{2})=1$. With $i=2$ and $\RW'=`0$, \eqref{eq:DPI1} reduces to~\eqref{eq:DPI:2:1}, while  \eqref{eq:DPI2} reduces to~\eqref{eq:DPI:2:2}.

\subsection{General lower bound}
\label{sec:LB}

The lower bound on $\RS$ will be stated and derived using the following single-letter expression that extends the partition Wyner common information~\eqref{eq:JWP}:
\begin{Definition}
  \label{def:JW} 
  For a finite set $U$ with size $\abs {U}>1$, random vector $(\RZ'_U,\RW')$ and $`l\in `L(U,2^U`/\Set{`0,U})$,
  \begin{subequations}
    \label{eq:JW}
    \begin{align}
      J_{\opW,`l}(\RZ'_U|\RW'):=\inf \{I(\RW \wedge \RZ'_U|\RW')\mid \\
      I_{`l}(\RZ'_U|\RW',\RW)=0 \}. 
    \end{align}
  \end{subequations}
  For any partition $\mcP\in \Pi'(U)$, $J_{\opW,\mcP}(\RZ'_U|\RW')$~\eqref{eq:JWP} is the special case when $`l$ satisfies~\eqref{eq:co-partition}. In the bivariate case $U=\Set{i,j}$ where $i\neq j$, it reduces to $J_{\opW}(\RZ'_i\wedge\RZ'_j|\RW')$~\cite{wyner75}.
\end{Definition}
A bound on the support size of $\RW$ similar to Wyner common information can be imposed to make the computation more tractable.
\begin{Proposition}
  \label{pro:|w|}
  It is admissible to have $\abs{W}\leq \abs{Z'_U}\abs {W'}$ in~\eqref{eq:JW}, in which the ``$\inf$" can be replaced by ``$\min$".
\end{Proposition}
\begin{Proof}
  This follows from Lemma~\ref{lem:|W|} and \eqref{eq:`GJW} in Appendix~\ref{sec:LB:proof}.
\end{Proof}
The desired lower bound on $\RS$ is:
\begin{Theorem}
  \label{thm:LB}
  For the general scenario $S\subsetneq A$, if we have
  \begin{subequations} \label{eq:`luh}
    \begin{align}
      &\CS=I_{`l}(\RZ_U|\RZ_D)\kern1em \text{for some $`l\in `L(U,\mcH)$ where}\\
      &U\subseteq V\text{ is such that } V`/D`/ S\subseteq U\subseteq V`/ D \kern1em \text{and}\\
      &\mcH:=\Set{B\subseteq U\mid `0\neq B\not\supseteq A\cap U},
    \end{align}
  \end{subequations}
  then the communication complexity is lower bounded as
  \begin{align}
    \RS&\geq \liminf_{n\to`8}\frac{1}{n}H(\RF|\tRZ_D) \notag\\
    &\geq J_{\opW,`l}(\RZ_U|\RZ_D)-I_{`l}(\RZ_U|\RZ_D), \label{eq:rs:lb}
  \end{align}
  which is in fact a lower bound on the total discussion rate of the trusted users, since $H(\RF|\tRZ_D)=H(\RF_{V`/D`/S}|\tRZ_D)$.
\end{Theorem}
\begin{Proof}
  See Appendix~\ref{sec:LB:proof}.
\end{Proof}

\subsection{With helpers}
\label{sec:helpers}

In this section, we specialize the results to the scenario $A\subseteq V$ but $S=D=`0$. This will be assumed throughout the section, unless otherwise stated.

\begin{Theorem}
  \label{thm:LB:A<V}
  Let $`L^*(A,\RZ_V)$ be the set of optimal fractional partitions in the characterization~\eqref{eq:Cs:CN08} of $\CS$ by $I_{`l}$, and 
  \begin{align}
    \mcH:=`1\{B,V`/B\mid B\in \op{supp}(`l^*), `l^*\in`L^*(A,\RZ_V)`2\}.  \label{eq:H:A<V}
  \end{align}
  Then, 
  \begin{subequations}
    \label{eq:LB:A<V}
    \begin{align}
      \RS&\geq \max_{`l^*\in`L^*(A,\RZ_V)} J_{\opW,`l^*}(\RZ_V)-\CS \label{eq:LB:A<V:JW}\\
      &\geq I_{`l}(\RZ_V)-\CS,  \label{eq:LB:A<V:I_`l}
    \end{align}
  \end{subequations}
  for any $`l\in`L(V,\mcH)$.
\end{Theorem}
\begin{Proof}
	See Appendix~\ref{sec:LB:A<V:proof}.
\end{Proof}
\begin{Theorem}
  \label{thm:OO:A<V}
  $\RS=\RCO$ if, for $\mcH$ defined in~\eqref{eq:H:A<V},
  \begin{align}
    \exists `l\in `L(V,\mcH),I_{`l}(\RZ_V)=H(\RZ_V),
  \end{align}
  i.e., $H(\RZ_B|\RZ_{V`/B})=0$ for all $B\in \op{supp}(`l)$.
\end{Theorem}
\begin{Proof}
This follows immediately from Theorem~\ref{thm:LB:A<V} by making use of Proposition~\ref{pro:Cs:CN08} with $D=S=`0$.
\end{Proof}
%\begin{Remark}
Note that~\eqref{eq:LB:A=V:JW} is the special case of~\eqref{eq:LB:A<V:JW} when $`l$ is chosen to be~\eqref{eq:co-partition} for the fundamental partition $\mcP^*(\RZ_V)$, and~\eqref{eq:LB:A=V:JD} is the special case of~\eqref{eq:LB:A<V:I_`l} when $`l$ is chosen to be~\eqref{eq:partition} for the fundamental partition $\mcP^*(\RZ_V)$. The sufficient condition~\eqref{eq:OO:A=V} in Theorem~\ref{thm:OO:A=V} also follows from Theorem~\ref{thm:OO:A<V} when $`l$ satisfies~\eqref{eq:partition} for the fundamental partition $\mcP^*(\RZ_V)$.
%\end{Remark}

The following is an example taken from~\cite[Example A.1]{chan15mi}. It has the property that the optimal $`l^*$ to~\eqref{eq:Cs:CN08} is not the co-partition (i.e., the divergence upper bound~\cite[(26) in Example~4]{csiszar04} is loose), unlike the case with no helpers in Theorem~\ref{thm:LB:A=V}. %Hence, one cannot apply Theorem~\ref{thm:LB:A=V} directly by Theorem~\ref{thm:LB:A<V}.
\begin{Example}
  \label{eg:fls}
  Let $\RZ_4,\RZ_5$ and $\RZ_6$ be independent uniformly random bits, and define 
  \begin{align*}
    \RZ_1&:=\kern2.2em\RZ_5\oplus\RZ_6\\
    \RZ_2&:=\RZ_4\kern2.2em\oplus\RZ_6\\
    \RZ_3&:=\RZ_4\oplus\RZ_5\kern1.5em
  \end{align*}
  With $V:=[6]$ and $A=[3]$, it can be shown that 
  \begin{align*}
    &`L^*(A,\RZ_V)=\Set{`l^*}\text{ where } \\
    &`l^*(B)\in\Set*{0,\frac 1 4}\text{ for } B\subseteq V`/\Set{`0} \text{ and } \\
    &\op{supp}(`l^*)=\Big\{\Set{2,3,4},\Set{1,3,5},\Set{1,2,6},\\
    &\kern6emV`/\Set{1},V`/\Set{2},V`/\Set{3}\Big\}.
  \end{align*}
  Consider the fractional partition $`l$ with
  \begin{align*}
    `l(B):=\begin{cases}
    \frac12 & \text{ if } V`/B\in \op{supp}(`l^*) \\
    0 & \text{ otherwise. } 
    \end{cases}
  \end{align*}
  It can be checked that $`l\in`L(V,\mcH)$ with $\mcH$ defined in~\eqref{eq:H:A<V}, using the fact that every $i\in V$ appears in exactly two subsets of supp($`l$), which is a subset of $\mcH$. We also have $I_{`l}(\RZ_V)=H(\RZ_V)$ since $H(\RZ_B|\RZ_{V`/B})=0$ for all $B\in \op{supp}(`l)$. It follows from Theorem~\ref{thm:OO:A<V} that $\RS=\RCO$, and so omniscience is optimal.  
\end{Example}

The following example shows that not only is the lower bound~\eqref{eq:LB:A<V} loose, but the sufficient condition is also not necessary, even for a simple PIN~(Definition~\ref{def:PIN}).
\begin{Example}
  \label{eg:pin:3}
  Let $\RX_{a}$ and $\RX_{b}$ be uniformly random and independent bits. With $V:=[3]$, let 
  \begin{align*}
    \RZ_1&:=\RX_{a}\\
    \RZ_2&:=(\RX_{a},\RX_{b})\\
    \RZ_3&:=\kern2.0em\RX_{b},
  \end{align*} 
  which is the same as the source in Example~\ref{eg:slack} but with $\RX_{\opc}$ removed.
  Consider $A=\Set{1,3}$, $S=D=`0$. Then, $\CS$ in~\eqref{eq:Cs:CN08} is 1, where the extremal\footnote{All other solutions can be expressed as convex combinations of the extremal solutions.} optimal solutions are $`l^{(1)}$ and $`l^{(2)}$ defined as 
  \begin{subequations}
    \begin{align*}
      &`l^{(1)}(B)=\begin{cases}
      1  & \text{ if } B\in \Set{\Set{1,2},\Set{3}} \\
      0  & \text{ otherwise, }
      \end{cases}\kern1em\text{and}\\
      &`l^{(2)}(B)=\begin{cases}
      1  & \text{ if } B\in \Set{\Set{1},\Set{2,3}} \\
      0  & \text{ otherwise. }
      \end{cases}
    \end{align*}
  \end{subequations}
  It can be achieved non-asymptotically with $n=1$ and $(\RK,\RF)=(\RX_{a},\RX_{a}\oplus\RX_{b})$. The support~\eqref{eq:H:A<V} for the optimal $`l$'s is $\mcH=\Set{\Set{1,2},\Set{2,3},\Set{1},\Set{3}}$. The lower bound on $\RS$ given by Theorem~\ref{thm:LB:A<V} is trivial since $`l^{(1)}$ and $`l^{(2)}$ are the only feasible choices supported by $\mcH$, i.e., it is easy to see that $`L(V,\mcH)=\Set{`l^{(1)},`l^{(2)}}$. However, by the result of \cite{chan17csr}, it can be shown that omniscience is indeed optimal in this case, i.e., $\RS=1$.
\end{Example}

\subsection{With Silent Users}
\label{sec:silent}

This section considers the scenario $S\subsetneq A=V$, i.e, all the users are active but some of them may be forced to be silent. This will be assumed throughout the section unless otherwise stated. We begin by providing an alternate characterization of the secrecy capacity~in \cite[Theorem~6]{amin10a}.

\begin{Proposition}
  \label{prop:CS:s}
  \begin{subequations}
    \label{eq:CS:s}
    \begin{alignat}{2}[left={\CS=}\empheqlbrace]
      & \min_{i\in S} I(\RZ_{V\setminus S}\wedge\RZ_i) & \kern1em &\text{if $\abs{V\setminus S}=1$} \label{s:alpha}\\
      & \min\{\alpha,I(\RZ_{V\setminus S})\} & & \text{if $\abs{V\setminus S}>1$} \label{s:alphaI},
    \end{alignat}
  \end{subequations}
  where $\alpha:=\min_{i\in S} I(\RZ_{V\setminus S}\wedge\RZ_i)$.
\end{Proposition}
\begin{Proof}
	See Appendix~\ref{sec:silent:proof}
\end{Proof}
The result can be easily extended to the case with untrusted helpers, i.e., $S\subsetneq A=V`/D$ with $D$ possibly non-empty. To be precise, we have
\begin{alignat*}{2}[left={\CS=}\empheqlbrace]
  &\min_{i\in S} I(\RZ_{(V\setminus D)\setminus S}\wedge\RZ_i|\RZ_D)&\kern1em&\text{if }\abs{(V\setminus D)\setminus S}=1 \notag\\ \bigskip
  &\min\{\alpha,I(\RZ_{(V\setminus D)\setminus S}|\RZ_D)\} && \text{if }\abs{(V\setminus D)\setminus S}>1 %\label{cs:s:d},
\end{alignat*}
where $\alpha:=\min_{i\in S} I(\RZ_{(V\setminus D)\setminus S}\wedge\RZ_i|\RZ_D)$.

We now turn our attention to lower bounding $\RS$ for the case with $S\subsetneq A=V$. For this, we introduce some convenient notation, starting with the definition
\begin{equation}
     S^*:=\Set{i\in S \mid I(\RZ_{V`/S}\wedge\RZ_i)=`a},
     \label{eq:S*}
\end{equation}
where $`a$ is as defined in Proposition~\ref{prop:CS:s}. We extend the notation introduced in Theorem~\ref{thm:LB:A=V}: for any $U \subseteq V$, the $\mcP^*$ in the subscripts of $J_{\opW,\mcP^*}(\RZ_U)$, $J_{\opD,\mcP^*}(\RZ_U)$ and $I_{\mcP^*}(\RZ_U)$  denotes the fundamental partition $\mcP^*(\RZ_U)$. 

Applying the lower bound in Theorem~\ref{thm:LB:A=V} with an appropriate choice of $U$ and $\mcP\in\Pi'(U)$ yields the following result.
\begin{Theorem}
  \label{thm:LB:S}
  %For the multiterminal source model with $S\subsetneq A=V$, we have
  \begin{subequations}
    \label{eq:RS:S}
    \begin{align}[left={\RS\geq}\empheqlbrace]
      &J_{\opW,\mcP^*}(\RZ_{V`/S})-I(\RZ_{V`/S})  \notag\\
      & \kern2em \text{ if } I(\RZ_{V`/S})<\alpha\text{ and } \abs{V`/S}>1, \label{RS:S:1}\\ 
      &\max_{i\in S^*}J_{\opW}(\RZ_{V`/S}\wedge\RZ_i)-\alpha,  \notag\\
      & \kern2em \text{ if } \abs{V`/S}=1,\notag\\
      & \kern2em  \text{ or, if } I(\RZ_{V`/S})>\alpha \text{ and } \abs{V`/S}>1, \label{RS:S:2}\\ 
      &\max_{i\in S^*} J_{\opW,\mcP^*}(\RZ_{(V`/S)\cup\{i\}})-\alpha, \notag\\
      & \kern2em  \text{ if } I(\RZ_{V`/S})= \alpha \text{ and } \abs{V`/S}>1, \label{RS:S:3}
    \end{align}
  where $S^*$ is as defined in \eqref{eq:S*}.
  \end{subequations}
\end{Theorem}
\begin{Proof}
	See Appendix~\ref{sec:silent:proof}
\end{Proof}
The lower bounds in Theorem~\ref{thm:LB:S} can be weakened by replacing $J_{\opW,\mcP}$ with the more easily computable $J_{\opD,\mcP}$. Using arguments similar to those in Section~\ref{sec:nohelper}, we arrive at the following sufficient condition for $\RS=\RCO$ to hold.

\begin{Theorem}
  \label{thm:OO:S}
  %For a multiterminal source model with $S\subsetneq A=V$, we have 
  $\RS=\RCO$ in either of the following scenarios:
  \begin{enumerate}[label=(\roman*)]
  \item\label{OO:S:1} $H(\RZ_C|\RZ_{V`/C})=0, \forall C\in\mcP^*(\RZ_{V`/S})$, when $\abs{V`/S}>1$ and $I(\RZ_{V`/S})<\alpha$,
  \item\label{OO:S:2} $\exists i\in S^*$ such that $H(\RZ_{V`/S}|\RZ_i)=0$, when $\abs{V`/S}=1$, or when $\abs{V`/S}>1$ and $I(\RZ_{V`/S})>\alpha$,
  \item\label{OO:S:3} $\exists i\in S^*$ such that $H(\RZ_C|\RZ_{V`/S`/C},\RZ_i)=0, \forall C\in\mcP^*(\RZ_{V`/S})\cup\{i\}$, when $\abs{V`/S}>1$ and $I(\RZ_{V`/S})=\alpha$,
  \end{enumerate}
  where $S^*$ is as defined in \eqref{eq:S*}.
\end{Theorem}
\begin{Proof}
	See Appendix~\ref{sec:silent:proof}
\end{Proof}
%Before setting out to prove Theorems~\ref{thm:LB:S} and \ref{thm:OO:S}, let us illustrate the results using a simple example.

\begin{Example}
  \label{ex:PIN:S}
  Consider the PIN in Example~\ref{eg:pin:3} with $A=V=[3]$. We consider the following cases:
  \begin{compactitem}
  \item $S=\{3\}$: It is easy to verify that $I(\RZ_{V`/S})=1$ with $\mcP^*(\RZ_{V`/S})=\{\{1\},\{2\}\}$, and $\alpha=I(\RZ_{\{1,2\}}\wedge\RZ_3)=1=I(\RZ_{V`/S})$. It is obvious that $S^*=S=\{3\}$. Therefore, the condition for Theorem~\ref{thm:OO:S}.\ref{OO:S:3} holds and so $\RS=\RCO$. 
  \item $S=\{2\}$:	
    Again, it is easy to verify that $I(\RZ_{V`/S})=I(\RZ_{\{1,3\}})=0$ and $\mcP^*(\RZ_{V`/S})=\{\{1\},\{3\}\}$. Also, $\alpha=I(\RZ_{\{1,3\}}\wedge\RZ_2)=2>I(\RZ_{V`/S})$. Now, as $H(\RZ_3|\RZ_1)=1>0$, Theorem~\ref{thm:OO:S}.\ref{OO:S:1} fails to confirm whether $\RS=\RCO$. However, it is easy to see that $\CS=0$ and $\RCO=2$, which follows using Theorem~6 of \cite{amin10a} and Proposition~\ref{prop:CS:s}. Therefore, $\RS=0$ holds trivially, and hence $\RS<\RCO$. 
  \item $S=\{1,3\}$:	
    In this case, we have $\abs{V`/S}=1$ and see that $\alpha=\min\{I(\RZ_2\wedge\RZ_3), I(\RZ_2\wedge\RZ_1)\}=1$, with $S^*=S=\{1,3\}$. However, it turns out that $H(\RZ_2|\RZ_i)=1>0, i=1,3$, and hence Theorem~\ref{thm:OO:S}.\ref{OO:S:2} is unable to conclude whether $\RS=\RCO$.
  \end{compactitem}
  We remark here that for the special case of a hypergraphical source (as defined in Definition~\ref{def:BN}), the sufficient conditions in Theorem~\ref{thm:OO:S} can be strengthened to a necessary and sufficient condition for $\RS=\RCO$. (See Theorem~\ref{thm:hyp:OO:S}.) Using the stronger result, we can show that $\RS=\RCO$ holds for the last case when $S=\Set{1,3}$.  
\end{Example}

\subsection{The Hypergraphical Source with Silent Users}
\label{sec:hypsilent}

In this section, we restrict our attention to the hypergraphical source with silent users, i.e, $S\subsetneq A=V$. The goal of this section is to strengthen the sufficient conditions for $\RS=\RCO$ given in Theorem~\ref{thm:OO:S}. We will show that the strengthened conditions are both necessary and sufficient for $\RS=\RCO$ to be valid, as promised in Example~\ref{ex:PIN:S}. 

The idea is based on the following observation.

\begin{Proposition}
  \label{prop:hypred}
  For any hypergraphical source, $(V,E,`x)$, $\CS,\RS$ and $\RCO$ remain unchanged by removing any hyperedge $e'\in E$ such that $`x(e')\subseteq S$.
\end{Proposition}
\begin{Proof}
	See Appendix~\ref{sec:proof:hypsilent}
\end{Proof}
Thanks to this fact we will assume that the hypergraphical sources considered later in this section satisfy
\begin{gather}
  \forall e\in E, `x(e)\not\subseteq S. \label{hypred}
\end{gather}

Using \eqref{hypred}, the lower bound in Theorem~\ref{thm:LB:S} can be strengthened to the following for the hypergraphical source.
\begin{Theorem}
  \label{thm:hyp:LB:S}
  For any hypergraphical source $(V,E,`x)$ with $S\subsetneq A=V$, we have
   \begin{subequations}
    \label{eq:RS:hypS}
  \begin{align}[left={\RS\geq}\empheqlbrace]
    &J_{\opW,\mcP^*}(\RZ_{V`/S})-I(\RZ_{V`/S})  \notag\\
    & \hspace{0.7cm}\text{ if } I(\RZ_{V`/S})<\alpha\text{ and } \abs{V`/S}>1, \label{RS:hypS:1}\\ %\bigskip
    &J_{\opW,(V`/S)\cup\{\{i\}\mid i\in S^*\}}(\RZ_{(V`/S)\cup S^*})-\alpha,  \notag\\
    & \hspace{0.7cm}\text{ if } \abs{V`/S}=1,\notag\\
    & \hspace{0.7cm} \text{ or, if } I(\RZ_{V`/S})>\alpha \text{ and } \abs{V`/S}>1, \label{RS:hypS:2}\\ %\bigskip
    &J_{\opW,\mcP^*(\RZ_{V`/S})\cup\{\{i\}\mid i\in S^*\}}(\RZ_{(V`/S)\cup S^*})-\alpha, \notag\\
    & \hspace{0.7cm} \text{ if } I(\RZ_{V`/S})=1\text{ and } \abs{V`/S}>1, \label{RS:hypS:3}
  \end{align}
  where $S^*$ is as defined in \eqref{eq:S*}.
    \end{subequations}
\end{Theorem}
\begin{Proof}
	See Appendix~\ref{sec:proof:hypsilent}
\end{Proof}
The results of Theorem~\ref{thm:hyp:LB:S} can be used to obtain sufficient conditions for $\RS=\RCO$ to hold, by following the same steps as in the proof of Theorem~\ref{thm:OO:S}. Fortunately, it turns out that those conditions are also necessary, a fact that can be proved using the idea of decremental secret key agreement highlighted in \cite{chan16isit}.

\begin{Theorem}
  \label{thm:hyp:OO:S}
  For any hypergraphical source $(V,E,`x)$ with $S\subsetneq A=V$, we have $\RCO=\RS$ iff
  \begin{enumerate}[label=(\roman*)] 
  \item\label{OO:hypS:1} $H(\RZ_C|\RZ_{V`/C})=0, \forall C\in\mcP^*(\RZ_{V`/S})$, when $\abs{V`/S}>1$ and $I(\RZ_{V`/S})<\alpha$,
  \item\label{OO:hypS:2} $H(\RZ_{V`/S}|\RZ_{S^*})=0$, when $\abs{V`/S}=1$ or, if $\abs{V`/S}>1$ and $I(\RZ_{V`/S})>\alpha$,
  \item\label{OO:hypS:3} $H(\RZ_C|\RZ_{((V`/S)\cup S^*)`/C})=0, \forall C\in\mcP^*(\RZ_{V`/S})$, when $\abs{V`/S}>1$ and $I(\RZ_{V`/S})=\alpha$.
  \end{enumerate}
\end{Theorem}
\begin{Proof}
	See Appendix~\ref{sec:proof:hypsilent}
\end{Proof}
%We conclude this section by stating the proofs of the above results.

\subsection{With Untrusted Users}
\label{sec:untrusted_users}

The lower bounds and sufficient conditions derived so far (Theorems~\ref{thm:OO:A=V}--\ref{thm:hypergraph} and Theorems~\ref{thm:LB:A<V}--\ref{thm:hyp:OO:S})  %(Theorems~\ref{thm:LB:A=V}, \ref{thm:OO:A=V}, \ref{thm:hypergraph}, \ref{thm:LB:A<V}, \ref{thm:OO:A<V}, \ref{thm:LB:S}, \ref{thm:OO:S}, \ref{thm:hyp:LB:S} and \ref{thm:hyp:OO:S}) 
can all be extended to the case with untrusted helpers by further conditioning on $\RZ_D$ in the entropies, as in Theorem~\ref{thm:LB}. For hypergraphical sources, this is equivalent to removing the hyperedges incident on $D$, as in Corollary~\ref{cor:DX}.

\section{Challenges}
\label{sec:challenge}

In this section, 
we conclude our work by explaining some challenges that remain and techniques that potentially improve the results derived so far. 
%We first show that, even in the case with no helpers or silent users, the sufficient condition in Theorem~\ref{thm:OO:A=V} for the optimality of omniscience may not be necessary, and the lower bound in Theorem~\ref{thm:LB:A=V} on $\RS$ may not be tight. Then, we will also explain some potential improvements of the lower bound by a change of scenario.

\subsection{Limitation}
\label{sec:limitation}

We first show that the sufficient condition in Theorem~\ref{thm:OO:A=V} for the optimality of omniscience may not be necessary for the following example from \cite{chan16itw}, resolving the conjecture therein.

\begin{Example}
  \label{eg:snn}
  Let $\RX_{a},\RX_{b},\RX_{\opc}$ and $\RX_{\opd}$ be uniformly random and independent bits, and define 
  \begin{align*}
    \RZ_1&:=\RX_{a}\\
    \RZ_2&:=\RX_{b}\\
    \RZ_3&:=\RX_{\opc}\\
    \RZ_4&:=(\RX_{a},\RX_{b},\RX_{\opc}\oplus\RX_{\opd})\\
    \RZ_5&:=(\RX_{a},\RX_{b},\RX_{\opd}).
  \end{align*}
  With $A=V:=[5]$ and $S=`0$,
  it can be shown that 
  \begin{align*}
    &\CS=I(\RZ_V)=1\text{ with }
    \mcP^*(\RZ_V)=\Set{\Set{1},\Set{2},\Set{3},\Set{4,5}}\\
    &\RCO=H(\RZ_V)-\CS=3\\
    &J_{\opW,\mcP^*}(\RZ_V)=J_{\opD,\mcP^*}(\RZ_V)=3<H(\RZ_V)=4
  \end{align*}
  with $\RW=(\RX_{\Set{a,b,\opc}})$.
  To achieve the capacity, we can choose for $n=1$ 
  \begin{align*}
    \RK&:=\RZ_1=\RX_{a}\\
    \RF_4&:=\RX_{\opc}\oplus \RX_{\opd}\\
    \RF_5&:=(\RX_{a}\oplus\RX_{b},\RX_{a}\oplus \RX_{\opd}),
  \end{align*}
  which also achieves omniscience at the minimum rate.
  
  Note that the sufficient condition~\eqref{eq:OO:A=V} for the optimality of omniscience does not hold because
  \begin{align*}
    H(\RZ_{\Set {4,5}}|\RZ_{\Set {1,2,3}}) &= H(\RX_{\opd},\RX_{\opc}\oplus \RX_{\opd}|\RX_{\opc}) = 1 >0.
  \end{align*}
  The following result will show that omniscience is indeed optimal for this example, and so the sufficient condition is not necessary. Furthermore, since the sufficient condition is derived from the lower bound~\eqref{eq:LB:A=V} on $\RS$, the bound is also loose for this example.
\end{Example}

\begin{Proposition}
  \label{pro:snn}
  For Example~\ref{eg:snn}, $\RS=\RCO$.
\end{Proposition}

\begin{Proof}
	See Appendix~\ref{sec:proof:snn}.
\end{Proof}

\subsection{Potential Improvements}
\label{sec:LB:scenario_change}

%\input{LB_scenario_change}

%In this section, we will introduce some general techniques to strengthen the upper bound on $\RS$. In particular, we will make use of the following monotonicity of $(\CS,\RS)$ with respect to certain change of scenario, namely the vector $(A,S,D,V,\RZ_V)$ of user sets and private source 
In this section, we give some potential improvements of the lower bound by a change of scenario.
%More precisely, in the presence of both silent users and trusted helpers, one may apply Theorem~\ref{thm:LB} directly. 
%Alternatively, similar to Section~\ref{sec:UB_scenario_change}, we can reduce it to the case with either silent users or trusted helpers by a change of scenario, and then apply the specialized bounds in Theorems~\ref{thm:LB:A<V} and~\ref{thm:LB:S}.
\begin{Theorem} 
  \label{thm:LB:user1}
  $\CS$ and $\RS$ remain unchanged by the following change of user sets:
  \begin{compactenum}[(i)]
  \item  A vocal untrusted user is turned into a silent untrusted user, and a new trusted helper with the same private source as the original vocal untrusted user is added. That is to say, $(S,V)$ becomes $(S\cup\Set{i}, V\cup\Set{i'})$ for some $i\in D`/S$ and with $i'\not\in V$ being a new user with private source $\RZ_{i'}=\RZ_{i}$ identical to that of $i$.
  \item A trusted helper $i\in V`/A`/S`/D$ with $H(\RZ_i|\RZ_j)=0$ for some vocal user $j\in V`/S$ is removed, i.e., $V$ becomes $V`/\Set{i}$.
  \end{compactenum}
\end{Theorem}
\begin{Proof}
	See Appendix~\ref{sec:proof:po:improve}
\end{Proof}
\begin{Theorem} 
  \label{thm:LB:user2}
  Suppose $(\CS, \RS)$ becomes $(\CS', \RS')$ by one of the following change of user sets:
  \begin{compactenum}[(i)]
  \item a silent user is removed, i.e., $(A,S,D,V)$ becomes $(A,S`/\Set{i},D`/\Set{i},V`/\Set{i})$ for some $i \in S\cap D$, or $(A`/\Set{i},S`/\Set{i},D,V`/\Set{i})$ for some $i\in A\cap S$.
  \item a silent active user is turned into a vocal active user, i.e., $S$ becomes $S`/\Set{i}$ for some $i\in A\cap S$.
  \end{compactenum}
  Then, $\CS'\geq\CS$. If equality holds, then $\RS'\leq\RS$.
\end{Theorem}
\begin{Proof}
	See Appendix~\ref{sec:proof:po:improve}
\end{Proof}
\begin{Example}
  \label{eg:hyp:4}
  Let $\RX_{a}$ and $\RX_{b}$ be independent uniformly random bits. Consider the PIN in Example~\ref{eg:pin:3} but with user $4$ added so that the private source consists of
  \begin{align*}
    \RZ_1&:=\RX_{a}\\
    \RZ_2&:=(\RX_{a},\RX_{b})\\
    \RZ_3&:=\kern2.0em\RX_{b}\\
    \RZ_4&:=\kern2.0em\RX_{b}
  \end{align*}
  Suppose $(A,S,D)=([3],\Set{1,3},`0)$. It can be shown that $\CS=\RCO=1$, which is achievable non-asymptotically with $n=1$ and $(\RK,\RF)=(\RZ_1,\RF_2)=(\RZ_a,\RZ_a\oplus\RZ_b)$. We can apply (ii) in Theorem~\ref{thm:LB:user1} to remove the trusted user 4, since $H(\RZ_4|\RZ_2)=0$ and $2\in V`/S$. With $V$ changed to $V'=\Set{1,2,3}$, the $\CS$ and $\RS$ remain unchanged. %i.e., $\min\Set{I(\RZ_2\wedge\RZ_1),I(\RZ_2\wedge\RZ_3)}=1=\CS$ by~\eqref{eq:Cs:S}. 
  Since the model is hypergraphical (in particular, a PIN), we can apply Theorem~\ref{thm:hyp:LB:S} to show that $\RS$ of the new scenario is at least 1, %$I_{\Set{1,2,3}}(\RZ_{\Set{1,2,3}})=1$,
  and so $\RS=\RCO=1$ in the original scenario by Theorem~\ref{thm:LB:user1}
\end{Example}
The following conjectures, if proven correct, can further improve the lower bound~\eqref{eq:rs:lb}. They are true if one can prove the stronger conjecture in~\cite{MKS16} that private randomization does not decrease $\RS$.
\begin{Conjecture}
  $\RS$ does not increase by 
  \begin{compactenum}[(i)]
  \item making a trusted helper active provided that the private source of the helper determines that of another active user.
  \item forcing a vocal active user silent if its source is determined by that of another vocal user.
  \end{compactenum}
  \label{conj:RS}
\end{Conjecture}
\begin{Example}
  \label{eg:pin:3'}
  Consider the PIN in Example~\ref{eg:pin:3} with $V=[3]$. Let $(A,S,D)=(\Set{1,3},`0,`0)$. As discussed in Example~\ref{eg:pin:3}, the lower bound~\eqref{eq:rs:lb} fails to show $\RS\geq 1$. However, if the conjecture above is proved, then we could apply (i) in the conjecture to turn the trusted helper into an active vocal user, in which case $\RS=1$ as described in the previous example for the new scenario.
\end{Example}

%\section{Conclusion}
%\label{sec:conclusion}
%\input{conclusion}

%\clearpage
%\newpage
\appendices

\makeatletter
\renewcommand{\thesubsection}{\thesectiondis-\arabic{subsection}}
\renewcommand{\thesubsectiondis}{\arabic{subsection}.}
\@addtoreset{equation}{section}
\@addtoreset{Theorem}{section}
\renewcommand{\theTheorem}{\thesection.\arabic{Theorem}}
\renewcommand{\theequation}{\thesection.\arabic{equation}}
\renewcommand{\theparentequation}{\thesection.\arabic{parentequation}}
\@addtoreset{Lemma}{section}
\renewcommand{\theLemma}{\thesection.\arabic{Lemma}}
\@addtoreset{Corollary}{section}
\renewcommand{\theCorollary}{\thesection.\arabic{Corollary}}
\@addtoreset{Example}{section}
\renewcommand{\theExample}{\thesection.\arabic{Example}}
\@addtoreset{Remark}{section}
\renewcommand{\theRemark}{\thesection.\arabic{Remark}}
\@addtoreset{Proposition}{section}
\renewcommand{\theProposition}{\thesection.\arabic{Proposition}}
\@addtoreset{Definition}{section}
\renewcommand{\theDefinition}{\thesection.\arabic{Definition}}
\@addtoreset{Claim}{section}
\renewcommand{\theClaim}{\thesection.\arabic{Claim}}
\@addtoreset{Subclaim}{Theorem}
\renewcommand{\theSubclaim}{\theLemma.\arabic{Subclaim}}
\makeatother

\section{Proof of Proposition~\ref{pro:USD}}
\label{sec:proof:USD}
  Consider $j\in D$ first. As will be useful to a later result, we will prove the stronger statement that $\RU_j$ neither increases $\CS$ nor decreases $\RS$ \emph{even when $\RU_j$ is a public randomization~\cite{chan10phd} observed by everyone in addition to the wiretapper}, i.e., with~\eqref{eq:F}  modified to have $\RF_i$ depend directly on $\RU_j$. To do so, it suffices to show that the recoverability~\eqref{eq:recover} and secrecy~\eqref{eq:secrecy} constraints continue to hold even if $\RU_j$ is chosen to be deterministic. 
More precisely, for any $`d>0$, let
\begin{subequations}
	\label{eq:USD:t1}
	\begin{align}
		\kern-.5em U_j(`d)&:=\bigg\{u\in U_j \mid \notag\\
		&\kern-1em\Pr(\exists i\in A, \RK\neq `q_i(\tRZ_i,\RF)\mid \RU_j=u)\leq `d, \label{eq:USD:t1a}\\
		&\kern-1em\frac1n `1[\log\abs {K}-H(\RK|\RF,\tRZ_D,\RU_j=u)`2]\leq `d\bigg\}.\label{eq:USD:t1b}\kern-.5em
	\end{align}	
\end{subequations}
We have the desired result if $U_j(`d_n)\neq `0$ for some $`d_n\to 0$ since, by choosing $\RU_j$ to be deterministically equal to any element in $U_j(`d_n)$, \eqref{eq:USD:t1a} and \eqref{eq:USD:t1b} implies \eqref{eq:recover} and \eqref{eq:secrecy} respectively. Indeed, not only can we show that $U_j(`d)\neq `0$, i.e., $\Pr(\RU_j\in U_j(`d))>0$, but also that
%\begin{Claim} 
	\begin{align}
		\lim_{n\to `8} \Pr(\RU_j\in U_j(`d))=1\kern1em \forall `d>0.
		\label{eq:claim:1}
	\end{align}
%\end{Claim}
%\begin{Proof}
	Let $U'_j(`d)$ be the set $U_j(`d)$ in \eqref{eq:USD:t1} with only \eqref{eq:USD:t1a} (but not \eqref{eq:USD:t1b}) imposed. Similarly, let $U''_j(`d)$ to be the set with only \eqref{eq:USD:t1b} imposed. It follows that 
	\begin{align*}
		U_j(`d)=U'_j(`d)\cap U''_j(`d)
	\end{align*}
	and so, by the union bound,
	\begin{align*}
		\Pr(\RU_j\in U_j({`d})) \geq 1- \Pr(\RU_j\not\in U'_j{(`d)})- \Pr(\RU_j\not\in U''_j{(`d)}).
	\end{align*}
	It suffices to show that the last two probabilities go to $0$ asymptotically in $n$. By the Markov inequality,
	\begin{align*}
		\Pr(\RU_j\not\in U'_j(`d))&\leq \tfrac{\Pr(\exists i\in A, \RK\neq `q_i(\tRZ_i,\RF))}{`d}\\
		\Pr(\RU_j\not\in U''_j(`d))&\leq \tfrac{\frac1n `1[\log\abs {K}-H(\RK|\RF,\tRZ_D,\RU_j)`2]}{`d}.
	\end{align*}
	The bounds go to zero as desired by \eqref{eq:recover} and \eqref{eq:secrecy}, hence completing the proof of \eqref{eq:claim:1}.
%\end{Proof}

Consider the remaining case $j\in S$. (Unlike the previous case, we do not consider $\RU_j$ is a public randomization here.)
Note that
\begin{align}
	I(\RU_j\wedge \tRZ_{V`/\Set{j}},\RF)=0   \label{eq:pvu=0}
\end{align}
because $\RF$ in~\eqref{eq:F} does not depend on $\RU_j$ as user~$j$ is silent, and the $\RU_j$ is independent of $\tRZ_{V`/\Set {j}}$ by the assumption~\eqref{eq:USD}. 
We will show that this implies that
%\begin{Claim}
	\begin{align}
		\lim_{n\to `8} \frac 1 n I(\RU_j\wedge \RK |\RF,\tRZ_{D})=0 \label{eq:pvuj=0}
	\end{align}
%\end{Claim}
%\begin{Proof}
	Since $\abs{A}\geq 2$, there exists another active user, say $i\in A`/\Set{j}$. By the recoverability condition~\eqref{eq:recover} for user~$i$ (which does not depend on $\RU_j$), we have
	\begin{align*}
		\lim_{n\to `8}\Pr\{\RK\neq  `q_i(\tRZ_i,\RF)\}=0
	\end{align*}
	which gives
	\begin{align*}
		I(\RU_j\wedge \RK|\RF,\tRZ_D)&\utag{a}\leq   I(\RU_j\wedge \tRZ_i,\RF|\RF,\tRZ_D)+n`d_n 
		%&\utag{b}\leq I(\RU_j\wedge \tRZ_{V`/\Set{j}},\RF)+n`d_n  \\
		%&
		\utag{b}=n`d_n
	\end{align*}
	for some $`d_n\to 0$. Here, \uref{a} follows from Fano's inequality, and \uref{b} is because 
	\begin{align*}
		I(\RU_j\wedge \tRZ_i,\RF|\RF,\tRZ_D)&\leq I(\RU_j\wedge \tRZ_i,\RF,\tRZ_D) \\
		&\leq I(\RU_j\wedge \tRZ_{V`/\Set{j}},\RF),
	\end{align*}
	which equals zero by~\eqref{eq:pvu=0}, completing the proof of \eqref{eq:pvuj=0}. 
%\end{Proof}

Now, by~\eqref{eq:secrecy},
\begin{align*}
	0&=\lim_{n\to `8}\frac 1 n `1[\log \abs{K}-H(\RK|\RF,\tRZ_D)`2] \\
	&=\lim_{n\to `8}\frac 1 n `1[\log \abs{K}-H(\RK|\RF,\tRZ_D,\RU_j)`2]\\
	&=\lim_{n\to `8}\frac 1 n `1[\log \abs{K}-\max_{\Ru\in \RU_j} H(\RK|\RF,\tRZ_D,\RU_j=\Ru)`2]
\end{align*}
where the second equality follows from~\eqref{eq:pvuj=0}.
Hence, by setting $\RU_j=\Ru$ deterministically, \eqref{eq:secrecy} remains to hold (since $\max_{\Ru\in \RU_j} H(\RK|\RF,\tRZ_D,\RU_j=\Ru)=H(\RK|\RF,\tRZ_D)$ in the above). Furthermore, \eqref{eq:recover} (without $\RU_j$) also hold by~\eqref{eq:pvuj=0}. This completes the proof of the proposition.

\section{Proofs for Section~\ref{sec:nohelper}}
\label{sec:A=V:proof}
%\input{nohelper_proof}
%In the following, we will present the proofs for Theorems~\ref{thm:LB:A=V} and \ref{thm:hypergraph}, and Proposition~\ref{pro:hyp:JW}.

\subsection{Proof of Theorem~\ref{thm:LB:A=V}}
\label{sec:LB:A=V:proof}
It is enough to prove \eqref{eq:LB:A=V:JW}, since \eqref{eq:LB:A=V:JD} then follows from \eqref{eq:DJH}.
Let $\RU_V$ be the optimal sequence of randomization that achieves $\RS$, and let $\RS^{\NR}(\tRZ_V)$ be the communication complexity when the source $\RZ_V$ is changed to $\tRZ_V$ instead (see \eqref{eq:tRZi} for the definition of $\tRZ_V$). Then, 
\begin{align*}
\RS(\RZ_V)\utag{a}\geq \frac{1}{n}\RS^{\NR}(\tRZ_V)  &\utag{b}\geq \frac{1}{n}[\CW(\tRZ_V)-I(\tRZ_V)]  \\
&\utag{c}\geq \frac{1}{n}[nJ_{\opW,\mcP^*}(\RZ_V)-I(\tRZ_V)]  \\
&\utag{d}= J_{\opW,\mcP^*}(\RZ_V)-I(\RZ_V)  
\end{align*}
\begin{compactitem}
	\item To explain \uref{a}, note that the secrecy capacity of the new scenario is $n\CS$, since randomization does not change the secrecy capacity~\cite{csiszar04}. Any optimal scheme that achieves $\RS$ for the original scenario can therefore be translated directly to a scheme that achieves $n\RS$ for the new scenario without randomization.
	\item \uref{b} is by Proposition~\ref{pro:rsk_pre_lb} with $\RZ_V$ replaced by $\tRZ_V$, and $\CW(\tRZ_V)$ denoting the corresponding multi-letter multivariate Wyner common information~\eqref{eq:A=V:CW}.
	\item \uref{c} follows from 
	\begin{align}
		\CW(\tRZ_V)\geq n J_{\opW,\mcP^*}(\RZ_V),\label{eq:CWJWP}
	\end{align}
	which will be argued in more detail later.
	\item To explain \uref{d}, note that for all $B \subseteq V$,
	\begin{align*}
		H(\tRZ_{B}) &=H(\RZ_{B}^n,\RU_B)\\
		&=nH(\RZ_{B})+H(\RU_B),
	\end{align*}
	which gives
	\begin{align*}
		I_{\mcP}(\tRZ_V)=nI_{\mcP}(\RZ_V)+I_{\mcP}(\RU_V)
	\end{align*}
	for all $\mcP\in \Pi'(V)$. Since $I_{\mcP}(\RU_V)=0$ by the fact that the $\RU_i$'s are mutually independent~\eqref{eq:U}, the above equation implies $I(\tRZ_V)=nI(\RZ_V)$ as desired.
\end{compactitem}
To explain \eqref{eq:CWJWP}, 
consider the optimal sequence in $n'$ of $\RL$ to $\CW(\tRZ_V)$. By standard arguments,
\begin{align*}
H(\RL)&\geq I(\tRZ^{n'}_V\wedge \RL)\geq I(\RZ^{nn'}_V\wedge \RL)\\
&=H(\RZ_V^{nn'})-H(\RZ_{V}^{nn'}|\RL)\\
&=\sum_{t=1}^{nn'}H(\RZ_{Vt})-\sum_{t=1}^{nn'} H(\RZ_{Vt}|\RZ_V^{t-1},\RL)
\end{align*}
where the second inequality follows from the usual data processing inequality (see~\eqref{eq:DPI:2:1}) since $\RZ_V^n$ is determined by $\tRZ_V$, and so, we have the Markov chain $\RL"-"\tRZ^{n'}_V"-"\RZ^{nn'}_V$.  
Let $\RJ$ be the usual time-sharing random variable uniformly distributed over $[nn']$ and independent of everything else, namely $(\tRZ_V^{n'},\RL)$, and define 
\begin{align*}
	\RW_\RJ:=(\RJ, \RZ_V^{\RJ-1},\RL).
\end{align*}
Then, the above inequality gives
\begin{align}
	\label{eq:CW:A=V:SL:O}
	\frac1{n'} H(\RL) &\geq n I(\RZ_{V\RJ}\wedge\RW_{\RJ}).
\end{align}
On the other hand, we can also bound $I_{\mcP^*}$ in the constraint~\eqref{eq:A=V:CW2} of $\CW$ as follows:
\begin{align*}
	I_{\mcP^*(\tRZ_V)}(\tRZ_V^{n'}|\RL) 
	&\geq I_{\mcP^*(\RZ_V)}(\RZ_V^{nn'}|\RL)\\
	&=\frac{1}{\abs{\mcP^*}-1}\bigg[\sum_{C\in \mcP^*}\underbrace{H(\RZ_C^{nn'}|\RL)}_{`(1)}-\underbrace{H(\RZ_V^{nn'}|\RL)}_{`(2)}\bigg]
\end{align*}
where, as in the statement of the theorem, $\mcP^*$ denotes $\mcP^*(\RZ_V)$ for convenience. In the above inequality, we have applied $\mcP^*(\tRZ_V)=\mcP^*(\RZ_V)$ and the data processing inequality~\cite[(5.20b)]{chan15mi} since $\RZ_i^n$ is determined by $\tRZ_i$. (See also~\eqref{eq:DPI1} with $I_{`l}$ reduces to $I_{\mcP}$ by restricting $`l$ to \eqref{eq:co-partition}.) %, and .
Expanding $`(1)$ and $`(2)$ by the chain rule,
\begin{align*}
	`(1) &= \kern-.2em \sum_{t=1}^{nn'} \kern-.2em H(\RZ_{Ct}|\RL,\RZ_{C}^{t-1}) \kern-.2em \\
&\geq \kern-.2em \sum_{t=1}^{nn'}H(\RZ_{Ct}|\RL,\RZ_{V}^{t-1})\kern-.1em =\kern-.1em nn' H(\RZ_{C\RJ}|\RW_{\RJ})\\
	`(2) &= \sum_{t=1}^{nn'}H(\RZ_{Vt}|\RL,\RZ_V^{t-1})= nn' H(\RZ_{V\RJ}|\RW_\RJ).
\end{align*}
Altogether, we have
\begin{align}
\frac{1}{n'} I_{\mcP^*(\tRZ_V)}&(\tRZ_V^{n'}|\RL) \notag \\
&\geq\kern-.2em \frac{n}{\abs{\mcP^*}-1}\kern-.2em `1[\sum_{C\in \mcP^*} \kern-.2em H(\RZ_{C\RJ}|\RW_{\RJ})-H(\RZ_{V\RJ}|\RW_\RJ)`2]  \notag \kern-3em \\
&=nI_{\mcP^*}(\RZ_{V\RJ}|\RW_{\RJ}),\label{eq:CW:A=V:SL:C}
\end{align}
Now, for $`d\geq 0$, define
\begin{align}
	`G(`d)&:=\sup_{P_{\RW|\RZ_V}: I_{\mcP^*}(\RZ_V|\RW)\leq `d} H(\RZ_V|\RW), \label{eq:`G:t}
\end{align}
where the supremum is over all possible choices of the conditional distribution $P_{\RW|\RZ_V}$. The expression depends implicitly on the distribution $P_{\RZ_V}$. It follows that
\begin{align*}
	`G`1(\tfrac{1}{nn'} I_{\mcP^*(\tRZ_V)}(\tRZ_V^{n'}|\RL)`2) &\geq H(\RZ_{V\RJ} | \RW_{\RJ}) 
\end{align*}
since $\RZ_{V\RJ}$ has the same distribution as $\RZ_{V}$ and so the conditional distribution $P_{\RW_{\RJ}|\RZ_{V\RJ}}$ is a feasible solution to \eqref{eq:`G:t} with $`d$ chosen appropriately from the bound \eqref{eq:CW:A=V:SL:C} on $I_{\mcP^*}(\RZ_{V\RJ}|\RW_{\RJ})$. Together with~\eqref{eq:CW:A=V:SL:O}, we have
\begin{align*}
	\CW(\tRZ_V) &\geq \lim_{n'\to `8} n `1[H(\RZ_{V\RJ})-`G`1(\tfrac{1}{nn'} I_{\mcP^*(\tRZ_V)}(\tRZ_V^{n'}|\RL)`2)`2]\\
	&= n `1[H(\RZ_V) -  \lim_{`d\to 0} `G(`d) `2]
\end{align*}
where the last equality is because $H(\RZ_{V\RJ})=H(\RZ_V)$ and $\tfrac{1}{n'} I_{\mcP^*(\tRZ_V)}(\tRZ_V^{n'}|\RL)$ goes to $0$ as $n'$ goes to $`8$ by the constraint \eqref{eq:A=V:CW2} for $\CW(\tRZ_V)$. It can be shown that $`G(`d)$ is continuous in $`d$ using the same argument as in \cite{wyner75}. For completeness, this is proved for the more general case in Lemma~\ref{lem:|W|} in Appendix~\ref{sec:LB:proof}. Hence,
\begin{align*}
\CW(\tRZ_V) &\geq n`1[H(\RZ_V)-`G(0)`2] \\
       &=nJ_{\opW,\mcP^*}(\RZ_V)
\end{align*}
by the definition~\eqref{eq:JWP} of $J_{\opW,\mcP}$.

\subsection{Proof of Theorem~\ref{thm:hypergraph}}
\label{sec:hypergraph:proof}

To prove Theorem~\ref{thm:hypergraph}, we use the idea of decremental secret key agreement~\cite[Theorem~4.2]{chan16isit}.
%To proof Theorem~\ref{thm:hypergraph}, we will rely on the following property of the MMI~\eqref{eq:I}. 
\begin{Proposition}[\mbox{\cite[Theorem~4.2]{chan16isit}}]\label{th:dska}
	If $\RZ_V$ can be rewritten for some $`0\neq T\subseteq C\in \mcP^*(\RZ_V)$ as
	\begin{align}\label{eq:excess}
		\RZ_i = \begin{cases}
			(\hat\RZ_i,\RX) & \forall i\in T\\
			 \hat\RZ_i & \forall i\in V`/T,
		\end{cases}
	\end{align}
	where $H(\RX)=H(\RX|\hat \RZ_V)>0$, then, we have
	\begin{align}
		\label{eq:reduce_excess}
		H(\RZ'_V)<H(\RZ_V)\kern1em \text{and} \kern1em I(\RZ'_V)=I(\RZ_V)
	\end{align}
	for some function 
	 $\RZ'_i=\vartheta_i(\RZ_i)$ for $i\in V$.
\end{Proposition}
Roughly speaking, when \eqref{eq:OO:A=V} fails for hypergraphical sources, we can identify and reduce excess randomness in the source without changing $\CS$, and so omniscience is not optimal in achieving $\RS$.

The ``if'' case of Theorem~\ref{thm:hypergraph} follows from Theorem~\ref{thm:OO:A=V} directly. To prove the ``only if'' part, suppose to the contrary that 
\begin{align*}
	H(\RZ_C|\RZ_{V\setminus C})>0 \kern1em \text{for some $C \in\mcP^*(\RZ_V)$.}
\end{align*}
For hypergraphical model, this means that 
\begin{align*}
	H(\RX_{e'}|\RZ_{V\setminus C})>0 \kern1em \text{for some $e' \in E$,}
\end{align*}
 i.e., $`x(e')\subseteq C$. Thus, \eqref{eq:excess} holds with $\RX:=\RX_{e'}$, $T:=`x(e')\subseteq C$ and 
 \begin{align*}
 	\hat \RZ_i:=(\RX_e\mid e\in E\setminus e', i\in`x(e)).
 \end{align*}
By Proposition~\ref{th:dska}, we have \eqref{eq:reduce_excess}. With $\RS'$ and $\RCO'$ denoting the communication complexity and the smallest rate of CO for the source $\RZ'_V$, we have 
\begin{align*}
\RS \utag{a}\leq \RS' \leq\RCO' &=H(\RZ'_V)-I(\RZ'_V) \\
&\utag{b}< H(\RZ_V)-I(\RZ_V)=\RCO(\RZ_V),
\end{align*}
where \uref{a} is due to the fact that processing $\RZ_i$'s individually cannot reduce the communication complexity $\RS$; and \uref{b} is by \eqref{eq:reduce_excess}. This completes the proof of Theorem~\ref{thm:hypergraph}. 

\subsection{Proof of Proposition~\ref{pro:hyp:JW}} 
\label{sec:hyp:JW:proof}

First, observe that with $\RW=(\RX_e\mid e \in E^*)$, using the assumption that the random variables $\RX_{e}$'s are mutually independent, we have 
\begin{align*}
\sum_{C\in \mcP^*}H(\RZ_{C}|\RW)&=\sum_{C\in \mcP^*}H(\RX_{\{e\in E \setminus E^*|`x(e)\subseteq C\}}) \\
&=H(\RX_{\{E\setminus E^*\}})=H(\RZ_V|\RW)
\end{align*}
Hence, $I_{\mcP^*}(\RZ_V|\RW)=0$, and so $\RW$ is a feasible solution to $J_{\opW,\mcP^*}(\RZ_V)$. Thus, $J_{\opW,\mcP^*}(\RZ_V)\leq H(\RX_{E^*})$. By~\eqref{eq:DJH},
On the other hand, we also have, by~\eqref{eq:DJH},
\begin{align*}
J_{\opW,\mcP^*}(\RZ_V)&\geq H(\RZ_V)- \sum_{C\in \mcP^*} H(\RZ_C|\RZ_{V`/C})\\
&=H(\RX_E)- \sum_{C\in \mcP^*} H(\RX_{\{e\in E \setminus E^*|`x(e)\subseteq C\}})\\
&=H(\RX_E)-H(\RX_{\{E\setminus E^*\}})\\
&=H(\RX_{E^*})
%\underbrace
\end{align*}
Thus, $J_{\opW,\mcP^*}(\RZ_V)= H(\RX_{E^*})$ with $\RW=(\RX_e\mid e \in E^*)$ being an optimal solution.

\section{Proofs for Section~\ref{sec:upperbound}}
\label{sec:upperbound:proof}

\subsection{Proof of Theorem~\ref{thm:CSRCO}}
\label{sec:CSRCO:proof}

\noindent\underline{Converse proof of $\CS$:}\\[.2em]
We first prove `$\leq$' for \eqref{eq:CSRCO} by making use of the following result that directly extends the technique of the converse proof of \cite[Theorem~2]{csiszar04} and \cite[Theorem~6]{amin10a}.
\begin{Lemma} 
	\label{lem:CSRCO1}
	For any $B\subseteq V`/D`/S$, we have
\begin{subequations}
	\label{eq:CS:*}
\begin{align}
	&\kern-1em \limsup_{n\to `8} \frac1n H(\RK|\RF,\tRZ_{V`/S`/B}) \geq H(\RZ_B|\RZ_{V`/S`/B}) - r(B)\kern-.5em \label{eq:CS:*1} \\
	\begin{split}
	& \text {with} \kern1em r_i:=\limsup_{n\to `8} \frac1n \bigg[ \sum\nolimits_{t\in [r]} H(\RF_{it}\mid \tRF_{it},\tRZ_D)\\
		&\kern4em +H(\tRZ_i|\tRZ_D,\tRZ_{[i-1]},\RK,\RF)-H(\RU_i)\bigg].
	\end{split} \label{eq:CS:*2} 
\end{align}
\end{subequations}
The inequality is satisfied with equality if $B=V`/D`/S$.
\end{Lemma} 
This completes the proof because, by the secrecy constraint~\eqref{eq:secrecy},
\begin{align*}
	\liminf_{n\to`8} \frac1n \log \abs {K} &\leq \limsup_{n\to `8} \frac1n H(\RK|\RF,\tRZ_D)\\
	&= H(\RZ_{V`/D`/S}|\RZ_D) - r(V`/D`/S)
\end{align*}
by the equality case of \eqref{eq:CS:*1}  with $B=V`/D`/S$. Moreover, $r_{V`/D`/S}$ satisfies \eqref{eq:`rSW} because, for any $j\in A$ and $B\subseteq V`/D`/S`/\Set {j}$, the limit in \eqref{eq:CS:*1} is $0$ by Fano's inequality and the recoverability constraint~\eqref{eq:recover} as $j\in V`/S`/B$. (Note that the constraints for $B\ni j$ are redundant.)
\begin{Proof}[Lemma~\ref{lem:CSRCO1}]
	By the assumption~\eqref{eq:U} of the private randomizations and the memorylessness of the private source,
	\begin{align*}
		H(\tRZ_B|\tRZ_{V`/S`/B}) = \sum_{i\in B} H(\RU_i) + n H(\RZ_B|\RZ_{V`/S`/B}).
	\end{align*}
	Alternatively, since $\RF$ is determined by $\tRZ_{V`/S}$ by~\eqref{eq:F}, we have
	\begin{align*}
		H(\tRZ_B|\tRZ_{V`/S`/B}) &= H(\RF,\tRZ_B|\tRZ_{V`/S`/B}) \notag \\
		&= \underbrace{H(\RK,\RF,\tRZ_B|\tRZ_{V`/S`/B})}_{`(1)} - n`d_n
	\end{align*}
	where $`d_n:= \frac1n H(\RK|\RF,\tRZ_{V`/S})$ goes to $0$ as $n\to `8$ by Fano's inequality because $\RK$ can be recovered from $(\RF,\tRZ_{V`/S})$ asymptotically by~\eqref{eq:recover}, due to the assumption $S\subsetneq A$ that there must be at least one vocal active user, i.e, $A\cap (V`/S)\neq `0$. Expanding the last entropy term $`(1)$ by the chain rule gives
	\begin{align*}
		`(1)&=\overbrace{H(\RF|\tRZ_{V`/S`/B})}^{`(2)}\kern-.2em + H(\RK|\RF\kern-.1em,\kern-.1em\tRZ_{V`/S`/B}) \kern-.2em +\overbrace{\kern-.2em H(\tRZ_B|\RK,\kern-.1em\RF\kern-.1em,\kern-.1em\tRZ_{V`/S`/B})}^{`(3)}\\
		`(2)&=\sum_{t\in [r]} \sum_{i\in V`/S} H(\RF_{it}|\tRF_{it},\tRZ_{V`/S`/B})\\
			&\utag{a}=\sum_{i\in B} \sum_{t\in [r]} H(\RF_{it}|\tRF_{it},\tRZ_{V`/S`/B})
			\utag{b}\leq \sum_{i\in B} \sum_{t\in [r]} H(\RF_{it}|\tRF_{it},\tRZ_D)\\
		`(3)&=\sum_{i\in B} H(\tRZ_i|\tRZ_{(V`/S`/B)\cup [i-1]},\RK,\RF)\\ 
			&\utag{c}\leq \sum_{i\in B} H(\tRZ_i|\tRZ_D,\tRZ_{[i-1]},\RK,\RF), 
	\end{align*}
	where \uref{a} is because the entropy terms for $i\not\in B$ are zero by \eqref{eq:F}. Rearranging the terms give \eqref{eq:CS:*} with the desired equality condition because inequalities \uref{b} and \uref{c} hold with equality if $B=V`/D`/S$.
\end{Proof}

\noindent\underline{Characterization of $\RCO$:}\\[.2em]
Next, we prove the characterization of $\RCO$ in \eqref{eq:RS<=RCO}. For each $j\in A$, let
\begin{subequations}
	\label{eq:pzR'}
\begin{align}
	&\rsfsR'(\RZ_{V`/D`/S}|\RZ_{D\cup \Set{j}}):=\{ r_{V`/D`/S} \in `R^{V`/D`/S} \mid \label{eq:pzR'ZV|D} \\ 
	&\kern 2.5em r(B)\geq H(\RZ_B|\RZ_{V`/S`/B},\RZ_j)\;\forall B\subseteq V\kern-.2em`/\kern-.2em D\kern-.2em`/\kern-.2em S\} \label{eq:pzR'ZV|D:SW}\\
	&\rsfsR'(\RZ_D|\RZ_j):=\{ r_D \in `R^D \mid \label{eq:pzR'ZD|j} \\ 
 & \kern2.5em r(B)\geq H(\RZ_B|\RZ_{D`/B},\RZ_j)\;\forall B\subseteq D \}\label{eq:pzR'ZD|j:SW}
\end{align}
\end{subequations}
Note that, by the standard result of independent source coding with side information, $\rsfsR'(\RZ_{V`/D`/S}|\RZ_{D\cup \Set{j}})$ is the set of achievable rate tuple for encoding each components of the source $\RZ_{V`/D`/S}$ independently so that they can be recovered from the codewords given the source $\RZ_{D\cup \Set{j}}$ as side information. The omniscience constraint~\eqref{eq:recover:O2} requires the recoverability simultaneously for all $j\in A$, and so the achievable rate region is 
\begin{align*}
	\bigcap_{j\in A} \rsfsR'(\RZ_{V`/D`/S}|\RZ_{D\cup \Set{j}})
\end{align*}
by the result of normal source network~\cite[Chapter~1]{csiszar2011information}. $`r$ in \eqref{eq:`rmin} is the minimum sum rate over this region because \eqref{eq:`rSW} is composed of \eqref{eq:pzR'ZV|D:SW} for all $j\in A$. Similarly, it can be argued that
\begin{align*}
	\bigcap_{j\in A} \rsfsR'(\RZ_{D}|\RZ_j) \cap \rsfsR(\RZ_D)
\end{align*}
(with $\rsfsR(\RZ_D)$ defined in \eqref{eq:rD})
is the achievable rate region for the omniscience constraint~\eqref{eq:recover:O1} together with the rate constraints~\eqref{eq:pzRZD}. $\bar{`r}$ in \eqref{eq:`rbarmin} is the minimum sum rate over this region. Since the above two rate constraints are separable, the total minimum sum rate is given by $`r+\bar{`r}$, which completes the proof.\footnote{As a side note, although the omniscience strategy here assumes non-interactive discussion, it can be shown as in \cite{csiszar04} that the characterization of $\RCO$ remains unchanged even if interactive discussion is allowed.}\\[.4em]

\noindent\underline{Achievability of $\CS$ via omniscience:}\\[.2em]
We first argue that an optimal solution $r_D$ to \eqref{eq:`rbarmin} exists, and so the omniscience strategy is feasible. (An optimal solution $r_{V`/D`/S}$ to \eqref{eq:`rmin} clearly exists.)
As in \eqref{eq:pzR'}, let
\begin{align*}
	\rsfsR'(\RZ_D) := \Set{r_D\in `R^D \mid r(B)\geq H(\RZ_B|\RZ_{D`/B})\; \forall B\subseteq D}
\end{align*}
which is the set of achievable rate tuples for encoding the components of $\RZ_D$ independently so that they can be recovered from the codewords (without any side-information).
\begin{Proposition}[\cite{schrijver02}]
	\label{pro:polymatroid}
	$\rsfsR(\RZ_D)$ is the downward hull of $\rsfsR(\RZ_D)\cap \rsfsR'(\RZ_D)$.
\end{Proposition}
\begin{Proof}
Since the entropy function is a normalized submodular function~\cite{fujishige78}, $\rsfsR(\RZ_D)$ defines an extended polymatroid and $\rsfsR(\RZ_D)\cap \rsfsR'(\RZ_D)$ is the base of the polymatroid~\cite{schrijver02}. The result follows immediately from the fact that an extended polymatroid is a downward hull of its base.
\end{Proof}
It follows that $\rsfsR(\RZ_D)\cap \rsfsR'(\RZ_D)$ is non-empty since its downward hull $\rsfsR(\RZ_D)$ is clearly non-empty. Furthermore,
\begin{align*}
	r(D)=H(\RZ_D) \kern1em \forall r_D\in \rsfsR(\RZ_D)\cap \rsfsR'(\RZ_D),
\end{align*}
which is the maximum and minimum possible sum rates over $\rsfsR(\RZ_D)$ and $\rsfsR'(\RZ_D)$ respectively.
An optimal solution to \eqref{eq:`rbar} exists because any $r_D\in \rsfsR(\RZ_D)\cap \rsfsR'(\RZ_D)$ is a feasible solution, i.e., for all $j\in A$ and $B\subseteq D$,
\begin{align*}
	r(B) &=r(D)-r(D`/B)\\
	&\geq H(\RZ_D)-H(\RZ_{D`/B})=H(\RZ_B|\RZ_{D`/B}),
\end{align*}
satisfying the constraint~\eqref{eq:`rbarSW}.

It remains to show that the omniscience strategy achieves $\CS$ in \eqref{eq:CS}. Consider $r^*_{V`/D`/S}$ optimal to \eqref{eq:`rmin} and any $r^*_D$ optimal to \eqref{eq:`rbarmin}. Note that $r^*_D\in \rsfsR(\RZ_D)$ by~\eqref{eq:`rbarmin}. Then, by Proposition~\ref{pro:polymatroid}, there exists a non-negative weight vector $`d_D\geq \M{0}$ such that $r^*_D+`d_D\in\rsfsR(\RZ_D)\cap \rsfsR'(\RZ_D)$, which is therefore in $\rsfsR'(\RZ_D)$.
By the usual source coding results~\cite{csiszar2011information}, there exists $(\RF,\RG_D)$ at rate $(r^*_{V`/S},`d_D)$ such that
\begin{align*}
	\lim_{n\to`8}\Pr(\RZ_D^n \neq `f(\RF_D,\RG_D) )=0
\end{align*}
in addition to satisfying the omniscience constraints~\eqref{eq:recover:O}. Note that $\RG_D$ is constructed only for the purpose of proof and will not be discussed in public ($\RF$ is the public discussion as usual). $\RG_D$ is the public discussion saving of our scheme~\eqref{eq:recover:O} compare to~\eqref{eq:recover:O:CN04}. It follows by Fano's inequality that the l.h.s.\ of the secrecy constraint~\eqref{eq:secrecy} can be rewritten as
\begin{align*}
	&\liminf_{n\to`8} \frac1n `1[\log\abs{K} - H(\RK|\RF,\RZ_D^n)`2] \\
	&\kern8em = \liminf_{n\to`8} \frac1n `1[\log\abs{K} - H(\RK|\RF,\RG_D)`2].
\end{align*}
By \cite[Lemma~B.2]{csiszar04}, the r.h.s.\ can be made equal to $0$ (satisfying~\eqref{eq:secrecy}) with 
\begin{align*}
	\lim_{n\to`8} \frac1n \log\abs {K} &\geq  H(\RZ_{V`/S}) - r^*(V`/S) - `d(D)\\
	&= H(\RZ_{V`/D`/S}|\RZ_D) - \overbrace{r^*(V`/D`/S)}^{=`r} \\
	&\kern1em +\underbrace{`1[H(\RZ_D)- r^*(D)- `d(D)`2]}_{=0}.
\end{align*}
This achieves the r.h.s.\ of \eqref{eq:CSRCO} as desired.

\subsection{Proofs of Theorem~\ref{thm:user} and its Corollaries}
\label{sec:user:proof}

\begin{Proof}[Theorem~\ref{thm:user}]
	We will argue that for both the cases (i)-(ii), a capacity achieving scheme for the new scenario is a valid SK generation scheme for the original scenario and hence $\CS'\leq\CS$. In particular, if $\CS'=\CS$, then the capacity achieving schemes for the changed scenario will be capacity achieving for the original scenario as well, and hence $\RS\leq\RS'$. 
	
	\emph{Case (i):}
	Consider turning an achievability scheme in the new scenario to that of the original scenario. To satisfy \eqref{eq:F}, the discussion by the new trusted helper can be performed by the original vocal active user. The original vocal active user can recover the key because the new silent active user can, and so \eqref{eq:recover} holds. Observe that \eqref{eq:secrecy} continues to hold as the untrusted users remain unchanged.
	
	\emph{Case (ii):}
	The constraint on \eqref{eq:F} becomes more stringent with the removal of a vocal helper, while the other constraints, namely, \eqref{eq:recover} and \eqref{eq:secrecy}, remain unchanged. Hence, any capacity achieving scheme for the new scenario continues to be an SK generation scheme for the original one.
\end{Proof}

\begin{Proof}[Corollary~\ref{cor:ub:s}]
	Suppose $\CS'=\CS$. The procedures (i) and (ii) correspond to the cases (i) and (ii) of Theorem~\ref{thm:user}, and so $\RS\leq\RS'$. Also, using \eqref{eq:RS<=RCO} we have $\RS'\leq\RCO'$. Suppose $(\rho,\bar{\rho})$ becomes $(\rho',\bar{\rho}')$ in the new scenario. Note that, $\bar{\rho}=\bar{\rho}'$ if the sets $(A,D)$ remain unchanged. We also have \eqref{eq:CSRCO}, that 
	$$
	\rho'=\rho-\underbrace{\left[H(\RZ_{(V`/D)`/S}|\RZ_D)-H(\RZ_{(V' `/ D)`/S'}|\RZ_D)\right]}_{\beta},
	$$
	by noting that $A`/S'=(V' `/D)`/S'$. Here,
	\begin{align*}
		\beta & = H(\RZ_{V`/S})-H(\RZ_{V'`/S'})\\
		& = H(\RZ_{(S' `/S)\cup(V`/V')}|\RZ_{V'`/S'}) \geq 0.
	\end{align*}
	Hence, by \eqref{eq:RS<=RCO}, $\RCO'=\bar{\rho}'+\rho'=\bar{\rho}+\rho-\beta\leq\bar{\rho}+\rho=\RCO$, which completes the proof of \eqref{ub:s}. Furthermore, $\RS=\RCO$ happens only if $\beta=0$, which is the same as \eqref{OO:O:NC}.
\end{Proof}

\begin{Proof}[Corollary~\ref{cor:ub:o}]
	Suppose, \eqref{ub:o} holds. Then, by \eqref{ub:o:cf}, every active user can recover $\RU^n$. By \cite[Lemma~B3]{csiszar04}, \eqref{eq:secrecy} holds for a choice of $\RK$ as a function of $\RU^n$ of rate $H(\RU|\RZ_D)$. Therefore, $\CS$ can be achieved without public discussion, i.e., $\RS=0$. Now, if \eqref{OO:O:NC} holds in addition, then \eqref{eq:recover:O} holds without discussion, i.e., $\RCO=0$. Conversely, suppose that \eqref{OO:O:NC} fails, i.e., for some $j\in A$, $0<H(\RZ_{V`/S}|\RZ_j)=H(\RZ_D|\RZ_j)+H(\RZ_{V`/S`/\{j\}`/D}|\RZ_j)$ holds. Then, either $H(\RZ_D|\RZ_j)>0$, in which case $\bar{\rho}>0$, or $H(\RZ_{V`/S`/\{j\}`/D}|\RZ_j)>0$, in which case $\rho>0$. In either case, $\RCO>0$ by \eqref{eq:RS<=RCO}.
\end{Proof}

\subsection{Proof of Theorem~\ref{thm:ub:sl}}
\label{sec:ul:sl:proof}

	The idea is to process the original source $\RZ_V$ to $\RZ_V^{(q)}$ possibly with different choices of $q$ at different times. We will show that \eqref{ub:sl:secr} ensures that secrecy in the new scenario guarantees secrecy in the original scenario. On the other hand, \eqref{ub:sl:cs} makes sure that the capacity does not diminish. 
	
	To proceed, divide the $n$-block of time instances into consecutive $n_q$-blocks for $q\in Q$, such that
	\begin{align}
		\sum_{q\in Q}n_q & = n\kern1em \text{ and }
		\lim_{n\to\infty}\frac{n_q}{n}=P_{\RQ}(q)\kern1em \forall q\in Q, \label{eq:nq}
	\end{align}
	where, $P_{\RQ}(\cdot)$ is the distribution of some random variable $\RQ$ taking values in a finite set $Q$. The source is processed block-by-block, with the source corresponding to the $q$-th block being processed to $\RZ_V^{(q)}$. Therefore, $\RZ_V^n$ becomes $\overline{\RZ}_V:=({\RZ_V^{(q)}}^{n_q}|q\in Q)$. There exists a public discussion $\RF$ at the rate $\RCO'$ for the active users to recover $\overline{\RZ}_V$, which can be argued using the \emph{strong law of large numbers} and \eqref{eq:nq}. 
	By Lemma~B3 of \cite{csiszar04}, a key $\RK$ of rate equal to the r.h.s.\ of \eqref{ub:sl:cs} can be recovered by the active users, which satisfies \eqref{eq:secrecy} with $\RZ_D$ replaced by $\overline{\RZ}_D$. 
	
	To complete the proof, we show that \eqref{eq:secrecy} is still valid with $\tRZ_D$. Recalling that $\tRZ_D = \RZ_D^n$, we have
	\begin{align}
		\kern-.8em \frac{1}{n}H(\RK|\RF,\tRZ_D) & = \frac{1}{n}H(\RK|\RF,\RZ_D^n) \notag \\ %\label{pr:ub:sl:1}\\
		& = \frac{1}{n}\left[H(\RK|\RF,\overline{\RZ}_D)-I(\RZ_D^n\wedge\RK|\RF,\overline{\RZ}_D)\right]\kern-.2em.\label{pr:ub:sl:2}\kern-.5em
	\end{align}
	%  where \eqref{pr:ub:sl:1} follows from the fact that randomization does not minimize $\RCO$. 
	Therefore, for some $`d_n\to 0$,
	\begin{align}
		\kern-.7em I(\RZ_D^n\wedge\RK|\RF,\overline{\RZ}_D) & \leq I(\RZ_D^n\wedge\overline{\RZ}_{V`/D},\RK|\RF,\overline{\RZ}_D)\notag\\
		& \utag{a}\leq I(\RZ_D^n\wedge \overline{\RZ}_{V`/D}|\RF,\overline{\RZ}_D)+n\delta_n\kern-.2em\notag\\
		& \utag{b}\leq I(\RZ_D^n\wedge \overline{\RZ}_{V`/D}|\overline{\RZ}_D)+n\delta_n\kern-.2em \notag\\% \label{pr:ub:sl:3}\\
		& =\sum_{q\in Q}n_qI(\RZ_D\wedge\RZ_{V`/D}^{(q)}|\RZ_D^{(q)})+n\delta_n\notag\\
		& \utag{c}= n\delta_n. \label{pr:ub:sl:4}
	\end{align}
	%where, the sequence $\delta_n\to 0$ as $n\to\infty$. 
	\uref{a} %\eqref{pr:ub:sl:3} 
	is by Fano's inequality because $\RK$ is recoverable asymptotically from $\overline{\RZ}_{V`/D}$ given $\overline{\RZ}_D$. \uref{b} is because $\RF$ is determined by $\overline{\RZ}_{V}$. \uref{c} %\eqref{pr:ub:sl:4} 
	follows directly from the assumption~\eqref{ub:sl:secr} in the theorem statement. Therefore, combining \eqref{pr:ub:sl:2} 
	and \eqref{pr:ub:sl:4}, 
	we have $\frac{1}{n}H(\RK|\RF,\tRZ_D)\geq\frac{1}{n}H(\RK|\RF,\overline{\RZ}_D)-\delta_n$, which combined with \eqref{eq:secrecy} with respect to\ $\overline{\RZ}_D$ gives us the desired result.

\section{Proofs for Section~\ref{sec:lowerbound}}
\label{sec:lowerbound:proof}

\subsection{Proof of Shearer-Type Lemma}
\label{sec:shearer:proof}

In this section, we prove a stronger version of Proposition~\ref{pro:shearer} below:

\begin{Lemma}
	\label{lem:Shearer}
	For any random vector $(\RZ'_U,\RW')$ and $`l\in `L(U,2^U`/\Set{`0,U})$,
	\begin{subequations}\label{eq:I_lambda_bound}
		\begin{align}
			I_{`l}(\RZ'_U|\RW')&\geq \max_{B\in 2^U`/\Set{`0,U}}`l(B)I(\RZ'_B\wedge\RZ'_{U`/ B}|\RW')\label{eq:I_`l_lb}\\ 
			I_{`l}(\RZ'_U|\RW')&\leq \sum_{B\in 2^U`/\Set{`0,U}}`l(B)I(\RZ'_B\wedge\RZ'_{U`/ B}|\RW')\label{eq:I_`l_ub}
		\end{align}
	\end{subequations}
	which are the lower and upper bounds of the fractional partition information in terms of Shannon's mutual information.
\end{Lemma}
Note that $I_{`l}(\RZ'_U|\RW')=0$ implies the lower bound~\eqref{eq:I_`l_lb} is zero, which implies~\eqref{eq:I_`l=0}. Conversely, $I_{`l}(\RZ'_U|\RW')=0$ if the upper bound~\eqref{eq:I_`l_ub} is zero, which is implied by~\eqref{eq:I_`l=0}.\footnote{It also follows from Lemma~\ref{lem:Shearer} that $I_{`l}(\RZ'_U|\RW')\rightarrow 0$ is equivalent to $\forall B\in \op{supp}(`l),I(\RZ'_B\wedge\RZ'_{U`/ B}|\RW')\rightarrow 0$, which is not covered by Proposition~\ref{pro:shearer} directly.}
\begin{Proof}
	Without loss of generality, let $U:=[m]$ for some integer $m>1$, and assume the optimal solution to~\eqref{eq:I_`l_lb} is $[l]$ for some $l\in [m]$. By definition~\eqref{eq:I_`l},
	\begin{align*}
		I_{`l}(\RZ'_U|\RW')&=\underbrace{H(\RZ'_U |\RW')}_{`(1)}-\kern-1em \sum_{B\in 2^U`/\Set{`0,U}} \kern-1em `l(B) \underbrace{H(\RZ'_B|\RZ'_{U`/ B},\RW')}_{`(2)}
	\end{align*}
	By the chain rule,
	\begin{align*}
		`(1) &= \sum_{i\in U}\overbrace{\sum_{\substack{B\in 2^U`/\Set{`0,U}:\\ i\in B}}`l(B)}^{\text {$=1$ by \eqref{eq:`l}}}H(\RZ'_i|\RZ'_{[i-1]},\RW')\\
		`(2) &= \sum_{i\in B}H(\RZ'_i|\RZ'_{[i-1]\cup (U`/ B)},\RW').
	\end{align*}
	Exchanging the summations in $`(1)$, substituting both $`(1)$ and $`(2)$ back to the original expression and simplify using the definition of mutual information, we have
	\begin{align*}
		I_{`l}(\RZ'_U|\RW')
		&= \sum_{B\in 2^U`/\Set{`0,U}} \kern-1em `l(B)\sum_{i\in B}I(\RZ'_i\wedge\RZ'_{U`/ B}| \RZ'_{[i-1]},\RW')\\
		&\utag{a}\leq\sum_{B\in 2^U`/\Set{`0,U}} \kern-1em `l(B)\sum_{i\in B}I(\RZ'_i\wedge\RZ'_{U`/ B}| \RZ'_{[i-1]\cap B},\RW')\\
		&\utag{b}=\sum_{B\in 2^U`/\Set{`0,U}} \kern-1em `l(B)I(\RZ'_B\wedge\RZ'_{U`/ B}| \RW')
	\end{align*}
	where \uref{a} follows from the fact that conditioning does not increase entropy, and the equality holds if $[i-1]\subseteq B$; \uref{b} follows from chain rule expansion. This gives the desired upper bound~\eqref{eq:I_`l_ub}. The lower bound~\eqref{eq:I_`l_lb} follows from the equality case when $B=[l]$, and the fact that all the other terms in the sum are non-negative.
\end{Proof}

\subsection{Proof of Lemma~\ref{lem:DPI}}
\label{sec:DPI:proof}

	Consider proving~\eqref{eq:DPI1} first.
	By definition~\eqref{eq:I_`l}, 
	\begin{align*}
		I_{`l}(\RZ'_U|\RW')-I_{`l}(\RZ''_U|\RW')
		&=\overbrace{H(\RZ'_U|\RW')-H(\RZ''_U|\RW')}^{`(1)}\\
		&\kern-9em -\kern-1em\sum_{B\in 2^U`/\Set {`0,U}}\kern-1em `l(B)\underbrace{`1[H(\RZ'_B|\RZ'_{U`/B},\RW')-H(\RZ''_B|\RZ''_{U`/B},\RW') `2]}_{`(2)} 
	\end{align*}
	Note that by the definition of $\RZ''_U$, we have for $B\ni i$ that,
	\begin{align*}
		`(1)=`(2)=H(\RZ'_i|\RZ'_{U`/\Set{i}},\RW')-H(\RZ''_i|\RZ'_{U`/\Set{i}},\RW').
	\end{align*}
	Since the value is independent of $B$, we have
	\begin{align*}
		I_{`l}(\RZ'_U|\RW')-I_{`l}(\RZ''_U|\RW') &= `(1)-`(1)\overbrace{\sum_{B\ni i}`l(B)}^{\text{$=1$ by \eqref{eq:`l}}}  - \sum_{B\not \ni i} `l(B) `(2)\\
		&= -\sum_{B\not \ni i} `l(B) `(2)
	\end{align*} 
	For $B\not \ni i$, it can be shown using standard arguments that
	\begin{align*}
		`(2)&=I(\RZ''_i\wedge\RZ'_B|\RZ'_{U`/B`/\Set{i}},\RW')-I(\RZ'_i\wedge\RZ'_B|\RZ'_{U`/B`/\Set{i}},\RW')\\
		&\leq I(\RZ''_i\wedge\RZ'_B|\RZ'_{U`/B`/\Set{i}},\RW',\RZ'_i)\\
		&\leq \underbrace{I(\RZ''_i\wedge\RZ'_{U`/\Set{i}}|\RW',\RZ'_i)}_{`(3)},
	\end{align*}
	the value of which is independent of $B$. Hence, 
	\begin{align*}
		I_{`l}(\RZ'_U|\RW')-I_{`l}(\RZ''_U|\RW') &\geq -\,`(3)\sum_{B\not \ni i} `l(B) \\
		&=  -\,`(3)`1[\sum_{B}`l(B)-\sum_{B\ni i}`l(B)`2]
	\end{align*}
	which simplifies to $-`d$ as desired by \eqref{eq:`l} and the fact that $\RZ_i''=\RY'$.
	
	Consider proving~\eqref{eq:DPI2}. By definition~\eqref{eq:I_`l}, 
	\begin{align*}
		&I_{`l}(\RZ'_U|\RW')-I_{`l}(\RZ'_U|\RW',\RY')\\
		&\kern3em = \underbrace{I(\RY'\wedge\RZ'_U|\RW')}_{`(4)} - \sum_{B}`l(B) \underbrace{I(\RY'\wedge\RZ'_B|\RZ'_{U`/B},\RW')}_{`(5)}
	\end{align*}
	For $B\not \ni i$, we have by standard techniques that 
	\begin{align*}
		`(5) &\leq I(\RY'\wedge\RZ'_{U`/\Set{i}}|\RW',\RZ'_i),
	\end{align*}
	the value of which is independent of $B$. Hence,
	\begin{align*}
		\sum_{B\not\ni i}`l(B) `(5) \leq `d.
	\end{align*}
	Hence, we have
	\begin{align*}
		I_{`l}(\RZ'_U|\RW')-I_{`l}(\RZ'_U|\RW',\RY') +`d \geq  `(4) - \sum_{B\ni i}`l(B)`(5)
	\end{align*}
	and so it suffices to prove that the r.h.s.\ is at least $`g$. By \eqref{eq:`l} again,
	\begin{align*}
		`(4) - \sum_{B\ni i}`l(B)`(5) 
		&=  \sum_{B\ni i}`l(B) `1[`(4) -`(5)`2]\\
		&=   \sum_{B\ni i}`l(B) I(\RY'\wedge\RZ'_{U`/B}|\RW') \\
		&\geq \sum_{B\ni i}`l(B) \max_{j\in U`/B}I(\RY'\wedge\RZ'_j|\RW')
	\end{align*}
	which is at least $`g$ as desired.

\subsection{Proof of Theorem~\ref{thm:LB}}
\label{sec:LB:proof}

We will show that for any $\CS$-achieving scheme,
\begin{align}
	\limsup_{n\to`8}\frac 1 n H(\RK,\RF|\tRZ_D)\geq J_{\opW,`l}(\RZ_U|\RZ_D) \label{eq:lbt1}
\end{align}
and so we have the desired lower bound~\eqref{eq:rs:lb} since 
\begin{align*}
	H(\RK,\RF|\tRZ_D)=H(\RF|\tRZ_D)+H(\RK|\RF,\tRZ_D)\kern1em \text {and}  \\
	\limsup_{n\to`8}\frac 1 n H(\RK|\RF, \tRZ_D)\geq\CS=I_{`l}(\RZ_U|\RZ_D)
\end{align*}
by~\eqref{eq:secrecy} and the assumption~\eqref{eq:`luh}. To prove \eqref{eq:lbt1}, we will rely on the following fundamental property of $I_{`l}$~\eqref{eq:I_`l} for secret key agreement:
\begin{Lemma}
	\label{lem:I_`l}
	If $\CS=I_{`l}(\RZ_U|\RZ_D)$ as in~\eqref{eq:`luh}, then
	\begin{align}
		\lim_{n\to `8}\frac1 n I_{`l}(\tRZ_U|\RK,\RF,\tRZ_D)=0 \label{eq:I_`l|KF=0}
	\end{align} 
	for any $\CS$-achieving scheme.
\end{Lemma}
It follows that $\RL=(\RK,\RF)$ for any $\CS$-achieving scheme is a feasible solution to
\begin{subequations}
	\label{eq:cw}
	\begin{align}
		C_{\op{W},`l}&:=\inf\limsup_{n\to `8}\frac 1 n H(\RL|\tRZ_D)  \kern1em \text{such that}\label{eq:cw:1} \\
		&\lim_{n\to`8}\frac{1}{n}I_{`l}(\tRZ_U|\RL,\tRZ_D)=0.  \label{eq:cw:2}
	\end{align}
\end{subequations}
In other words,
\begin{align*}
	\limsup_{n\to `8}\frac1 n H(\RK,\RF|\tRZ_D)\geq C_{\op{W},`l}.
\end{align*} 
and the proof is completed by showing that: 
\begin{Lemma}
	\label{lem:SL}
	\begin{align}
	C_{\op{W},`l}=J_{\opW,`l}(\RZ_U|\RZ_D), \label{eq:lbt2}
	\end{align}
	which is a single-letterization of \eqref{eq:cw}.
\end{Lemma}

\begin{Proof}[Lemma~\ref{lem:I_`l}]
	We will show using the data processing inequalities in Lemma~\ref{lem:DPI} that
	\begin{subequations}
		\label{eq:I_`lt}
		\begin{align}
			\frac1 n I_{`l}(\tRZ_U|\RF,\tRZ_D)\leq I_{`l}(\RZ_U|\RZ_D) \kern1em \text {and}\label{eq:I_`lt2}
		\end{align}
		\vspace{-1em}
		\begin{align}
			\lim_{n\to `8}\frac1 n `1\{\log \abs{K}-`1[I_{`l}(\tRZ_U|\RF,\tRZ_D)-I_{`l}(\tRZ_U|\RK,\RF,\tRZ_D)`2]`2\}\leq 0. \label{eq:I_`lt1}
		\end{align}
	\end{subequations}
	Then, for any $\CS$-achieving scheme,
	\begin{align*}
		\lim_{n\to `8}\frac 1 n `1[\log \abs{K}-I_{`l}(\tRZ_U|\RF,\tRZ_D)`2]\geq 0
	\end{align*}
	by~\eqref{eq:I_`lt2} and that the key rate is $\CS=I_{`l}(\RZ_U|\RZ_D)$ by assumption. Applying this to \eqref{eq:I_`lt1} gives $\leq$ in \eqref{eq:I_`l|KF=0}, and the reverse inequality follows from Proposition~\ref{pro:shearer}.
	
	We first show \eqref{eq:I_`lt2}. Applying~\eqref{eq:DPI2} with
	\begin{align*}
		\RZ'_U=\tRZ_U,\; \RY'=\RF_{it}, \kern1em \text {and}\kern1em \RW'=(\tRZ_D,\tRF_{it})
	\end{align*}
	for $ i\in V`/S$ and $t\in [r]$
	gives
	\begin{align}
		\label{eq:eq:I_`lt3}
		I_{`l}(\tRZ_{U}|\tRZ_D,\tRF_{it},\RF_{it})\leq I_{`l}(\tRZ_U|\tRZ_D,\tRF_{it}),
	\end{align}
	because $`g\geq 0$ and $`d=0$ in \eqref{eq:DPI2} as
	\begin{align*}
		I(\RY'\wedge\RZ'_{U`/\Set{i}}|\RW',\RZ'_i)&\leq H(\RY'|\RZ'_i,\RW') \\
		&=H(\RF_{it}|\tRZ_i,\tRZ_D,\tRF_{it})=0
	\end{align*}
	by~\eqref{eq:F}.
	Applying~\eqref{eq:eq:I_`lt3} repeatedly for different $(i,t)$ yields
	\begin{align}
	\label{eq:I_`lt3:chain}
	I_{`l}(\tRZ_U|\tRZ_D)&\geq I_{`l}(\tRZ_U|\tRZ_D,\RF_{11})\notag\\
					&\geq I_{`l}(\tRZ_U|\tRZ_D,\RF_{21})\notag\\
					&\geq\dots \notag\\
					&\geq I_{`l}(\tRZ_U|\tRZ_D,\RF).
	\end{align}
	On the other hand, note that for all $B \subseteq U$, by~\eqref{eq:tRZi},
	\begin{align*}
		H(\tRZ_{B}|\tRZ_D) &=H(\RZ_{B}^n,\RU_B|\RZ_D^n)\\
		&=nH(\RZ_{B}|\RZ_D)+H(\RU_B),
	\end{align*}
	which gives
	\begin{align*}
		I_{`l}(\tRZ_U|\tRZ_D)=nI_{`l}(\RZ_U|\RZ_D)+I_{`l}(\RU_U)
	\end{align*}
	for all $`l\in `L(U,2^U`/\Set{`0,U})$. Since $I_{`l}(\RU_U)=0$ by \eqref{eq:U} that $\RU_i$'s are mutually independent, the above equation implies $I_{`l}(\tRZ_U|\tRZ_D)=nI_{`l}(\RZ_U|\RZ_D)$.
	This together with~\eqref{eq:I_`lt3:chain} give the desired~\eqref{eq:I_`lt2}.
	
	To show~\eqref{eq:I_`lt1}, we again apply~\eqref{eq:DPI2} but with 
	\begin{align*}
		\RZ'_U=\tRZ_U,\; \RY'=\RK, \kern1em \text {and}\kern1em \RW'=(\tRZ_D,\RF) 
	\end{align*}     
	and any $i\in A\cap U$, which is feasible by the assumption $S\subsetneq A$ that there is at least one active vocal user and $U\supseteq V`/D`/S$ from \eqref{eq:`luh}. This gives 
	\begin{align}
		I_{`l}(\tRZ_{U}|\tRZ_D,\RF)\geq I_{`l}(\tRZ_U|\RK,\tRZ_D,\RF)+H(\RK|\tRZ_D,\RF)-n`d_n  \label{eq:eq:I_`lt4}
	\end{align}
	for some $`d_n\to 0$ as $n\to `8$, because
	\begin{compactitem}
		\item the term $`d$ in \eqref{eq:DPI2} goes to $0$ because	\begin{align}
			I(\RY'\wedge\RZ'_{U`/\Set{i}}|\RZ'_i,\RW') \leq H(\RK|\tRZ_i,\tRZ_D,\RF)\leq n`d'_n   \label{eq:I_`lt5}
		\end{align} 
		for some $`d'_n\to 0$ as $n\to `8$ by~\eqref{eq:recover} and Fano's inequality;
	\item the term $`g$ in \eqref{eq:DPI2} can be bounded as follow:
	\begin{align*}
		&\min_{B\in \op{supp}(`l) : B\ni i}\max_{j\in U`/ B}I(\RY'\wedge\RZ'_j|\RW') \\
		&\kern2em\utag{a}\geq \min_{j\in A}I(\RY'\wedge\RZ'_j|\RW') \\
		&\kern2em=\min_{j\in A}I(\RK\wedge\tRZ_j|\RF,\tRZ_D)  \\
		&\kern2em=\min_{j\in A}`1[H(\RK|\RF,\tRZ_D)-H(\RK|\tRZ_j,\RF,\tRZ_D) `2]  \\
		&\kern2em\utag{b}\geq \min_{j\in A}H(\RK|\RF,\tRZ_D)-n`d'_n
	\end{align*}
	where \uref{a} is due to $(U`/B)\cap A\neq `0, \forall B\in 2^U`/\Set {`0,U}$, \uref{b} is by~\eqref{eq:I_`lt5} (with $j$ in place of $i$).
	\end{compactitem}
	\eqref{eq:eq:I_`lt4} implies~\eqref{eq:I_`lt1} by~\eqref{eq:secrecy} as desired. Although not essential for the proof of the lemma here, the reverse inequality $\geq$ of \eqref{eq:I_`lt1} also holds more generally by the definition of $I_{`l}$:
	\begin{align*}
		&\kern-2em I_{`l}(\tRZ_U|\RF,\tRZ_D)-I_{`l}(\tRZ_U|\RK,\RF,\tRZ_D)\\
		&= H(\RK|\RF,\tRZ_D) - \sum_{B} `l(B) H(\RK|\tRZ_{U`/B},\RF,\tRZ_D)\\
		&\leq \log \abs{K}.
	\end{align*}
	Hence, \eqref{eq:I_`lt1} is indeed satisfied with equality.
\end{Proof}

\begin{Proof}[Lemma~\ref{lem:SL}]
	We single-letterize $C_{\op{W},`l}$ as in~\cite{wyner75}:
	\begin{align}
		\label{eq:cw:obj:lb}
		\kern-1.5emH(\RL|\tRZ_D)  & \geq I(\RZ_U^n\wedge \RL|\tRZ_D) \notag\\
		&=H(\RZ_U^n|\tRZ_D)-H(\RZ_U^n|\tRZ_D,\RL)  \notag\\
		&=\sum_{t=1}^nH(\RZ_{Ut}|\RZ_{Dt})-\sum_{t=1}^{n}H(\RZ_{Ut}|\RZ_{U}^{t-1},\tRZ_D,\RL)  \notag\kern-2em\\
		&=\sum_{t=1}^nH(\RZ_{Ut}|\RZ_{Dt})-\sum_{t=1}^{n}H(\RZ_{Ut}|\RZ_{U}^{t-1},\tRZ_D,\RL,\RZ_{Dt})  \notag\kern-4em\\
		&=nI(\RZ_{U\RJ}\wedge \RW_{\RJ}|\RZ_{D\RJ})
	\end{align}
	where $\RJ$ is the usual time-sharing random variable uniformly distributed over $[n]$ and independent of $(\RZ_U,\tRZ_D,\RL)$, and
	\begin{align*}
		\RW_{\RJ}:=(\RJ,\RZ_U^{\RJ-1},\RL,\tRZ_D).
	\end{align*}	
	We can also bound $I_{`l}$ in the constraint~\eqref{eq:cw:2} of $C_{\op{W},`l}$:
	\begin{align*}
		I_{`l}(\tRZ_U|\RL,\tRZ_D) &\geq \underbrace{I_{`l}(\RZ_U^n|\RL,\tRZ_D)}_{`(1)}
	\end{align*}
	by the data processing inequality~\eqref{eq:DPI1} since $\RZ_i^n$ is determined by $\tRZ_i$. By definition~\eqref{eq:I_`l}	
	\begin{align}
		`(1) &=\underbrace{H(\RZ_U^n|\RL,\tRZ_D)}_{`(2)}-\sum_{B}`l(B) \underbrace{H(\RZ_B^n|\RZ_{U`/B}^{n},\RL,\tRZ_D)}_{`(3)}
	\end{align}
	Using the fact that $\tRZ_D = \RZ_D^n$, the r.h.s.\ can be further expanded as follows:
	\begin{align*}
		\label{eq:cw:cons:lb:1}
		`(2)&=\sum_{t=1}^nH(\RZ_{Ut}|\RZ_U^{t-1},\RL,\RZ_D^n) \notag\\
		&=\sum_{t=1}^nH(\RZ_{Ut}|\RZ_U^{t-1},\RL,\RZ_D^n,\RZ_{Dt}) \notag\\
		&=nH(\RZ_{U\RJ}|\RW_\RJ,\RZ_{D\RJ})
	\end{align*}
	\begin{align*}
		\label{eq:cw:cons:lb:2}
		`(3)&=\sum_{t=1}^{n}H(\RZ_{Bt}|\RZ_{B}^{t-1},\RZ_{U`/B}^{n},\RL,\RZ_D^n,\RZ_{Dt}) \notag  \\
		&\leq \sum_{t=1}^{n}H(\RZ_{Bt}|\RZ_{U}^{t-1},\RZ_{\Set{U`/B} t},\RL,\RZ_D^n,\RZ_{Dt})  \notag \\
		&=nH(\RZ_{B\RJ}|\RW_\RJ,\RZ_{\Set{U`/B}\RJ},\RZ_{D\RJ})
	\end{align*}
	Altogether, we have the inequality
	\begin{align}
		\label{eq:jw:cons:ub}
		I_{`l}(\RZ_{U\RJ}|\RW_\RJ,\RZ_{D\RJ})\leq\frac 1n I_{`l}(\tRZ_U|\RL,\tRZ_D). 
	\end{align}
	Similar to the arguments in the proof of Theorem~\ref{thm:LB:A=V} in 
	Appendix~\ref{sec:LB:A=V:proof},
	by~\eqref{eq:cw:obj:lb} and~\eqref{eq:jw:cons:ub}, and the fact that $\RZ_{U\RJ}$ has the same distribution as $\RZ_U$, we have
	\begin{align}
		&C_{\op{W},`l}\geq H(\RZ_U|\RZ_D)-\lim_{`d\to 0}`G(`d) \kern1em \text{where}\notag\\
		&`G(`d):=\sup_{\substack{P_{\RW|\RZ_{U\cup D}}:\\ I_{`l}(\RZ_U|\RW,\RZ_D)\leq `d}} \kern-1em H(\RZ_U|\RZ_D,\RW).\label{eq:`G}
	\end{align}
	(In fact, the above inequality is satisfied with equality.\footnote{The reverse inequality holds by the fact $\RW^n$ i.i.d.\ generated according to the solution $P_{\RW|\RZ_{U\cup D}}$ to \eqref{eq:`G} is a feasible solution to \eqref{eq:cw}.}) Note that
	\begin{align}
		H(\RZ_U|\RZ_D)-`G(0) = J_{\opW,`l}(\RZ_U|\RZ_D) \label{eq:`GJW}
	\end{align}
	and so the proof is completed by showing that $`G(`d)$ is continuous at $`d=0$.
	To show this, we will prove the following  support-type lemma that extends Proposition~\ref{pro:|w|}, following essentialy the same argument as in \cite{wyner75}.
\end{Proof}

\begin{Lemma}
	\label{lem:|W|}
	It is admissible to impose in~\eqref{eq:`G} that 
	\begin{align}
		\label{eq:|W|}
		\abs{W}\leq \begin{cases}
			\abs{Z_{U\cup D}}+1 & `d> 0  \\
			\abs{Z_{U\cup D}}     & `d=0, 
		\end{cases}
	\end{align}
	and so $\sup$ in~\eqref{eq:`G} can be replaced by $\max$ and $`G(`d)$ is continous in $`d$.\footnote{As in \cite{wyner75}, it is also possible to argue that $`G(`d)$ is non-decreasing and concave in $`d$.}
\end{Lemma}
\begin{Proof}[Lemma~\ref{lem:|W|}]
	Pick any $\Rz'_{U\cup D}\in\RZ_{U\cup D}$, and define $\rmS$ as the set of all possible vectors of values for
	\begin{align*}
		&\big(H(\RZ_U|\RZ_D,\RW=w),I_{`l}(\RZ_U|\RZ_D,\RW=w),\\
		&P_{\RZ_{U\cup D}|\RW=w}(\Rz_{U\cup D})\mid \Rz_{U\cup D}\in \RZ_{U\cup D}`/\Set{\Rz'_{U\cup D}}\big).
	\end{align*}
	There is a one-to-one mapping between the choice of $P_{\RZ_{U\cup D}|\RW=w}$ and the choice of $\Mv(w)\in\rmS$, noting that 
	\begin{align*}
		P_{\RZ_{U\cup D}|\RW=w}(\Rz'_{U\cup D}) 
		 &= 1-\sum_{\Rz\in \RZ_{U\cup D}`/\Set{\Rz'_{U\cup D}}} \kern-1em \kern-1em P_{\RZ_{U\cup D}|\RW=w}(\Rz_{U\cup D}).
	\end{align*}
	Thus, a feasible solution to~\eqref{eq:`G} corresponds to a choice of a set $W$, a distribution $P_{\RW}$ over $W$, and a vector $\Mv(w)$ for every $w\in\RW$, such that 
	\begin{align}
		\sum P_{\RW}(w)\Mv(w) &=(H(\RZ_U|\RZ_D,\RW),I_{`l}(\RZ_U|\RZ_D,\RW),\\
		&\kern-4em P_{\RZ_{U\cup D}|\RW}(\Rz_{U\cup D})\mid \Rz_{U\cup D}\in \RZ_{U\cup D}`/\Set{\Rz'_{U\cup D}}).
	\end{align}
	By the Fenchel-Eggleston-Carath{\'e}odory theorem~\cite{eggleston1958convexity}, it is admissible to choose $\abs{W}$ equal to the length of $\Mv(w)$ plus $1$, i.e., $\abs{Z_{U\cup D}}+1$ as desired in~\eqref{eq:|W|} for $`d\geq0$. If $`d=0$, i.e., one requires $I_{`l}(\RZ_U|\RZ_D,\RW)=0$, then $I_{`l}(\RZ_U|\RZ_D,\RW=w)=0$ for all $w\in W$ since $I_{`l}$ is non-negative by Proposition~\ref{pro:shearer}. In other words, the constraint is on individual choice of $P_{\RZ_{U\cup D}|\RW=w}$ and so we can redefine $\rmS$ without having $I_{`l}(\RZ_U|\RZ_D,\RW=w)$ as a component of $\Mv(w)$, i.e., which gives the smaller bound in~\eqref{eq:|W|}. 
	
	Suppose there is a sequence in $k$ of choices of $(P_{\RW_{k}},P_{\RZ_{U\cup D}|\RW_k})$ that attains $`G(`d)$ in the limit as $k\to`8$ while satisfying the constraint in \eqref{eq:`G}, i.e.,
	\begin{align*}
		I_{`l}(\RZ_U|\RZ_D,\RW_{k})\leq `d
	\end{align*}
	By imposing~\eqref{eq:|W|} such that $W$ is finite with size independent of $k$, the feasible choices of $(P_{\RW_{k}},P_{\RZ_{U\cup D}|\RW_k})$ form a compact set. Hence, there exists a subsequence $\Set{k_j}_{j=1}^{`8}$ such that
	\begin{align}
		P_{\RW}=\lim_{j\to `8}P_{\RW_{k_j}}\kern.5em\text {and}\kern.5em P_{\RZ_{U\cup D}|\RW}=\lim_{j\to `8}P_{\RZ_{U\cup D}|\RW_{k_j}}. \label{eq:proof:LB:t2}
	\end{align}
	By the continuity of entropy~\cite{csiszar2011information}, we also have 
	\begin{subequations}
		\label{eq:proof:LB:t3}
	\begin{align}
	I_{`l}(\RZ_U|\RZ_D,\RW)&=\lim_{j\to`8} I_{`l}(\RZ_U|\RZ_D,\RW_{k_j}),\; \text{and}\label{eq:proof:LB:t3:a}\\	H(\RZ_U|\RZ_D,\RW)&=\lim_{j\to`8}H(\RZ_U|\RZ_D,\RW_{k_j}).
	\label{eq:proof:LB:t3:b}
	\end{align}
	\end{subequations}
	Note that the r.h.s.\ of \eqref{eq:proof:LB:t3:a} is upper bounded by $`d$ since each term in the limit is. Furthermore, the r.h.s.\ of \eqref{eq:proof:LB:t3:b} attains $`G(`d)$ by assumption. Hence, the supremum in~\eqref{eq:`G} is achieved by the above choice of $\RW$, i.e., the $\sup$ in~\eqref{eq:`G} can be replaced by $\max$. 
	
	Consider proving the continuity of $`G(`d)$. Consider any sequence $\Set {`d_k}_{k=1}^{`8}$ such that $`d_k>`d$ and $`d_k\downarrow `d$ as $k\uparrow `8$. Since $`G(`d)$ is non-decreasing in $`d$, we have
	\begin{align}
		`G(`d) &\leq  \lim_{k\to`8} `G(`d_k). \label{eq:proof:LB:t4}
	\end{align}
	Let $(P_{\RW_{k}},P_{\RZ_{U\cup D}|\RW_k})$ be the optimal solution for $`G(`d_k)$. Then, as argued previously, $(P_{\RW},P_{\RZ_{U\cup D}|\RW})$ exists satisfying \eqref{eq:proof:LB:t2} and \eqref{eq:proof:LB:t3} for some subsequent $\Set{k_j}_{j=1}^{`8}$. Furthermore, the r.h.s.\ of \eqref{eq:proof:LB:t3:a} is equal to $\lim_{k\to`8}`d_k=`d$, and so $\RW$ is a feasible solution to \eqref{eq:`G}. The l.h.s.\ of \eqref{eq:proof:LB:t3:b} is therefore upper bounded by $`G(`d)$ and so
	\begin{align*}
		`G(`d) &\geq  \lim_{k\to`8} `G(`d_k),
	\end{align*}
	which is satisfied with equality by  \eqref{eq:proof:LB:t4}, implying that $`G(`d)$ is continuous in $`d$.
\end{Proof}

\subsection{Proof of Theorem~\ref{thm:LB:A<V}}
\label{sec:LB:A<V:proof}

	\eqref{eq:LB:A<V:JW} follows from Theorem~\ref{thm:LB} directly since $\CS=I_{`l^{*}}(\RZ_V)$ for all $\displaystyle{`l^*\in`L^*(A,\RZ_V)}$. 
	To show~\eqref{eq:LB:A<V:I_`l}, choose $`l^{*}\in`L^*(A,\RZ_V)$ such that
	\begin{align*}
		\op{supp}(`l^{*})=\bigcup_{`l'\in`L^{*}(A,\RZ_V)}\op{supp}(`l').
	\end{align*}
	This is possible, for instance, by choosing $`l^*$ as the average of the extreme elements in $`L^*(A,\RZ_V)$, which are the vertices of the feasible set in~\eqref{eq:Cs:CN08}, and so there are only a finite number of them by~\eqref{eq:`l}. Let $\RW$ be the optimal solution to $J_{\opW,`l^*}(\RZ_V)$, and consider $`l\in`L(V,\mcH)$ with $\mcH$ defined in~\eqref{eq:H:A<V}, we then have
	\begin{align*}
		J_{\opW,`l^*}(\RZ_V)&=I(\RZ_V\wedge\RW)\\
		&=H(\RZ_V)-H(\RZ_V|\RW) \\
		&\utag{a}\geq H(\RZ_V)-\sum_{B\in \mcH}`l(B)H(\RZ_B|\RW)  \\
		&\utag{b}= H(\RZ_V)-\sum_{B\in \mcH}`l(B)H(\RZ_B|\RZ_{V`/B},\RW)  \\
		&\geq H(\RZ_V)-\sum_{B\in \mcH}`l(B)H(\RZ_B|\RZ_{V`/B})
	\end{align*}
	which gives $I_{`l}(\RZ_V)$ as desired by \eqref{eq:I_`l}.
	The inequality~\uref{a} is because of the Shearer-type Lemma~\cite{madiman10} stated in a slightly different form than Proposition~\ref{pro:shearer}:
	\begin{align*}
		\sum_{B}`l(B)H(\RZ_B|\RW)&=\sum_{B}`l(B)\sum_{i\in B}H(\RZ_i|\RZ_{[i-1]\cap B},\RW) \\
		&\geq\sum_{B}`l(B)\sum_{i\in B}H(\RZ_i|\RZ_{[i-1]},\RW)  \\
		&=\sum_{i\in V}\sum_{i\in B}`l(B)H(\RZ_i|\RZ_{[i-1]},\RW) \\
		&=\sum_{i\in V}H(\RZ_i|\RZ_{[i-1]},\RW) \\
		&=H(\RZ_V|\RW).
	\end{align*}
	The equality \uref{b} is because the definition of $J_{\opW,`l^*}(\RZ_V)$ requires $I_{`l^*}(\RZ_V|\RW)=0$, which by~Proposition~\ref{pro:shearer}, results in $I(\RZ_B\wedge\RZ_{V`/B}|\RW)=0$ for all $B\in \op{supp}(`l^*)$, and hence, for all $B \in \mcH$.

\subsection{Proofs for Section~\ref{sec:silent}}
\label{sec:silent:proof}

\begin{Proof}[Proposition~\ref{prop:CS:s}]
	Applying Theorem~\ref{thm:CSRCO} to the current case $S\subsetneq A=V$, \eqref{eq:CSRCO} becomes
	\begin{equation}
	\CS=H(\RZ_{V\setminus S})-\RCO, \label{csrco:s}
	\end{equation}
	where $\displaystyle\RCO=`r=\min_{r_{V\setminus S}}r(V\setminus S)$ subject to the constraints
	\begin{subequations}
		\label{SW}
		\begin{alignat}{2}
			&r(B)  \geq H(\RZ_B|\RZ_{(V\setminus S)\setminus B}) &\kern1em& \forall B\subsetneq V\setminus S: B\neq\emptyset \label{SW0}\\\smallskip
			&r(V\setminus S)  \geq H(\RZ_{V\setminus S}|\RZ_i) && \forall i\in S, \label{SW1}
		\end{alignat}
	\end{subequations}
	where we have used a similar argument as in the proof of Corollary~\ref{cor:CSRCO:CN04} to derive \eqref{SW0}. Note also that the set of constraints are equivalent to the those in Corollary~\ref{cor:CSRCO:Amin10} but stated in a convenient form for the current proof. 
	We proceed to prove \eqref{s:alpha} and hence assume $\abs{V\setminus S}=1$. Observe that this condition renders \eqref{SW0} obsolete and hence using \eqref{csrco:s} we have $\CS=H(\RZ_{V\setminus S})-\max_{i\in S}H(\RZ_{V\setminus S}|\RZ_{i})=\alpha$ as desired. 
	
	To complete the proof of Proposition~\ref{prop:CS:s} we consider the case when $\abs{V\setminus S}>1$. Again, we shall prove this in a case by case basis. First, consider the case when \eqref{SW1} are redundant, and hence $\RCO\geq \max_{i\in S}H(\RZ_{V\setminus S}|\RZ_i)$. Also, observe that since $\RCO=\displaystyle\min_{r_{V\setminus S}} r(V\setminus S)$, where $r_{V\setminus S}$ is constrained by the first set of constraints in \eqref{SW0}, we have $H(\RZ_{V\setminus S})-\RCO=I(\RZ_{V\setminus S})$ using Proposition~\ref{pro:I}. Therefore, using \eqref{csrco:s}, we have $\CS=I(\RZ_{V\setminus S})$. Also, from the fact that $\RCO\geq \max_{i\in S}H(\RZ_{V\setminus S}|\RZ_i)$, we have $\CS=H(\RZ_{V\setminus S})-\RCO\leq\alpha$, and hence \eqref{s:alphaI} is satisfied. We finish the proof by looking at the remaining case, i.e., when there exists some $i\in S$ such that \eqref{SW1} is not redundant. An immediate consequence of this is $\RCO=H(\RZ_{V\setminus S}|\RZ_i)$ and hence using \eqref{csrco:s} we have $\CS=\alpha$. Also, defining $\RCO'=\displaystyle\min_{r_{V\setminus S}} r(V\setminus S)$, where $r_{V\setminus S}$ is constrained by \eqref{SW0}, we see that $\RCO\geq\RCO'$. Therefore, using Proposition~\ref{pro:I}, we have $I(\RZ_{V\setminus S})\geq H(\RZ_{V\setminus S})-\RCO=\CS$. Hence, we have $\CS=\min\{\alpha,I(\RZ_{V\setminus S})\}$ as desired.
\end{Proof} 

\begin{Proof}[Theorem~\ref{thm:LB:S}]
	We first consider the case when the conditions for \eqref{RS:S:1} hold. The proof is carried out by exactly following the same steps as in the proof of Theorem~\ref{thm:LB:A=V} with the choice $\mcP=\mcP^*(\RZ_{V`/S})$. This is possible as in this case $\CS=I(\RZ_{V`/S})$ by \eqref{s:alphaI}. Similarly, we prove the result for the case when the conditions for \eqref{RS:S:2} hold, by using $\CS=I(\RZ_{V`/S}\wedge\RZ_i)$, for some $i\in S^*$, which follows from \eqref{s:alpha}.
	
	For the remaining case when $\abs{V`/S}>1$ and $\alpha=I(\RZ_{V`/S})$, we observe using \eqref{s:alphaI} that every $i\in S^*$ satisfies 
	\begin{equation}
	\CS=I_{\mcP^*}(\RZ_{V`/S})=I(\RZ_{V`/S}\wedge\RZ_i). \label{LB:S:1}
	\end{equation} 
	Corollary~5.3 of \cite{chan15mi}, says that there exists some $\theta\in(0,1)$ which satisfies $I_{\mcP}(\RZ_{(V`/S)\cup\{i\}}) = \theta I_{\mcP^*}(\RZ_{V`/S})+(1-\theta)I(\RZ_{V`/S}\wedge\RZ_i)$, with $\mcP =  \mcP^*(\RZ_{V`/S})\cup\{i\}$.
	Hence, using \eqref{LB:S:1}, we have $\CS=I_{\mcP}(\RZ_{(V`/S)\cup\{i\}})$ for every $i\in S^*$. The result now follows by proceeding along the same steps as in the proof of Theorem~\ref{thm:LB:A=V}, with the choice $\mcP=\mcP^*(\RZ_{V`/S})\cup\{i\}$, for any $i\in S^*$.
\end{Proof}

\begin{Proof}[Theorem~\ref{thm:OO:S}]
	The proof technique is similar to the proof of Theorem~\ref{thm:OO:A=V}. We use the hypothesis of Theorem~\ref{thm:OO:S} to show that the lower bound to $\RS$ obtained in Theorem~\ref{thm:LB:S} evaluates to $\RCO$. This, in conjunction with the trivial upper bound $\RS\leq\RCO$, gives us the result.
	
	We first observe that the conditions in \ref{OO:S:1} imply that $J_{\opD,\mcP^*}(\RZ_{V`/S})=H(\RZ_{V`/S})$. Hence, via \eqref{RS:S:1} and the inequality $J_{\opW,\mcP^*}(\RZ_{V `/ S}) \ge J_{\opD,\mcP^*}(\RZ_{V `/ S})$, we have $\RS \ge H(\RZ_{V`/S}) - I(\RZ_{V`/S}) = \RCO$.
	% the lower bound in \eqref{RS:S:1} evaluates to $H(\RZ_{V`/S})-I(\RZ_{V`/S})=\RCO$ using \eqref{csrco:s} and \eqref{s:alphaI}. Thus, we have $\RS=\RCO$.}
	
	Next we consider the case when the conditions in \ref{OO:S:2} hold. Therefore, there exists $i\in S^*$ satisfying $J_{\opD,\{V`/S,\{i\}\}}(\RZ_{V`/S},\RZ_i)=H(\RZ_{V`/S}.\RZ_i)-H(\RZ_i|\RZ_{V`/S})=H(\RZ_{V`/S})$. Using \eqref{csrco:s} and Proposition~\ref{prop:CS:s}, the bound in \eqref{RS:S:2} evaluates to $\RS\geq\RCO$. 
	
	To complete the proof, we look at the scenario described in \ref{OO:S:3}. Observe that there exists $i\in S^*$, such that $J_{\opD,\mcP^*(\RZ_{V`/S})\cup\{i\}}(\RZ_{(V`/S)\cup\{i\}})=H(\RZ_{V`/S},\RZ_i)-\sum_{C\in\mcP^*(\RZ_{V`/S})}H(\RZ_C|\RZ_{(V`/S)`/C},\RZ_i)-H(\RZ_i|\RZ_{V`/S})=H(\RZ_{V`/S})$. Hence, the lower bound to $\RS$ in \eqref{RS:S:3} evaluates to $\RCO$ by \eqref{s:alphaI} and \eqref{csrco:s}. Therefore, we have $\RS=\RCO$ as required.
\end{Proof}

\subsection{Proofs for Section~\ref{sec:hypsilent}}
\label{sec:proof:hypsilent}

\begin{Proof}[Proposition~\ref{prop:hypred}]
	Choose any vocal active user $j\in A\cap(V`/S)$. Observe that by \eqref{eq:recover}, it is admissible to choose the secret key $\RK=`q_j(\tRZ_j,\RF)$ for some function $`q_j$. Assume there is a hyperedge $e'$ such that $`x(e')\subseteq S$. Then, the sequence of random variables $\RX_{e'}^n$ associated with the hyperedge $e'$ is independent of $(\RK,\RF,(\RX_e^n\mid e\in E`/\{e'\}),\RU_{V`/S})$. This is because $\RX_{e'}$ is not observed by any vocal user, including $j$, who generate $\RK,\RF$ entirely from $((\RX_e^n\mid e\in E`/\{e'\}),\RU_{V`/S})$. Similarly, it can be argued that $\RX_{e'}^n$ does not play any part in recovering $\RZ_{V`/S}^n$, as it is independent of $\RX_{e'}^n$. Therefore, removing the hyperedge $e'$ does not affect $\CS,\RS$ and $\RCO$.
\end{Proof}

\begin{Proof}[Theorem~\ref{thm:hyp:LB:S}]
	Proposition~\ref{prop:hypred} ensures it is enough to prove the results for hypergraphs satisfying \eqref{hypred}. Observe that \eqref{RS:hypS:1} follows directly from \eqref{RS:S:1}. We only need to verify the other two scenarios. 
	
	We begin by arguing the following claim, that $I(\RZ_j\wedge\RZ_{(V`/S)\cup S'})=\alpha$, for all $j\in S^*$, and all $S' \subseteq S^* \setminus \{j\}$. First, assume to the contrary that we have a strict inequality ($>$) instead of an equality for some $i\in S^*$ and some $S' \subseteq S^* \setminus \{i\}$. Then, there exists a hyperedge $e' \in E$ that contributes to $I(\RZ_j\wedge\RZ_{(V`/S)\cup S'})=H(\RX_{E'})$, but not to $I(\RZ_i\wedge\RZ_{V`/S})=H(\RX_{E^{''}})$, i.e., $e' \in E' `/E^{''}$ and $E'\supseteq E^{''}$. It immediately implies that $j\in`x(e')$ and $`x(e')\subseteq S$, which violates \eqref{hypred}. Hence, we must have $I(\RZ_j\wedge\RZ_{(V`/S)\cup S'})=\alpha$, for all $j\in S^*$ and all $S' \subseteq S^* \setminus \{j\}$.
	
	Using the above claim, we proceed to prove \eqref{RS:hypS:3}. Consider any $j\in S^*$, and observe that $\alpha=I(\RZ_{V`/S}\wedge\RZ_j)=I_{\mcP^*}(\RZ_{V`/S})$, using the hypothesis of \eqref{RS:hypS:3}. Now, using Corollary~5.3 of \cite{chan15mi}, there exists $\theta\in(0,1)$ such that $I_{\mcP^*(\RZ_{V`/S})\cup\{j\}}(\RZ_{(V`/S)\cup\{j\}})=\theta I_{\mcP^*}(\RZ_{V`/S})+(1-\theta)I(\RZ_{V`/S}\wedge\RZ_j)=\alpha$. We can continue with this process inductively to show that $I_{\mcP^*(\RZ_{V`/S})\cup\{\{i\}\mid i\in S^*\}}(\RZ_{(V`/S)\cup S^*})=\alpha=\CS$. Using this, one can proceed along similar steps as in the proof of Theorem~\ref{thm:LB:A=V} to obtain \eqref{RS:hypS:3}.
	
	The proof of \eqref{RS:hypS:2} follows using a similar inductive argument and we omit the details.
\end{Proof}

\begin{Proof}[Theorem~\ref{thm:hyp:OO:S}]
	To begin with, we restrict our attention to hypergraphs satisfying \eqref{hypred}. This is because of Proposition~\ref{prop:hypred} and the fact that none of the entropy terms in \ref{OO:hypS:1}-\ref{OO:hypS:3} are affected by the removal of some hyperedge $e$ satisfying $`x(e)\subseteq S$.
	
	We omit the proof of the fact that $\RS=\RCO$ if the required condition from \ref{OO:hypS:1}-\ref{OO:hypS:3} hold, by noting that the proof follows from Theorem~\ref{thm:hyp:LB:S} by the same steps as in the proof of Theorem~\ref{thm:OO:S}. We focus on proving the fact that $\RS=\RCO$ implies that the required condition from \ref{OO:hypS:1}-\ref{OO:hypS:3} hold. We proceed according to a case by case basis.
	
	\emph{Case I}: $\abs{V`/S}>1$ and $I(\RZ_{V`/S})<\alpha$. 
	
	We assume that \ref{OO:hypS:1} does not hold. We will show that $\RS<\RCO$. Then, there exists $e'\in E$ such that $`x(e')`/S\subseteq C$, for some $C\in\mcP^*(V`/S)$. We use the idea of \emph{decremental secret key agreement} as in \cite{chan16isit} to reduce $H(\RX_{e'})$ by an amount $\epsilon\in(0,\alpha-I(\RZ_{V`/S}))$. Whereas, this operation does not affect $I(\RZ_{V`/S})$, we note that $\alpha$ changes by at most $\epsilon$, thereby keeping $\CS$ unaffected. However, $H(\RZ_{V`/S})$ does decrease by $\epsilon$, and the fact that $\CS$ remains unchanged implies that $\RCO$ reduces by $\epsilon$ using \eqref{csrco:s}. Thus, we must have $\RS$ being strictly less than the $\RCO$ before the reduction by $\epsilon$.
	
	\emph{Case II}: $\abs{V`/S}=1$ or, when $\abs{V`/S}>1$ and $I(\RZ_{V`/S})>\alpha$.

	Here, we drop the case when $\abs{V`/S}=1$ as the condition holds by default. 
	
	Again, assume \ref{OO:hypS:2} does not hold. Then, there exists a hyperedge $e'\in E$ such that $`x(e')\subseteq (V`/S^*)$. We can reduce the entropy of $\RX_{e'}$ by some $`e>0$ small enough without affecting the secrecy capacity using decremental secret key agreement of \cite{chan16isit}. If 
	$\abs{V`/S}=1$, we can choose any $`e\in (0,\min_{i\in S/S^*} I(\RZ_{V`/S}\wedge \RZ_i)-`a)$ as the reduction in entropy will not affect the set $S^*$ of optimal solutions and therefore $`a$. In the other case  $\abs{V`/S}>1$ and $I(\RZ_{V`/S})>\alpha$, we impose an additional constraint that 
	$`e < I(\RZ_{V`/S})-\alpha$. Then, $\alpha$ remains unaffected after the reduction in entropy, whereas $I(\RZ_{V`/S})$ decreases by at most $\epsilon$. Thus, $\CS$ remains unchanged. Moreover, the fact that \eqref{hypred} holds implies $H(\RZ_{V`/S})$ reduces by $\epsilon$, and so does $\RCO$ using \eqref{csrco:s}. Therefore, we must have $\RS<\RCO$ before reduction. 
	
	\emph{Case III}: $\abs{V`/S}>1$ and $I(\RZ_{V`/S})=\alpha$
	
	Assume \ref{OO:hypS:3} is invalid and hence, there exists $e'\in E$ such that $`x(e')\subseteq C$ for some $C\in\mcP^*(\RZ_{V`/S})$. We reduce the entropy of $\RX_e$ by some amount of $\epsilon>0$. While $\alpha$ remains unaffected by the operation, the decremental secret key agreement detailed in \cite{chan16isit} ensures that choosing $\epsilon$ sufficiently small not affect $I(\RZ_{V`/S})$ either. Thus, $\CS$ is unaffected. However, clearly $H(\RZ_{V`/S})$ reduces by $\epsilon$ and so does $\RCO$. Hence, $\RS<\RCO$ before reduction as required.
\end{Proof}

\section{Proof for Section~\ref{sec:challenge}}

\subsection{Proof of Proposition~\ref{pro:snn}}
\label{sec:proof:snn}

	%Since $H(\RZ_1|\RZ_4)=H(\RZ_2|\RZ_4)=0$, users~$1$ and $2$ need not discuss as they can be simulated by user $4$. 
	To prove the desired result, we will make use of the following independence relation satisfied by the private source:
	\begin{align}
		\kern-.5em 0=I(\RZ_1\wedge \RZ_2)=I(\RZ_3\wedge \RZ_{\Set{1,2,4}})=I(\RZ_3\wedge \RZ_{\Set{1,2,5}}).\kern-.5em \label{eq:snn:src}
	\end{align}
 	The desired conclusion will be proved by showing the stronger result that
	\begin{align}
		\limsup_{n\to `8} \frac1n `1[ H(\RF_{\Set{4,5}}) - 3H(\RK)`2]\geq 0
		\label{eq:snn:FK}
	\end{align}
	which implies $\RS\geq 3\CS=3=\RCO$ as desired.
	
	To prove the above, define
	\begin{subequations}\label{eq:at_bt_ct}
		\begin{align}
			a_t:&=I(\tRZ_1\wedge\tRZ_2|\RF^t_V)-I(\tRZ_1\wedge\tRZ_2|\RF^{t-1}_V)\label{eq:a_t}\\
			b_t:&=I(\tRZ_3\wedge\tRZ_{\Set{1,2,4}}|\RF^t_V)-I(\tRZ_3\wedge\tRZ_{\Set{1,2,4}}|\RF^{t-1}_V)\label{eq:b_t}\\
			c_t:&=I(\tRZ_3\wedge\tRZ_{\Set{1,2,5}}|\RF^t_V)-I(\tRZ_3\wedge\tRZ_{\Set{1,2,5}}|\RF^{t-1}_V)\label{eq:c_t}
		\end{align}\notag
	\end{subequations}
	By definition of \eqref{eq:at_bt_ct}, we have 
	\begin{align*}
		&\sum_{t=1}^{r}(a_t+b_t+c_t)\\
		&=I(\tRZ_1\wedge\tRZ_2|\RF)+I(\tRZ_3\wedge\tRZ_{\Set{1,2,4}}|\RF)+I(\tRZ_3\wedge\tRZ_{\Set{1,2,5}}|\RF)\\
		&\geq3H(\RK)-3n\delta_n
	\end{align*}
	for some $`d_n\to 0$ as $n\to `8$. Here, the inequality follows from the recoverability~\eqref{eq:recover} and secrecy~\eqref{eq:secrecy} requirement, for instance, $I(\tRZ_1\wedge\tRZ_2|\RF)\geq I(\tRZ_1,\RK\wedge\tRZ_2,\RK|\RF)-\frac{n`d_n}2\geq H(\RK)-n\delta_n$. Then, it suffices to show that
	\begin{align}
		H(\RF_{\Set{4,5}})\geq \sum_{t=1}^{r}(a_t+b_t+c_t).
	\end{align}
	To achieve this, we will bound $a_t, b_t$ and $c_t$ one by one. We first bound $a_t$ as follows:
	\begin{align*}
		a_t&\utag{a}=I(\RF_{Vt}\wedge\tRZ_{2}|\RF^{t-1}_V,\tRZ_1)-I(\RF_{Vt}\wedge\tRZ_{2}|\RF^{t-1}_V)\\
		&\utag{b}= I(\RF_{Vt}\wedge\tRZ_{\Set{1,2}}|\RF^{t-1}_V)-I(\RF_{Vt}\wedge\tRZ_{1}|\RF^{t-1}_V)\\
		&\kern1em-I(\RF_{Vt}\wedge\tRZ_{2}|\RF^{t-1}_V)\\
		&\utag{c}= I(\RF_{\Set{1,2}t}\wedge\tRZ_{\Set{1,2}}|\RF^{t-1}_V)\\
		&\kern1em+I(\RF_{\Set{3,4,5}t}\wedge\tRZ_{\Set{1,2}}|\RF^{t-1}_V,\RF_{\Set{1,2}t})\\
		&\kern1em-I(\RF_{Vt}\wedge\tRZ_{1}|\RF^{t-1}_V)-I(\RF_{Vt}\wedge\tRZ_{2}|\RF^{t-1}_V)\\
		&\utag{d}= H(\RF_{\Set{1,2}t}|\RF^{t-1}_V)+I(\RF_{\Set{3,4,5}t}\wedge\tRZ_{\Set{1,2}}|\RF^{t-1}_V,\RF_{\Set{1,2}t})\\
		&\kern1em-I(\RF_{Vt}\wedge\tRZ_{1}|\RF^{t-1}_V)-I(\RF_{Vt}\wedge\tRZ_{2}|\RF^{t-1}_V)\\
		&\utag{e}\leq I(\RF_{\Set{3,4,5}t}\wedge\tRZ_{\Set{1,2}}|\RF^{t-1}_V,\RF_{\Set{1,2}t})
	\end{align*}
	where \uref{a} is due to the fact that 
		\begin{align*}
		&I(\tRZ_1,\RF_{Vt}\wedge\tRZ_{2}|\RF^{t-1}_V)\notag\\
		&=I(\tRZ_1\wedge\tRZ_{2}|\RF^{t-1}_V)+I(\RF_{Vt}\wedge\tRZ_{2}|\RF^{t-1}_V,\tRZ_1)\notag\\
		&=I(\RF_{Vt}\wedge\tRZ_{2}|\RF^{t-1}_V)+I(\tRZ_1\wedge\tRZ_{2}|\RF^{t}_V),
	\end{align*}
	\uref{b} and \uref{c} are due to the chain rule expansion,
	\uref{d} is due to the fact that
	\begin{align*}
	I(\RF_{\Set{1,2}t}\wedge\tRZ_{\Set{1,2}}|\RF^{t-1}_V)=H(\RF_{\Set{1,2}t}|\RF^{t-1}_V)
	\end{align*}
	by \eqref{eq:F}, \uref{e} is due to the fact that
	\begin{align*}
	&I(\RF_{Vt}\wedge\tRZ_{1}|\RF^{t-1}_V)+I(\RF_{Vt}\wedge\tRZ_{2}|\RF^{t-1}_V)\\
	&\geq I(\RF_{1t}\wedge\tRZ_{1}|\RF^{t-1}_V)+I(\RF_{\Set{1,2}t}\wedge\tRZ_{2}|\RF^{t-1}_V)\\
	&\geq I(\RF_{1t}\wedge\tRZ_{1}|\RF^{t-1}_V)+I(\RF_{2t}\wedge\tRZ_{2}|\RF^{t-1}_V,\RF_{1t})\\
	&=H(\RF_{1t}|\RF^{t-1}_V)+H(\RF_{2t}|\RF^{t-1}_V,\RF_{1t})\\
	&=H(\RF_{\Set{1,2}t}|\RF^{t-1}_V)
	\end{align*}
	We then bound $b_t$ as follows:
	\begin{align*}
		b_t&\utag{a}=I(\RF_{Vt}\wedge\tRZ_{\Set{1,2,4}}|\RF^{t-1}_V,\tRZ_3)-I(\RF_{Vt}\wedge\tRZ_{\Set{1,2,4}}|\RF^{t-1}_V)\\
		&\utag{b}=I(\RF_{Vt}\wedge\tRZ_{\Set{1,2,3,4}}|\RF^{t-1}_V)-I(\RF_{Vt}\wedge\tRZ_{3}|\RF^{t-1}_V)\\
		&\kern1em-I(\RF_{Vt}\wedge\tRZ_{\Set{1,2,4}}|\RF^{t-1}_V)\\
		&\utag{c}\leq H(\RF_{Vt}|\RF^{t-1}_V,\tRZ_{\Set{1,2,4}})-I(\RF_{Vt}\wedge\tRZ_{3}|\RF^{t-1}_V)\\
		&\utag{d}= H(\RF_{3t}|\RF^{t-1}_V,\tRZ_{\Set{1,2,4}},\RF_{\Set{1,2}t})\\
		&\kern1em+H(\RF_{4t}|\RF^{t-1}_V,\tRZ_{\Set{1,2,4}},\RF_{\Set{1,2,3}t})\\
		&\kern1em+H(\RF_{5t}|\RF^{t-1}_V,\tRZ_{\Set{1,2,4}},\RF_{\Set{1,2,3,4}t})-I(\RF_{Vt}\wedge\tRZ_{3}|\RF^{t-1}_V)\\
		&\utag{e}=H(\RF_{3t}|\RF^{t-1}_V,\tRZ_{\Set{1,2,4}},\RF_{\Set{1,2}t})\\
		&\kern1em+H(\RF_{5t}|\RF^{t-1}_V,\tRZ_{\Set{1,2,4}},\RF_{\Set{1,2,3,4}t})-I(\RF_{Vt}\wedge\tRZ_{3}|\RF^{t-1}_V)\\
		&\utag{f}\leq H(\RF_{3t}|\RF^{t-1}_V,\tRZ_{\Set{1,2}},\RF_{\Set{1,2}t})\\
		&\kern1em+H(\RF_{5t}|\RF^{t-1}_V,\tRZ_{\Set{1,2}},\RF_{\Set{1,2,3,4}t})-I(\RF_{Vt}\wedge\tRZ_{3}|\RF^{t-1}_V)\\
		&\utag{g}\leq H(\RF_{3t}|\RF^{t-1}_V,\tRZ_{\Set{1,2}},\RF_{\Set{1,2}t})\\
		&\kern1em+H(\RF_{5t}|\RF^{t-1}_V,\tRZ_{\Set{1,2}},\RF_{\Set{1,2,3,4}t})-H(\RF_{3t}|\RF^{t-1}_V,\RF_{\Set{1,2}t})
	\end{align*}
	where \uref{a} is due to the fact that 
	\begin{align*}
		&I(\tRZ_3,\RF_{Vt}\wedge\tRZ_{\Set{1,2,4}}|\RF^{t-1}_V)\notag\\
		&=I(\tRZ_3\wedge\tRZ_{\Set{1,2,4}}|\RF^{t-1}_V)+I(\RF_{Vt}\wedge\tRZ_{\Set{1,2,4}}|\RF^{t-1}_V,\tRZ_3)\notag\\
		&=I(\RF_{Vt}\wedge\tRZ_{\Set{1,2,4}}|\RF^{t-1}_V)+I(\tRZ_3\wedge\tRZ_{\Set{1,2,4}}|\RF^{t}_V),
	\end{align*}
	\uref{b} is due to the chain rule expansion,
	\uref{c} is due to the fact that 
	\begin{align*}
		I(\RF_{Vt}\wedge\tRZ_{\Set{1,2,3,4}}|\RF^{t-1}_V)\leq H(\RF_{Vt}|\RF^{t-1}_V),
	\end{align*}
	\uref{d} is due to the chain rule expansion and the fact that
	\begin{align*}
		H(\RF_{\Set{1,2}t}|\RF^{t-1}_V,\tRZ_{\Set{1,2,4}})=0
	\end{align*}
	by \eqref{eq:F}, Similarly, \uref{e} follows from \eqref{eq:F} that
	\begin{align*} 
		&H(\RF_{4t}|\RF^{t-1}_V,\tRZ_{\Set{1,2,4}},\RF_{\Set{1,2,3}t})=0
	\end{align*}
	\uref{f} follows from the fact that conditioning cannot increase entropy,
	\uref{g} is because
	\begin{align*}
	I(\RF_{Vt}\wedge\tRZ_{3}|\RF^{t-1}_V)&\geq I(\RF_{\Set{1,2,3}t}\wedge\tRZ_{3}|\RF^{t-1}_V) \\
		&\geq I(\RF_{3t}\wedge\tRZ_{3}|\RF^{t-1}_V,\RF_{\Set{1,2}t})\notag\\
		&=H(\RF_{3t}|\RF^{t-1}_V,\RF_{\Set{1,2}t})
	\end{align*} 
	by \eqref{eq:F}.
	
	Following similar steps as above, $c_t$ is also upper bounded by
	\begin{align*}
		c_t&\leq H(\RF_{\Set{3,4}t}|\RF^{t-1}_V,\tRZ_{\Set{1,2,5}},\RF_{\Set{1,2}t})\\
		&\kern1em+H(\RF_{5t}|\RF^{t-1}_V,\tRZ_{\Set{1,2,5}},\RF_{\Set{1,2,3,4}t}) \\
		&\kern1em-H(\RF_{3t}|\RF^{t-1}_V,\RF_{\Set{1,2}t})\\
		&\leq H(\RF_{\Set{3,4}t}|\RF^{t-1}_V,\tRZ_{\Set{1,2}},\RF_{\Set{1,2}t})-H(\RF_{3t}|\RF^{t-1}_V,\RF_{\Set{1,2}t})
	\end{align*}
	Therefore, we have $a_t+b_t+c_t$
	\begin{align*}
		&\leq I(\RF_{\Set{3,4,5}t}\wedge\tRZ_{\Set{1,2}}|\RF^{t-1}_V,\RF_{\Set{1,2}t})\\
		&\kern1em+H(\RF_{3t}|\RF^{t-1}_V,\tRZ_{\Set{1,2}},\RF_{\Set{1,2}t})\\
		&\kern1em+H(\RF_{5t}|\RF^{t-1}_V,\tRZ_{\Set{1,2}},\RF_{\Set{1,2,3,4}t})\\
		&\kern1em+H(\RF_{\Set{3,4}t}|\RF^{t-1}_V,\tRZ_{\Set{1,2}},\RF_{\Set{1,2}t})-2H(\RF_{3t}|\RF^{t-1}_V,\RF_{\Set{1,2}t})\\
		&\utag{a}\leq H(\RF_{\Set{3,4,5}t}|\RF^{t-1}_V,\RF_{\Set{1,2}t})-H(\RF_{3t}|\RF^{t-1}_V,\RF_{\Set{1,2}t})\\
		&=H(\RF_{\Set{4,5}t}|\RF^{t-1}_V,\RF_{\Set{1,2,3}t})
	\end{align*} 
	where~\uref{a} is because
	\begin{align*}
		I(\RF_{\Set{3,4,5}t}\wedge&\tRZ_{\Set{1,2}}|\RF^{t-1}_V,\RF_{\Set{1,2}t})\\
		&=H(\RF_{\Set{3,4,5}t}|\RF^{t-1}_V,\RF_{\Set{1,2}t})\\
		&\kern1em-H(\RF_{\Set{3,4,5}t}|\RF^{t-1}_V,\RF_{\Set{1,2}t},\tRZ_{\Set{1,2}}), 
	\end{align*}
	\begin{align*}
		H(\RF_{3t}|\RF^{t-1}_V,\tRZ_{\Set{1,2}},\RF_{\Set{1,2}t})&\leq H(\RF_{3t}|\RF^{t-1}_V,\RF_{\Set{1,2}t}),
	\end{align*}
	\begin{align*}
		H(\RF_{\Set{3,4,5}t}|&\RF^{t-1}_V,\RF_{\Set{1,2}t},\tRZ_{\Set{1,2}})\\
		&=H(\RF_{5t}|\RF^{t-1}_V,\tRZ_{\Set{1,2}},\RF_{\Set{1,2,3,4}t})\\
		&\kern1em+H(\RF_{\Set{3,4}t}|\RF^{t-1}_V,\tRZ_{\Set{1,2}},\RF_{\Set{1,2}t})
	\end{align*} 
	Finally,
	\begin{align*}
		H(\RF_{\Set{4,5}})&=\sum_{t=1}^{r}H(\RF_{\Set{4,5}t}|\RF_{\Set{4,5}}^{t-1})\\
		&\geq \sum_{t=1}^{r}H(\RF_{\Set{4,5}t}|\RF^{t-1}_V,\RF_{\Set{1,2,3}t})\\
		&\geq \sum_{t=1}^{r}(a_t+b_t+c_t),
	\end{align*}
	which completes the proof.

\subsection{Proofs of Theorems~\ref{thm:LB:user1} and \ref{thm:LB:user2}}
\label{sec:proof:po:improve}

\begin{Proof}[Theorem~\ref{thm:LB:user1}]
	We prove the cases one by one:
	\begin{compactenum}[(i)]
		\item We first show that an achieving scheme for the original scenario is an achieving scheme for the new scenario. To satisfy~\eqref{eq:F}, the discussion by the original vocal untrusted user $i$ can be done by the new vocal trusted helper $i'$. \eqref{eq:recover} and~\eqref{eq:secrecy} still hold because there is no change to $(A,D)$. Hence, $\CS$ does not decrease and $\RS$ does not increase. 
		
		To prove the reverse inequalities, consider an achieving scheme for the new scenario. By Proposition~\ref{pro:USD}, it suffices to show that the scheme can be applied to the original scenario, with private randomization allowed for the untrusted user. To satisfy~\eqref{eq:F}, the discussion and private randomization by the new user $i'$ can be done by the original vocal untrusted user. \eqref{eq:recover} and~\eqref{eq:secrecy} continue to hold trivially.
		\item Similar to the above case, the vocal user $j$ can play the role of the removed trusted helper $i$ in terms of private randomization and public discussion, and so~\eqref{eq:F} can be satisfied.  \eqref{eq:recover} and~\eqref{eq:secrecy} remain unchanged since $(A,D)$ remains unchanged.
	\end{compactenum}
\end{Proof}
\begin{Proof}[Theorem~\ref{thm:LB:user2}]
	It suffices to show that an achieving scheme for the original scenario can be applied to the new scenario. 
	\begin{compactenum}[(i)]
		\item \eqref{eq:F} continues to hold as the set $V`/ S$ of vocal users remains unchanged. \eqref{eq:recover} and~\eqref{eq:secrecy} also hold as they can only be less stringent with $(A,D)$ diminished.
		\item \eqref{eq:F} continues to hold because the set $V`/ S$ of vocal users becomes larger.\eqref{eq:recover} and~\eqref{eq:secrecy} remain unchanged trivially.
	\end{compactenum}
\end{Proof}

%\section*{Acknowledgment} 
%%\addcontentsline{toc}{section}{Acknowledgment}
%
%%\input{ack}
%

%\bibliographystyle{hieeetr}

\bibliographystyle{IEEEtran}
\bibliography{IEEEabrv,ref}

\end{document}